\pdfoutput=1
\documentclass{emulateapj}
\usepackage{natbib}
\usepackage{epsfig}

\newcommand{\acounits}{\mbox{M$_\odot$ pc$^{-2}$ (K km s$^{-1}$)$^{-1}$}}

\shorttitle{The Molecular ISM Fueling the NGC 253 Starburst}
\shortauthors{Leroy et al.}

\begin{document}

\slugcomment{Accepted for publication in the Astrophysical Journal}
\title{ALMA Reveals the Molecular Medium Fueling the Nearest Nuclear Starburst}

\author{Adam K. Leroy\altaffilmark{1}, 
Alberto D. Bolatto\altaffilmark{2,3}, 
Eve C. Ostriker\altaffilmark{4}, 
Erik Rosolowsky\altaffilmark{5}, 
Fabian Walter\altaffilmark{6}, 
Steven R. Warren\altaffilmark{2}, 
Jennifer Donovan Meyer\altaffilmark{1}, 
Jacqueline Hodge\altaffilmark{1,7}, 
David S. Meier\altaffilmark{8,9},
J\"{u}rgen Ott\altaffilmark{9}, 
Karin Sandstrom\altaffilmark{10},
Andreas Schruba\altaffilmark{11},
Sylvain Veilleux\altaffilmark{2}, 
Martin Zwaan\altaffilmark{12}}

\altaffiltext{1}{National Radio Astronomy Observtory, 520 Edgemont Road, Charlottesville, VA 22903, USA}
\altaffiltext{2}{Department of Astronomy, Laboratory for Millimeter-wave Astronomy, and Joint Space Institute, University of Maryland, College Park, Maryland 20742, USA}
\altaffiltext{3}{Visiting Humboldt Fellow, Max-Planck Institute for Astronomy, Heidelberg, Germany}
\altaffiltext{4}{Department of Astrophysical Sciences, Princeton University, Princeton, New Jersey 08544, USA}
\altaffiltext{5}{Department of Physics, University of Alberta, Edmonton, AB, Canada}
\altaffiltext{6}{Max Planck Institute f\"ur Astronomie, K\"onigstuhl 17, 69117, Heidelberg, Germany}
\altaffiltext{7}{Jansky Fellow}
\altaffiltext{8}{New Mexico Institute of Mining \& Technology, 801 Leroy Place, Socorro, NM 87801, USA}
\altaffiltext{9}{National Radio Astronomy Observatory, PO Box O, 1003 Lopezville Road, Socorro, New Mexico 87801, USA}
\altaffiltext{10}{Steward Observatory, University of Arizona, 933 North Cherry Avenue, Tucson, AZ 85721, USA}
\altaffiltext{11}{Max-Planck-Institut fŸr Extraterrestrische Physik (MPE), Giessenbachstr., D-85748 Garching, Germany}
\altaffiltext{12}{European Southern Observatory, Karl-Schwarzschild-Strasse 2, 85748 Garching, Germany}

\begin{abstract}
We use new ALMA observations to derive the mass, length, and time scales associated with the disk and
the star-forming clouds in the starburst at the heart of the nearby spiral galaxy NGC~253. This region
forms $\sim 2$~M$_{\odot}$~yr$^{-1}$ of stars and resembles other starburst systems in star formation-CO-HCN scaling relations,
with star formation consuming the gas reservoir at a normalized rate $10$ times higher than in normal galaxy disks.
 We present sensitive $\sim 35$~pc resolution observations of the bulk gas tracers CO and C$^{17}$O and the high critical density transitions 
HCN (1-0), HCO$^+$ (1-0), and CS (2-1) and their isotopologues. The starburst is fueled by a highly inclined distribution of dense 
gas with vertical extent $< 100$~pc and radius $\sim 250$~pc. Within this region, we identify ten starburst giant molecular clouds that appear
as both peaks in the dense gas tracer cubes and enhancements in the HCN-to-CO ratio map. These clouds appear as massive 
($\sim 10^7$~M$_\odot$) structures with sizes ($\sim 30$~pc) similar to GMCs in other systems, but when compared to a large literature
compilation they show very high line widths ($\sigma \sim 20$--$40$~km~s$^{-1}$) given their size, with implied Mach numbers as high as
$\mathcal{M} \sim 90$. The clouds also show very high surface densities ($\sim 6,000$~M$_\odot$~pc$^{-2}$) and volume densities 
($n_{\rm H2} \sim 2,000$~cm$^{-3}$). The self gravity from such high densities is sufficient to explain the high line widths and the short free fall 
time $\tau_{\rm ff} \sim 0.7$~Myr in the clouds may explain the more efficient star formation in NGC 253. We also consider the starburst region
as a whole. The geometry is confused by the high inclination, but we show that simple models support a non-axisymmetric, bar-like geometry
with a compact, clumpy region of high gas density embedded in an extended CO distribution. Even when considering the region as a whole,
the surface density still vastly exceeds that of a typical disk galaxy GMC. As in the clouds, timescales in the disk as a whole are short compared
to those in normal galaxy disks. The orbital time ($\sim 10$~Myr), disk free fall time ($\lesssim 3$~Myr), and disk crossing 
time ($\lesssim 3$~Myr) are each an order of magnitude shorter than in a normal spiral galaxy disk. We compare to simple models with mixed success, showing
that some but not all aspects of the structure correspond to the predictions from assuming vertical dynamical equilibrium or a marginally
stable thin gas disk. Finally, the CO-to-H$_2$ conversion factor implied by our cloud calculations is approximately Galactic, contrasting
with results showing a low value for the whole starburst region. The contrast provides resolved support for the idea of mixed molecular ISM phases in starburst
galaxies.
\end{abstract}

\keywords{}

\section{Introduction}
\label{sec:intro}

Stars form out of clouds composed mostly of molecular gas, H$_2$, so that the
molecular ISM represents the immediate reservoir for star formation in galaxies
\citep[e.g.,][]{SCHRUBA11}. The rate of star formation per unit mass of molecular gas appears to vary
weakly across the disks of normal galaxies in the local universe  but shows clear,
substantial enhancements in merging galaxies and the high surface brightness central regions
of many galaxies \citep[][and references therein]{SAINTONGE12,LEROY13}. A similar dichotomy appears at high redshift, with observations
indicating a higher rate of star formation per unit molecular gas in submillimeter galaxies
than ``main sequence'' star-forming spirals \citep[e.g., see the review by][]{CARILLI13}. 

A higher rate of star formation per unit molecular gas implies real physical changes in
the star formation process, not just a one-to-one scaling of star formation with
the available mass reservoir. These changes have wide-ranging implications for how galaxies grow and evolve. The shorter time 
to consume the gas reservoir has implications for the evolution of the galaxy population \citep[e.g.,][]{BAUERMEISTER10,KRUMHOLZ12}.
Differences in the character of molecular clouds during early universe starbursts have
been invoked as drivers for both bottom-heavy \citep{CONROY12} and top-heavy \citep{SHIM11}
stellar initial mass functions. Galactic winds with significant mass loading factors, apparently driven by intense star formation, 
have been observed at both low \citep[e.g.,][]{CHUNG11,BOLATTO13A} and high \citep[e.g.,][]{NEWMAN12} redshift. Such
winds deplete the available gas reservoir and enrich the intergalactic medium.

Stars form in bound, dense clouds and the contrast between clouds in starburst galaxies and their counterparts in more
quiescent disk galaxies must drive many of the differences between the two classes of systems. However,
major mergers and massive starbursts are relatively rare in the local universe, and thus typically distant. This distance,
combined with the limited sensitivity and resolution of millimeter-wave telescopes, means that
so far we have only a limited understanding of the {\em resolved} properties of the star-forming molecular structures that fuel
starburst galaxies. The Atacama Large Millimeter/submillimeter Array (ALMA) changes this, at last offering the power
to resolve the star-forming structures in many transitions in many starburst systems. 

To this end, we have used ALMA to observe the star-forming clouds in the nuclear starburst at the center of the
nearby \citep[$d=3.5$~Mpc,][]{REKOLA05} spiral galaxy NGC 253. In this paper, we present ALMA observations of
a suite of molecular lines in NGC 253 and use these to contrast the star-forming structures in NGC 253 with 
clouds in other systems. Based on these comparisons, we address the following questions:

\begin{itemize}
\item What are the mass, surface density, volume density, size, and line widths of the clouds driving the starburst in NGC 253? 
 How do these contrast with the values found in other environments? What is the degree of turbulence in these clouds, 
 gauged from the size-line width relation and their Mach number? Do they appear bound? Does external pressure play a large role?
 Given our estimated masses, what are the mass-to-light ratios (``conversion factors'') for individual lines?

\item What is the structure of the nuclear starburst? What are its geometry and kinematics, mass and surface density,
and how do the associated orbital, crossing, and free-fall times compare to the gas depletion time for the starburst?
How does the structure visible in CO contrast with that seen observing dense gas tracers? Do these agree with
theoretical expectations \citep[e.g.,][]{OSTRIKER11,KRUMHOLZ12B}?

\item  How do the properties of the disk and clouds in the starburst contrast with those in more quiescent star-forming 
environments? We compare the molecular gas depletion time to the dense gas fraction, the 
orbital timescale, the crossing time, and free-fall time in the disk and clouds. These 
quantities have been argued to set the maximum rate of star formation in molecular gas, and thus 
--- in combination with an efficiency factor determined by turbulent processes --- to drive the gas depletion time
\citep[see][among others]{GAO04,LADA12,GENZEL10,KRUMHOLZ12B}.
\end{itemize}

Addressing these questions involves estimating mass, time, and length scales of the molecular medium in the starburst
from two points of view:  a ``cloud view'' and a ``disk view.'' This paper presents both views. After providing some background on
nuclear starbursts and establishing NGC 253 as a representative of the class (\ref{sec:burstpop}), we present our new ALMA 
maps of NGC 253 (\S \ref{sec:data}). We then present results for the ``cloud'' and ``disk'' views in succession. In \S \ref{sec:clouds},
we treat the observed emission as a collection of molecular clouds. In \S \ref{sec:starburst}, we treat the nuclear disk as a more 
continuous distribution. In \S \ref{sec:aco_results} we discuss the implications of our work for using CO to estimate molecular masses.
We combine these views and present our conclusions in \S \ref{sec:discussion}.

This paper builds on a number of studies, including \citet{PAGLIONE01}, \citet{KNUDSEN07}, and \citet{SAKAMOTO11}, who all consider 
conditions in the NGC 253 starburst as revealed by molecular line observations of high critical density tracers. Our parameterization of the starburst 
complements \citet{JOGEE05} and \citet{SHETH05}, who considered similar topics in a sample of nuclear gas concentrations at much coarser 
resolution. Many of the same topics are investigated in a series of papers by the Nuclei of Galaxies (NUGA) collaboration 
\citep[][and following]{GARCIABURILLO03}.

\subsection{NGC 253 as a Nuclear Starburst}
\label{sec:burstpop}

\begin{figure*}
\plottwo{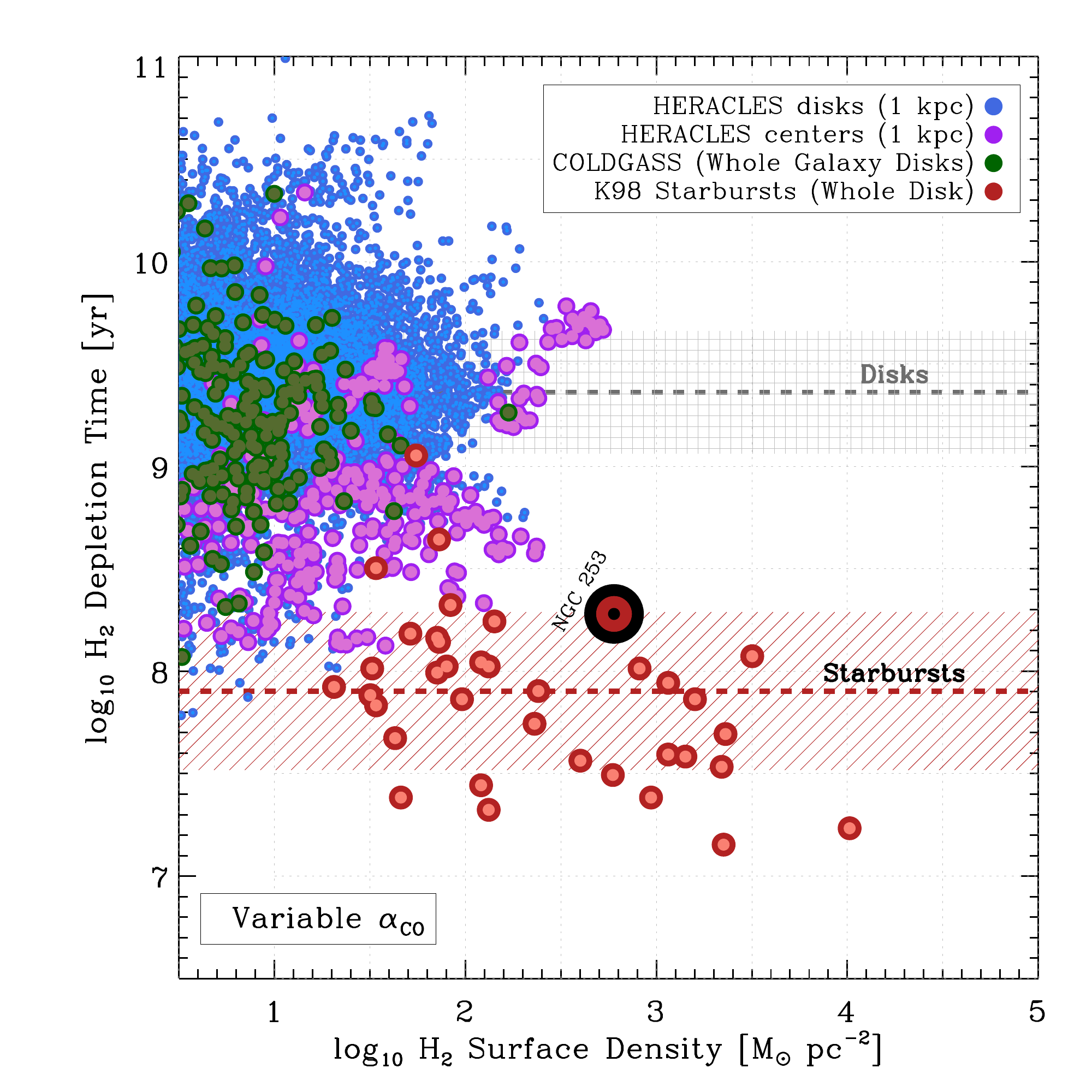}{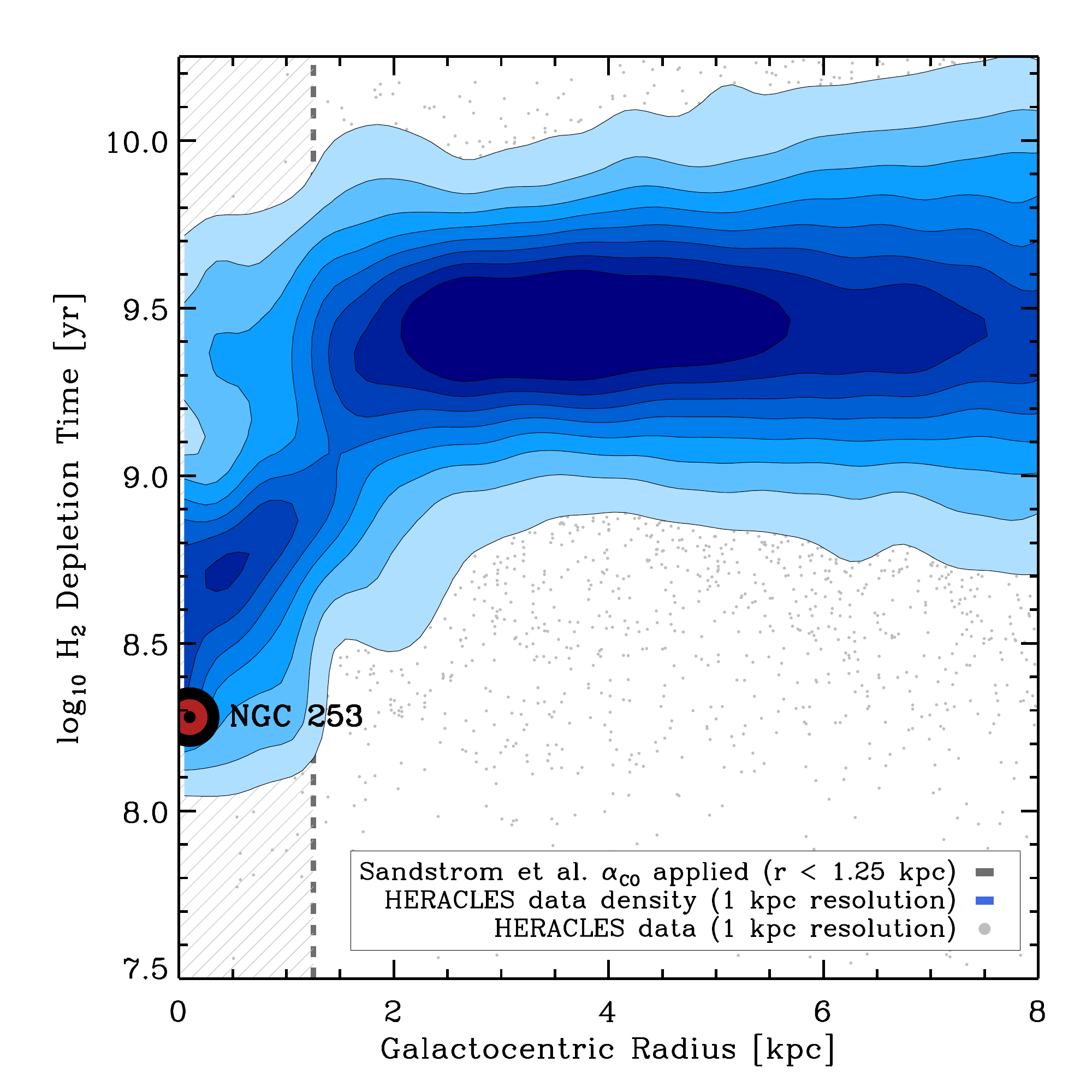}
\plottwo{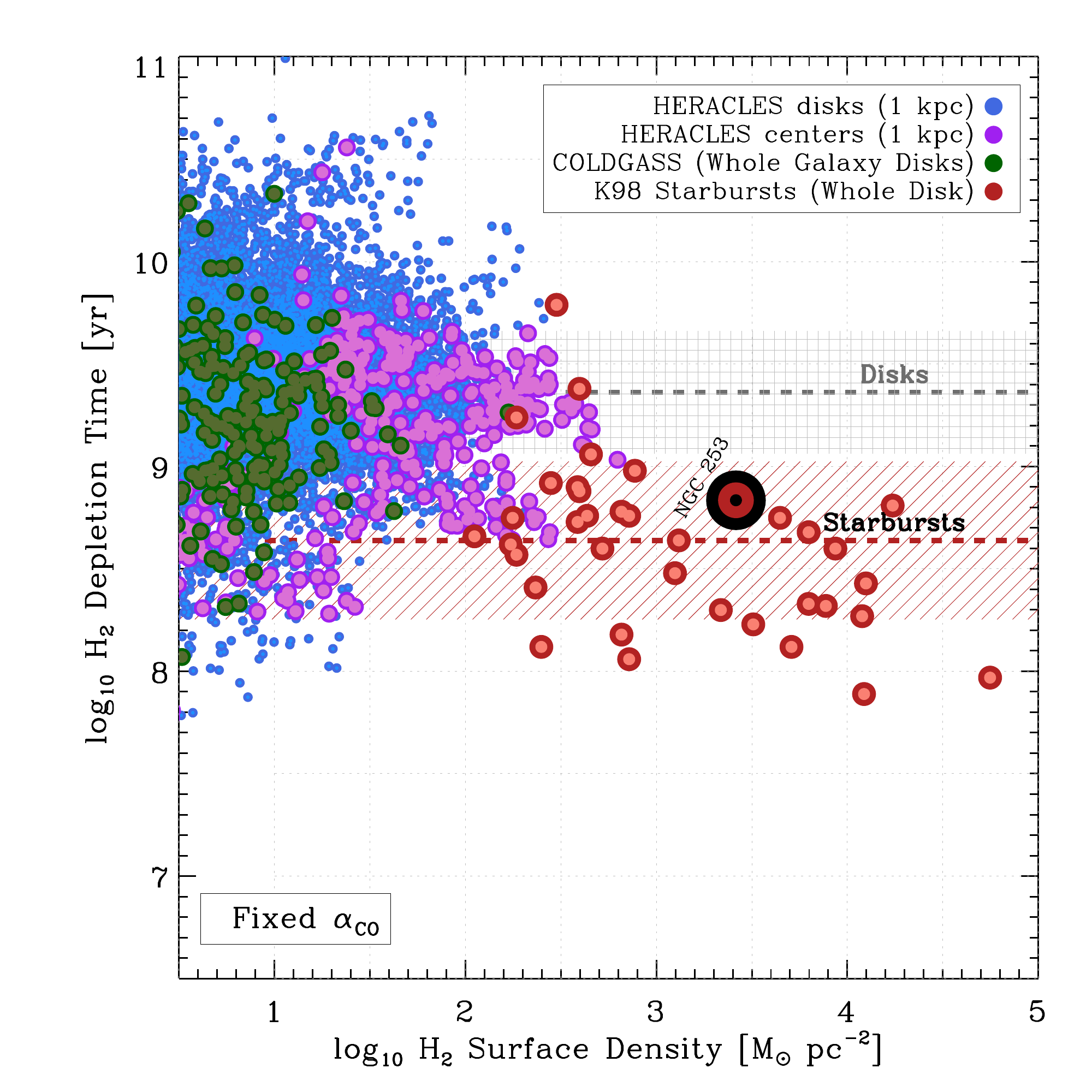}{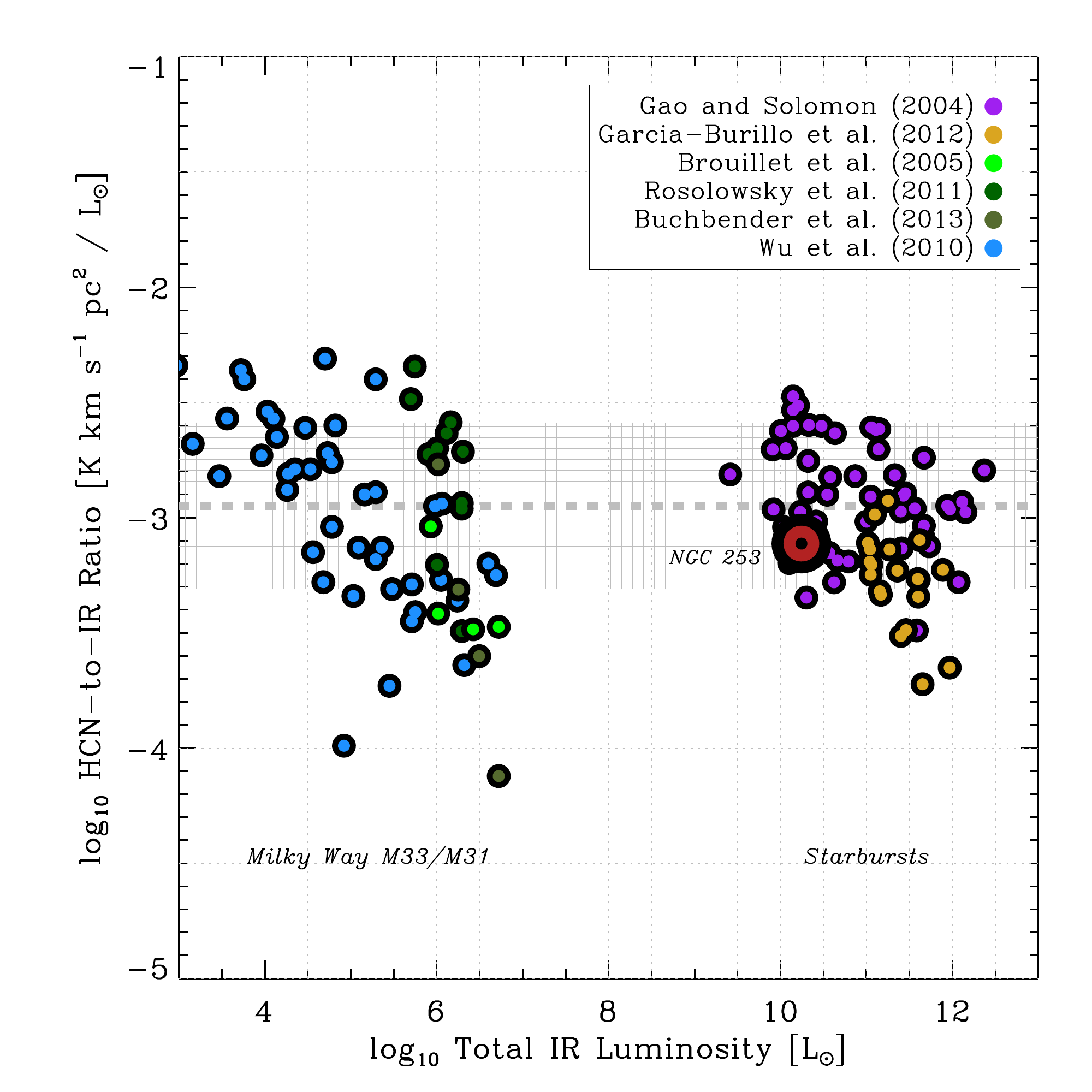}
\caption{NGC 253 as a nuclear starburst in context. ({\em top left}) H$_2$ depletion time, $\tau_{\rm dep}^{\rm mol}$, the time for star 
formation to consume the molecular gas reservoir, as a function of H$_2$ surface density for star-forming disk galaxies and starburst 
galaxies (note that the $x$-axis values blend a variety of physical scales). In the top left panel,
we apply best estimates of different conversion factors for different populations (separating disks, galaxy centers, and starburst galaxies; see text
for details). ({\em bottom left}) The same plot with a single $\alpha_{\rm CO}$ applied to all of the literature data. In both the top left and bottom left 
panel the gray and red region show the median and rms scatter for
disks and starbursts. ({\em top right}) NGC 253 as a nuclear starburst; $\tau_{\rm dep}^{\rm mol}$
as a function of radius (with density normalized at each radius) and corrections to $\alpha_{\rm CO}$ based on \citet{SANDSTROM13} applied
to galaxy centers (with the central region indicated in gray). The center of NGC 253 is coincident with the population of HERACLES galaxy centers, 
but would be among the most extreme starbursts in the sample. ({\em bottom right}) Ratio of HCN-to-IR luminosity as a function of IR luminosity for star forming systems from Galactic 
cores through starburst galaxies. Based on these plots, we argue that NGC 253 is a good example of the broader population of
starburst galaxies. Specifically, it is a good prototype of nuclear starburst regions, which 
appear intermediate between disks and merger-induced starbursts in surface brightness and H$_2$ depletion time but still appear 
to convert dense gas to stars at a normalized rate similar to most star-forming systems.
Unlike most starbursts, NGC 253 also lies at a distance where ALMA can resolve individual star-forming clouds.
\label{fig:tdep}
}
\end{figure*}

Figure \ref{fig:tdep} places NGC 253 in the context of the relationship between star formation, molecular gas, and dense gas from several recent surveys.
The top row plots the H$_2$ depletion time, $\tau_{\rm dep}^{\rm mol} = M_{\rm mol} / SFR$, the time that it would take present day star formation to 
consume the molecular gas mass as a function of molecular gas surface density. We combine points from COLDGASS \citep[green; CO 1--0][]{SAINTONGE12},
the most complete unresolved survey of molecular gas in the $z=0$ population, 1-kpc resolution apertures from the HERACLES survey \citep[CO 2--1][]{LEROY09,LEROY13}
with the central kpc (purple) colored differently from the disks (blue), and starburst disks drawn from \citet[red][treating the starburst as a single unresolved disk]{KENNICUTT98A}. 
These samples have disparate spatial scales, so $x$-axis variations among the populations should not be over-interpreted. The main information is contained in the vertical 
distribution of points, which show the typical ``disk galaxy'' $\tau_{\rm dep}^{\rm mol} \approx 1$--$2 \times 10^9$~yr (blue and green), along with
the lower ``starburst'' depletion time $\tau_{\rm dep}^{\rm mol} \sim 10^8$~yr (red). The separation between these populations when best-estimate
conversion factors are applied, which is also seen at high $z$, has fueled the idea of two ``sequences of star formation'' \citep[essentially a bimodal 
$\tau_{\rm dep}^{\rm mol}$][]{GENZEL10,DADDI10,CARILLI13}. Perhaps
less appreciated, Figure \ref{fig:tdep} shows that galaxy centers (purple) fill in the intermediate regime between the two populations, showing clear
enhancements in star formation activity relative to galaxy disks \citep[e.g.,][]{LEROY13,KENNICUTT98A} but often not reaching the extremes found in 
merger-induced starbursts \citep[e.g.,][]{SANDERS96}. Reinforcing the physical difference between disks and galaxy centers, the
purple points in Figure \ref{fig:tdep} also show systematically higher CO excitation, gauged from their (2--1)/(1--0) ratios \citep{LEROY09,LEROY13}.

The nucleus of NGC 253 \footnote{We plot the average $\left< \Sigma_{\rm mol} \right>$ for NGC 253 inside $r=250$~pc, for a comparison to the HERACLES points divide this value
by a factor of 4 (or subtract 0.6 dex in log).} shows a $\tau_{\rm dep}^{\rm mol}$ among the lowest (most active) found for any galaxy center in the sample. NGC 253 overlaps 
also the population of local starburst galaxies. The $\tau_{\rm dep}^{\rm mol}$ in the nucleus of NGC 253 appears $5$--$25$ shorter than the typical disk value (shown
in gray in Figure \ref{fig:tdep}, a contrast that we will bear in mind when considering characteristic timescales for the molecular emission (\S \ref{sec:discussion}). Note 
that, modulo inflow from the disk,  the future duration of the burst will be even shorter than these values because feedback-driven molecular outflows appear to deplete the 
gas at $\gtrsim 3$ times the SFR \citep{BOLATTO13A}.

The top right panel of Figure \ref{fig:tdep} highlights the radial component of the behavior seen in the top left panel. We plot $\tau_{\rm dep}^{\rm mol}$ as
a function of radius for the whole HERACLES sample \citep{LEROY13} at 1~kpc resolution (the plot shows data density normalized to the amount of data
at a given radius, which accounts for the larger area at large radii). The contours show a relatively narrow distribution of $\tau_{\rm dep}^{\rm mol}$ in galaxy disks
(large radii) and a wide range of values, often including significantly shorter $\tau_{\rm dep}^{\rm mol}$ in galaxy centers \citep[see][and Sandstrom et al. in prep.
for more discussion]{LEROY13,SANDSTROM13}. NGC 253 exhibits a low $\tau_{\rm dep}^{\rm mol}$ associated with its nuclear concentration of molecular gas,
reinforcing that we study a strong example of the nuclear starburst phenomenon.

The depletion times and surface densities in Figure \ref{fig:tdep} depend on the adopted CO-to-H$_2$ conversion factor, $\alpha_{\rm CO}$. In the top left
panel we adopt our best estimates, with galaxy centers corrected according to the recent work of \citet{SANDSTROM13}. They trace
total ISM mass using dust emission solve for $\alpha_{\rm CO}$ locally by combining dust, CO, and {\sc Hi} measurements. A main conclusion of that 
paper is that $\alpha_{\rm CO}$ is often (much) lower in the central parts of star-forming galaxies than in their disks. Similarly, the top left panel applies a a ``starburst'' conversion 
factor, $\alpha_{\rm CO} = 0.8$~M$_\odot$~(K~km~s$^{-1}$~pc$^{2})^{-1}$, for the red points \citep[][]{DOWNES98}, which is intended to capture the lower mass-to-light ratio
of CO in the wide-line width, excited ISM of starburst galaxies. Though this is a coarse correction, the idea of a lower $\alpha_{\rm CO}$ in starburst galaxies
is widely accepted. However, the details remain uncertain and a subject of intense research \citep[for more discussion see][]{BOLATTO13B}. For comparison,
the bottom left panel of Figure \ref{fig:tdep} shows the same collection of data with a single, fixed conversion factor, a ``Milky Way'' 
$\alpha_{\rm CO} = 4.35$~M$_\odot$~(K~km~s$^{-1}$~pc$^{2})^{-1}$, for all data with a line ratio correction of $1/0.7$ to convert CO 2--1 to CO 1--0 \citep[but
as already noted the variable excitation between disks, centers, and bursts introduce an additional complication see][]{LEROY09,KODA12,LEROY13}. 
The lower $\tau_{\rm dep}^{\rm mol}$ in galaxy centers and starbursts are still visible and NGC 253 remains an intermediate system, but the strength of the contrast
among populations is substantially reduced. This reinforces the point made by a number of authors \citep[e.g.,][]{BOUCHE07,OSTRIKER11,NARAYANAN12} that
the slope, or even the existence, of any simple ``star formation law'' depends sensitively on the approach to the $\alpha_{\rm CO}$. Less appreciated, the 
Figures also show that the conversion factor is also essential to understand whether starbursts at galaxy centers reflect mainly increased gas concentrations or
more efficient star formation \citep[the results of][imply the latter; see Sandstrom et al. in prep.]{SANDSTROM13}. We will return to the topic of the conversion 
factor in \S \ref{sec:aco_results}.

The first few panels show variations in the star formation rate per unit bulk molecular gas. A number of studies have shown that the more
direct relationship between {\em dense} molecular gas and star formation leads to a simpler correlation between tracers of recent star formation and 
tracers of dense gas, e.g., HCN 1--0 or HCO$^{+}$ 1--0 line emission, than between recent star formation and bulk molecular gas
\citep[e.g.,][]{GAO04,GARCIABURILLO12,LADA10}. At the simplest level, the main evidence for this claim is an approximately fixed apparent ratio of
dense gas to recent star formation even in the regime where $\tau_{\rm dep}^{\rm mol}$ varies. The final panel in Figure \ref{fig:tdep} shows
the HCN-to-IR ratio in NGC 253 compared to that in Galactic cores \citep{WU10}, Local Group GMCs \citep{BROUILLET05,ROSOLOWSKY11,BUCHBENDER13},
and starburst galaxies \citep{GAO04,GARCIABURILLO12}. Despite systematic disparities in $\tau_{\rm dep}^{\rm mol}$, NGC 253 exhibits roughly the same 
HCN-to-IR ratio as nearby luminous infrared galaxies, clouds in M33, and Milky Way cores. Compared to other starbursts, NGC 253 does appear 
slightly enhanced in IR relative to HCN, though the enhancement is well within the $1\sigma$ scatter (our HCN luminosity may be biased somewhat
low, though we do not find evidence for a large fraction of missing flux; see Section \ref{sec:data}), NGC 253 may be at a later stage in its evolution, 
HCN-to-dense gas conversion factor variations may be at play, or NGC 253 may simply be slightly more efficient than other starbursts. Other recent
evidence does point to systematic variations in how rapidly dense gas forms stars within and among galaxies 
\citep[][and Usero et al. (submitted)]{GARCIABURILLO12,KEPLEY14}. In this paper, we consider the scales too small and available star formation rate
tracers too coarse to do more than place a single point on Figure \ref{fig:tdep}. For purposes of framing this work, the bottom right panel in 
Figure \ref{fig:tdep} shows that NGC 253 forms stars from dense gas at approximately the same normalized rate as other star-forming systems.

These comparisons collectively argue that NGC 253 represents a good case study to understand the physics driving starbursts,
particularly the common nuclear starburst phenomenon. The nucleus of this galaxy agrees with integrated literature measurements
in its ratio of HCN to infrared luminosity, resembling not only other starbursts, but Milky Way cores and clouds in M33 and M31.
This supports the idea that dense gas (the structures traced by HCN, HCO$^{+}$, and CS in our data) may be considered the key star-forming units.
Comparing star formation to bulk molecular gas traced by CO emission, it shows a short depletion time in its central kpc, lying near the 
lower edge of the envelope of $\tau_{\rm dep}^{\rm mol}$ measured for the inner kpc of HERACLES galaxies and near the upper edge of 
the sample starbursts compiled by \citet{KENNICUTT98A}. As an integrated point, NGC 253 appears in these scaling relations as an 
extreme nuclear starburst, but not a strong outlier from lower resolution observations.

\subsection{Adopted Properties of NGC 253}
\label{sec:props}

\begin{deluxetable}{lc}
\tablecaption{Adopted Properties of NGC 253} 
\tablehead{ 
\colhead{Property} & 
\colhead{Value}
}
\startdata
Distance & 3.5~Mpc \citep{REKOLA05} \\
Nuclear Position Angle & $55\arcdeg$ (ALMA and {\em Spitzer} images) \\
Center Position & \citep{MULLERSANCHEZ10} \\
É $\alpha_{2000}$ & 00$^{\rm h}$ 47$^{\rm m}$ 33.14$^{s}$ \\
É $\delta_{2000}$ &  -25$\arcdeg$ 17$\arcmin$ 17.52$\arcsec$ \\
Star formation rate & $4.2$~M$_\odot$~yr$^{-1}$; L$_{\rm IR}$ \citep{SANDERS03} \\
$...$ fraction in starburst & $\sim 0.5$ (IR and radio images) \\
Inclination & $76^{\circ}$ \citep[e.g.][]{MCCORMICK13} \\
\hline
\multicolumn{2}{c}{Abundances ($\log_{10}$ A/B)} \\
\hline
CO/H$_2$ & $-4$ \\
C$^{17}$O/CO & $\sim -3$ (following Meier et al. (submitted)) \\
C$^{18}$O/C$^{17}$O & $0.6$ (W05, L04, W94) \\
C$^{18}$O/CO & $\sim -2.4$ \\
HCN/H$_2$  & $-8.3$ (M06; P04; range $-7.7$ to $-8.5$\tablenotemark{a}) \\
H$^{13}$CN/H$_2$ & $-9.9$ (M06; range $<-9.4$) \\
HCO$^{+}$/H$_2$ & $-8.8$ (M06; range $-8.1$ to $-8.8$) \\
H$^{13}$CO$^{+}$/H$_2$ & $-10.4$ (M06; range $-9.5$ to $-11.4$) \\
CS/H$_2$ & $-8.2$ (M06; range $-8.2$ to $-9.0$)
\enddata
\tablecomments{L04: \citet{LADD04}, M06: \citet{MARTIN06}, P04: \citet{PAGLIONE04}, 
W94: \citet{WILSON94}, W05: \citet{WOUTERLOOT05} }
\tablenotetext{a}{Abundance range across a wide set of environments from \citet{MARTIN06}.}
\label{tab:assumptions}
\end{deluxetable}

Table \ref{tab:assumptions} provides some of the numbers used in the scaling plots and other properties of the NGC 253 starburst
adopted in this paper. We derive the star formation rate for the whole system from the $L_{\rm TIR} \approx 3.5 \times 10^{10}$~L$_\odot$ 
reported by \citet[][this has been adjusted to our adopted distance]{SANDERS03} and the IR-to-SFR conversion of \citet{KENNICUTT12}. 
From inspection of {\em Herschel} images (P.I. S.~Veilleux) we estimate that $\sim 50\%$ of the star formation activity occurs in the nuclear 
region of NGC 253. Inspection of IRAC, MIPS, and VLA images suggest that similar fractions of mid-IR and radio emission 
($\sim 35$--$50\%$) arise from the ALMA field of view \citep{DALE09,CONDON87}. Assuming that this emission all arises from star formation,
this yields a SFR$\sim 2$~M$_\odot$~yr$^{-1}$ for the burst.  In Figure \ref{fig:tdep}, we draw CO and HCN luminosities and surface 
densities for the nucleus from our new observations \citep[][\S \ref{sec:data} of this paper]{BOLATTO13A}.

Table \ref{tab:assumptions} include abundances and isotopic ratios needed to interpret our data. These come mostly from \citet{MARTIN06}, 
who model the physical conditions in the center of NGC 253 based on single-dish spectroscopy. They also compile ranges of abundance determinations from 
other environments, which we note in the table and view as a useful estimate of the uncertainty associated in any calculation that invokes
an abundance. We adopt the $^{18}$O/$^{17}$O from \citet{WILSON94} who note the lack of a gradient with Galactocentric radius in the Milky Way.
Following Meier et al. (submitted), we take a fiducial CO-to-C$^{17}$O ratio of $1,000$, which resembles the value found in the Galactic center
\citep[e.g.,][]{WILSON94} but is twice as high as the value in the Solar Neighborhood, consistent with $^{17}$O being a secondary product enhanced
in regions of recent high mass star formation. For consistency, these two factors determine our C$^{18}$O/CO ratio to be $250$, which approximately agrees with
\citet{HENKEL93}.

Figure \ref{fig:coverage} shows the larger context for our observations \citep[left: $8\mu$m emission showing
the distribution of the ISM, right: $3.6\mu$m emission showing red starlight]{DALE09}. NGC 253 is a heavily inclined 
\citep[$i \approx 76\arcdeg$, e.g.,][]{MCCORMICK13} late-type spiral galaxy. The inner few kpc host a bar that is visible in both ISM
and stellar tracers. That bar has apparently funneled gas to the inner regions of the galaxy \citep[see][]{SORAI00}, creating a region of high density
and star formation activity in the inner few hundred pc \citep[a fairly common occurrence in barred spirals, e.g.,][]{SHETH05}. In fact
this region is so active that on a linear stretch (rather than the log stretch that we use), Figure \ref{fig:coverage} would show only the nuclear starburst.
The main starburst region appears to lie inside the inner Lindblad resonance \citep[which is estimated to be 
at $r_{\rm gal} \sim 0.3$--$0.4$~kpc][]{SORAI00,PAGLIONE04,IODICE14}, so that the starburst region studies in this paper may be considered a dynamically
distinct region from the larger bar.

\section{ALMA Observations of NGC 253}
\label{sec:data}

\begin{figure*}
\plottwo{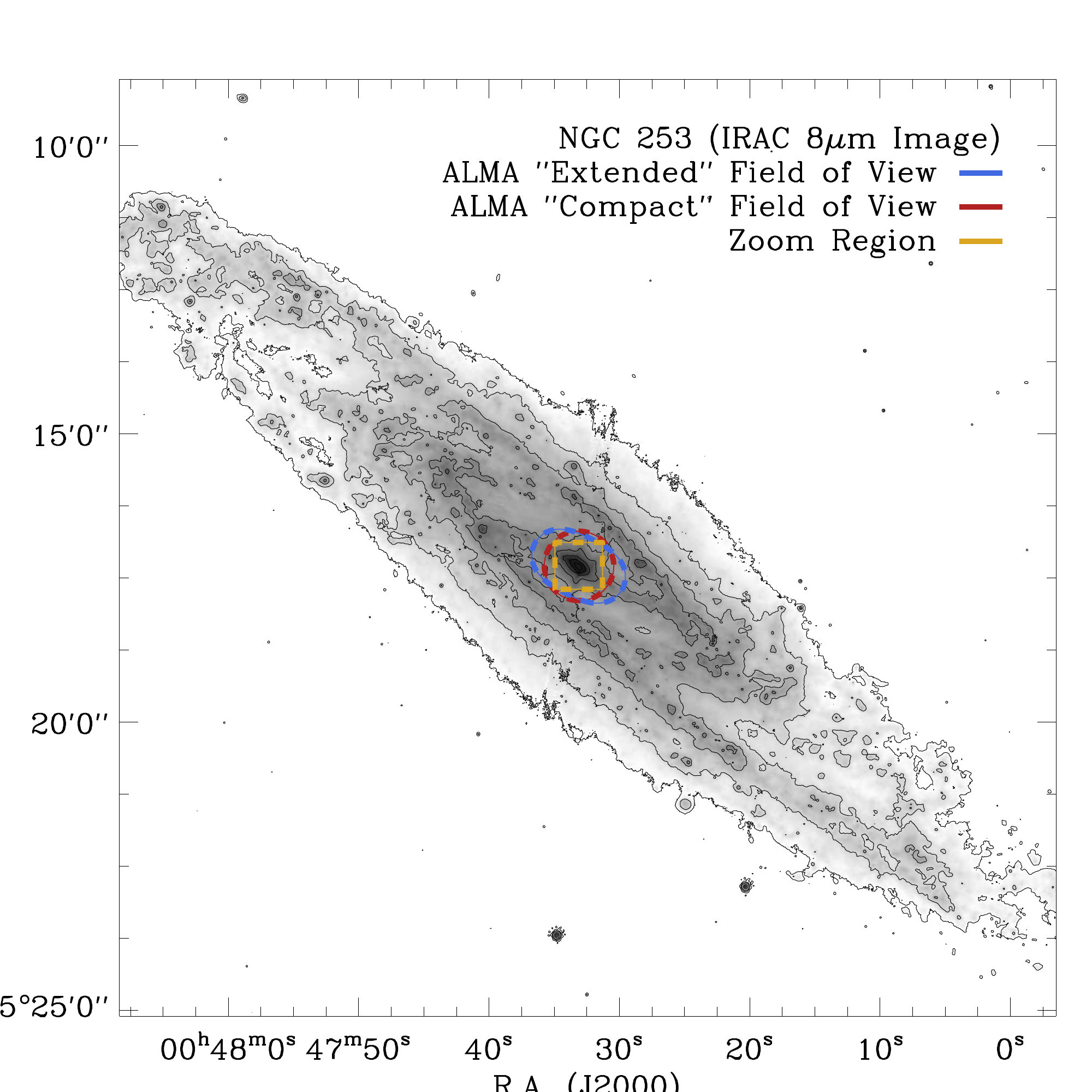}{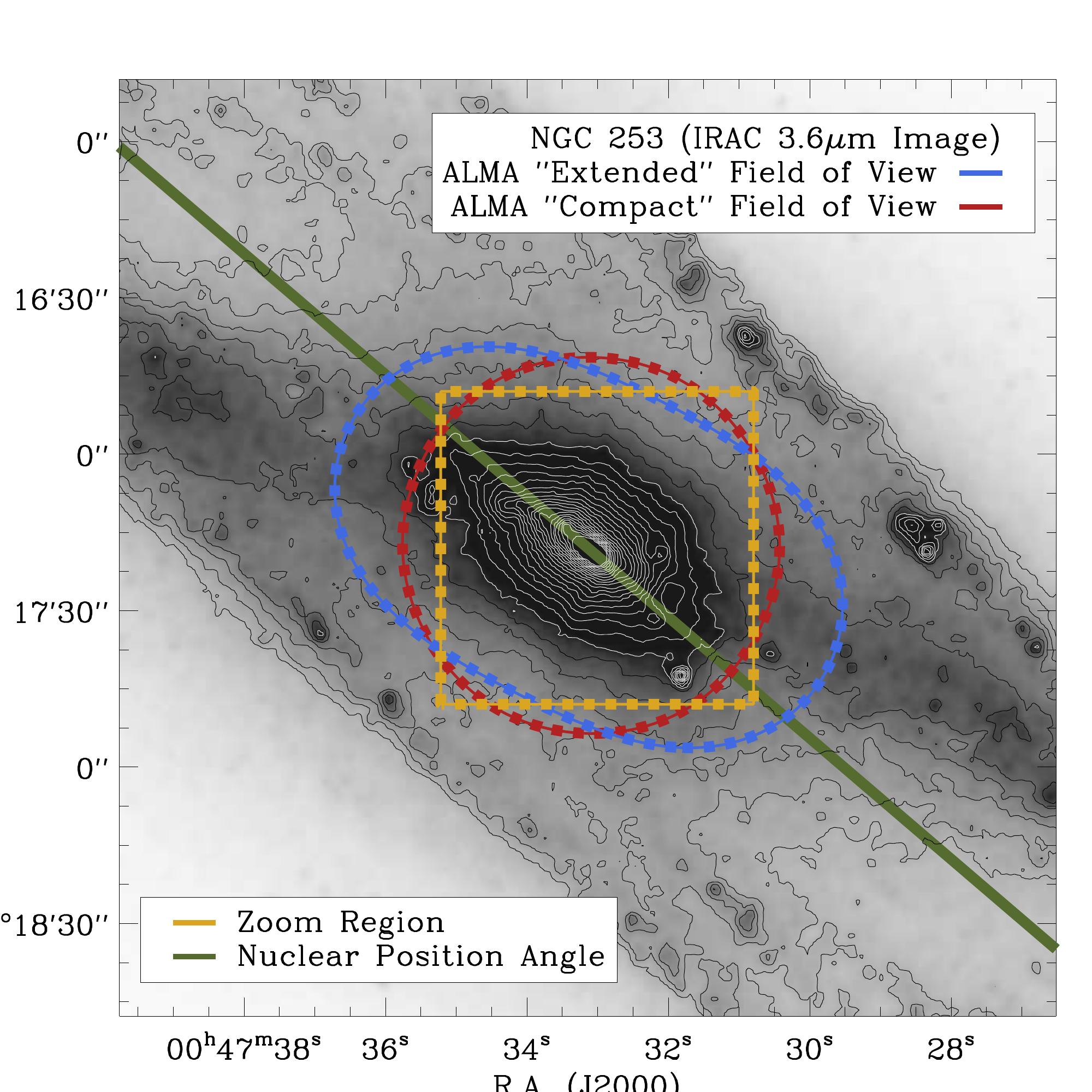}
\caption{({\em left}) Coverage of ALMA observations of the nuclear starburst in NGC 253. The {\em left} image shows 8$\mu$m emission
on a logarithmic stretch \citep{DALE09}. Red contours show the region of half maximum sensitivity for the 7-point mosaic 
used to observe CO and C$^{17}$O. Blue contours show coverage of the 3-point mosaic used to observe HCN, HCO$^{+}$, and CS. 
Roughly half the star formation in the galaxy ($\sim 35$--$50\%$ of the infrared and radio flux) originates from inside the ALMA field of view. The gold
box indicates the region shown in more detail in Figures \ref{fig:pp_maps} and \ref{fig:pv_maps}. ({\em right})
Situation of the field of view (and nuclear starburst) relative to the stellar bar, shown via the 3.6$\mu$m emission \citep{DALE09}. A line in the
right panel illustrates the nuclear position angle, which also describes the molecular gas distribution well. Black and white contours both trace
the 3.6$\mu$m emission to highlight stellar structure.}
\label{fig:coverage}
\end{figure*}

In this paper, we present new ALMA observations of the high critical density tracers HCN (1-0), HCO$^+$ (1-0), CS (2-1) 
and their optically thin isopotologues H$^{13}$CO$^+$ (1-0) and H$^{13}$CN (1-0). 
To obtain these maps, we used ALMA in its ``extended'' Early Science (``Cycle 0'') configuration to 
observe a three-point mosaic covering NGC 253's nuclear starburst. The lower sideband covers HCN (1-0), HCO$^{+}$ (1-0), 
H$^{13}$CN (1-0), and H$^{13}$CO$^{+}$ (1-0) and the upper sideband includes CS (2-1). Both sidebands cover many other 
transitions, which are discussed in Meier et al. (submitted). Figure \ref{fig:coverage} shows the half-power target area of 
our observations plotted over near-infrared images of the galaxy.

We also present a mixture of new and previously published CO (1-0) and C$^{17}$O data, observed in a 7-point mosaic
covering roughly the same area. \citet{BOLATTO13A} presented the first version of the $^{12}$CO (1-0) data, which they
combined with MOPRA single dish data (P.I. Ott). Those CO observations were mainly intended to recover the molecular outflow 
\citep{BOLATTO13A} and so used ALMA's ``compact'' Cycle 0 configuration in order to emphasize surface brightness sensitivity over 
resolution.  Here we combined these with new CO data obtained by ALMA in its second observing campaign (``Cycle 1''). These include observations 
with the 7m portion of the Atacama Compact Array (ACA), which recover emission with large spatial scales, and new 12m observations observed in a 
configuration closely matched to that used to observe the high critical density lines. 

In each case, we reduced the data following the standard ALMA procedure, including subtraction of continuum emission in the $u-v$ plane
(see more discussion in Meier et al. submitted). In the case of the CO data, which combine three distinct data sets (Cycle 0, Cycle 1 ACA, and 
Cycle 1 main array), we also found it necessary to set the weights in the $u-v$ data based on the rms noise (using the CASA task {\tt statwt}) 
before any imaging. We imaged each transition over a redshift range from $v_{\rm LSR} = -50$ to $550$~km~s$^{-1}$ with 5~km~s$^{-1}$ wide channels.
In the case of the $^{12}$CO (1-0) data only, we incorporated single dish data from MOPRA (P.I. Ott) by combining the single dish data with the interferometer
data in the fourier plane using the MIRIAD task {\tt immerge} and using this combined image as the initial model for subsequent CASA imaging. We adopt this
approach because it incorporates information from the single dish image but also gives preference to the higher quality ALMA data in the part of the 
$u-v$ plane where both data sets contribute.

Because NGC~253 is very bright, we found resulting images to be dynamic range limited. That is, calibration errors and not thermal noise drive the 
scatter in the channel maps where we observe emission. To improve the situation, we carried out a short timescale
self-calibration using the integrated emission in each spectral window. Averaging the data across a full spectral window to create a single image, we
cleaned, then used the cleaned image as a model in order to solve for phase variations on the timescale of individual integrations ($\sim 6$s). 
We then reimaged the data and repeated the self-calibration process. After two iterations, we also solved for short timescale amplitude variations and offsets among 
the spectral windows. This procedure also helped resolve any lingering calibration offsets among multiple executions or, in the case of the 
CO observations, observations with different arrays. The noise in the resulting images approach the expected thermal value, typically $\sim 0.2$~K for the high critical density lines
and lower for CO and C$^{17}$O (Table \ref{tab:data}). The final resolution of the cubes, imaged with a Briggs (parameter 0.5) weighting 
scheme, varied slightly with transition. Table \ref{tab:data} summarizes the data cubes treated in this paper, the synthesized beam and RMS noise in Kelvin 
of each cube before convolution to a common resolution. For almost all analysis in this paper, we use a set of cubes
on matched astrometric grids all convolved to share a synthesized beam with FWHM $2 \times 2\arcsec$.

\begin{figure*}
\plottwo{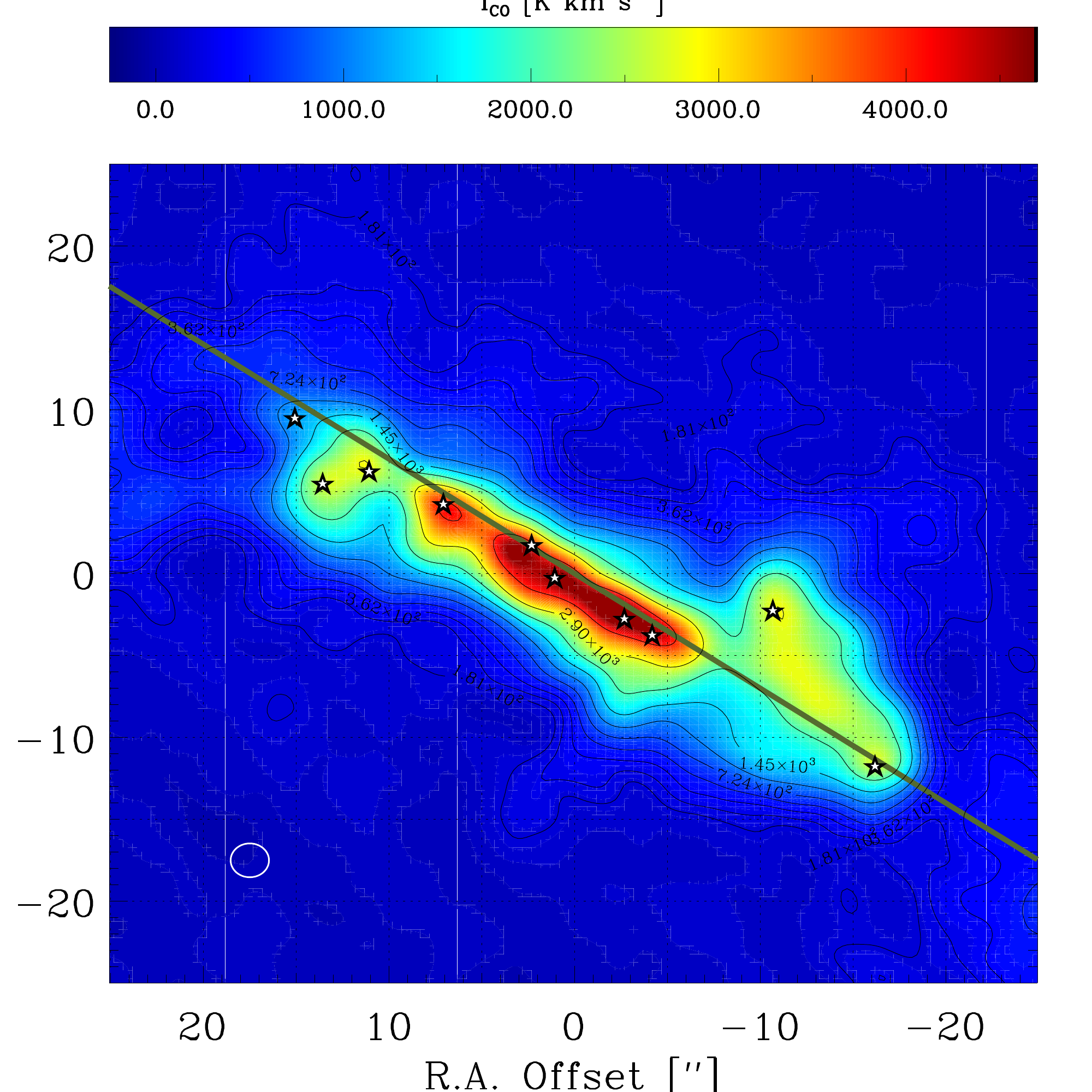}{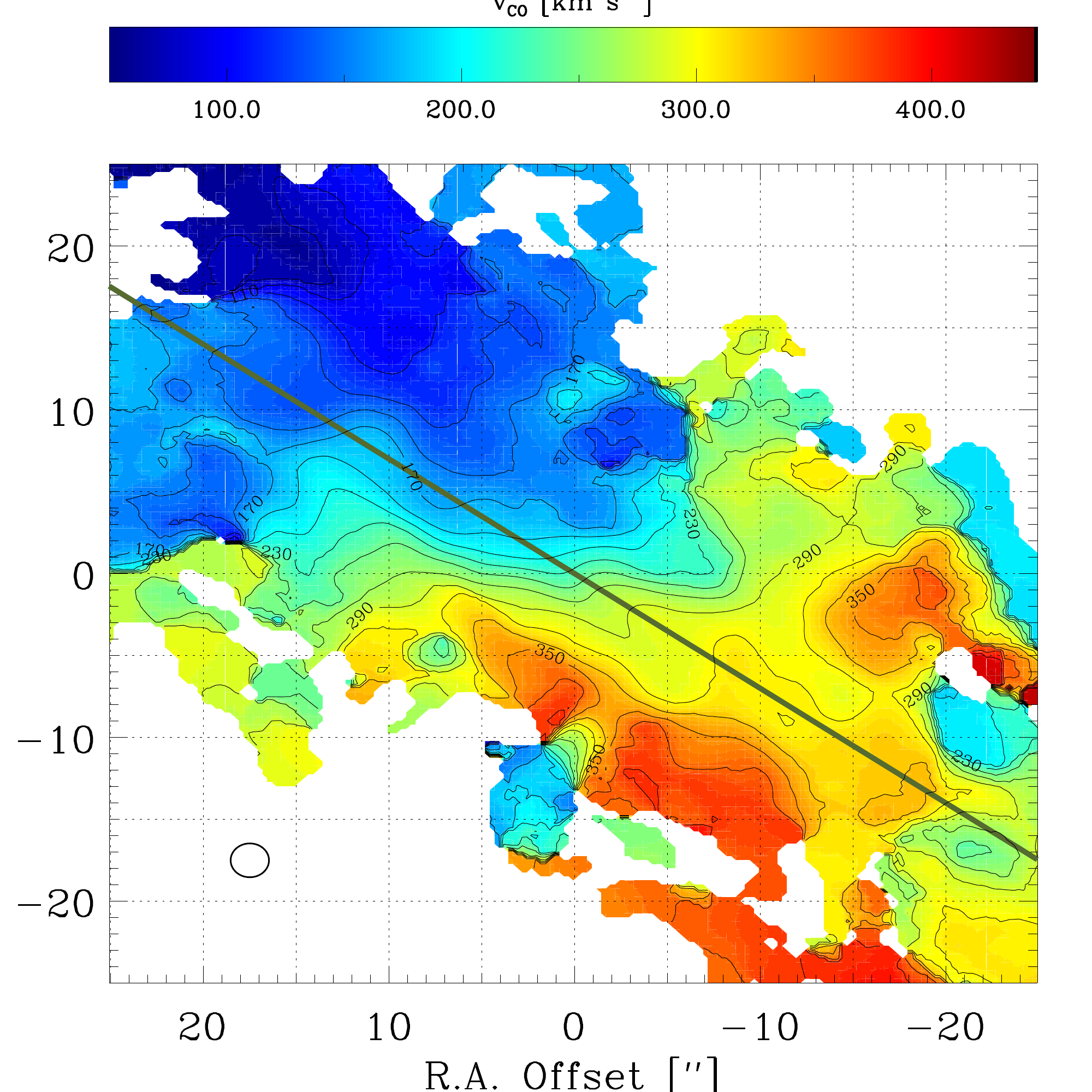}
\plottwo{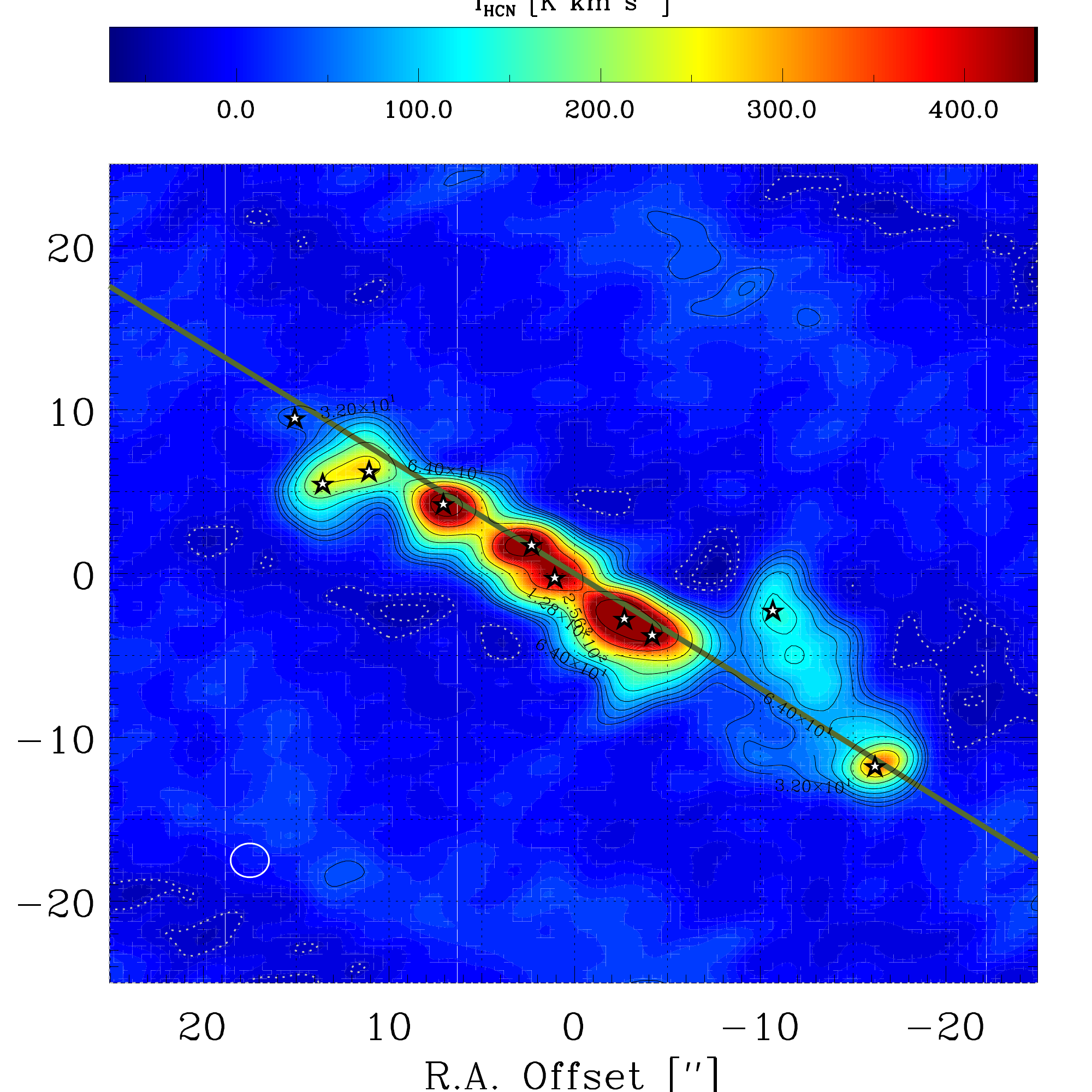}{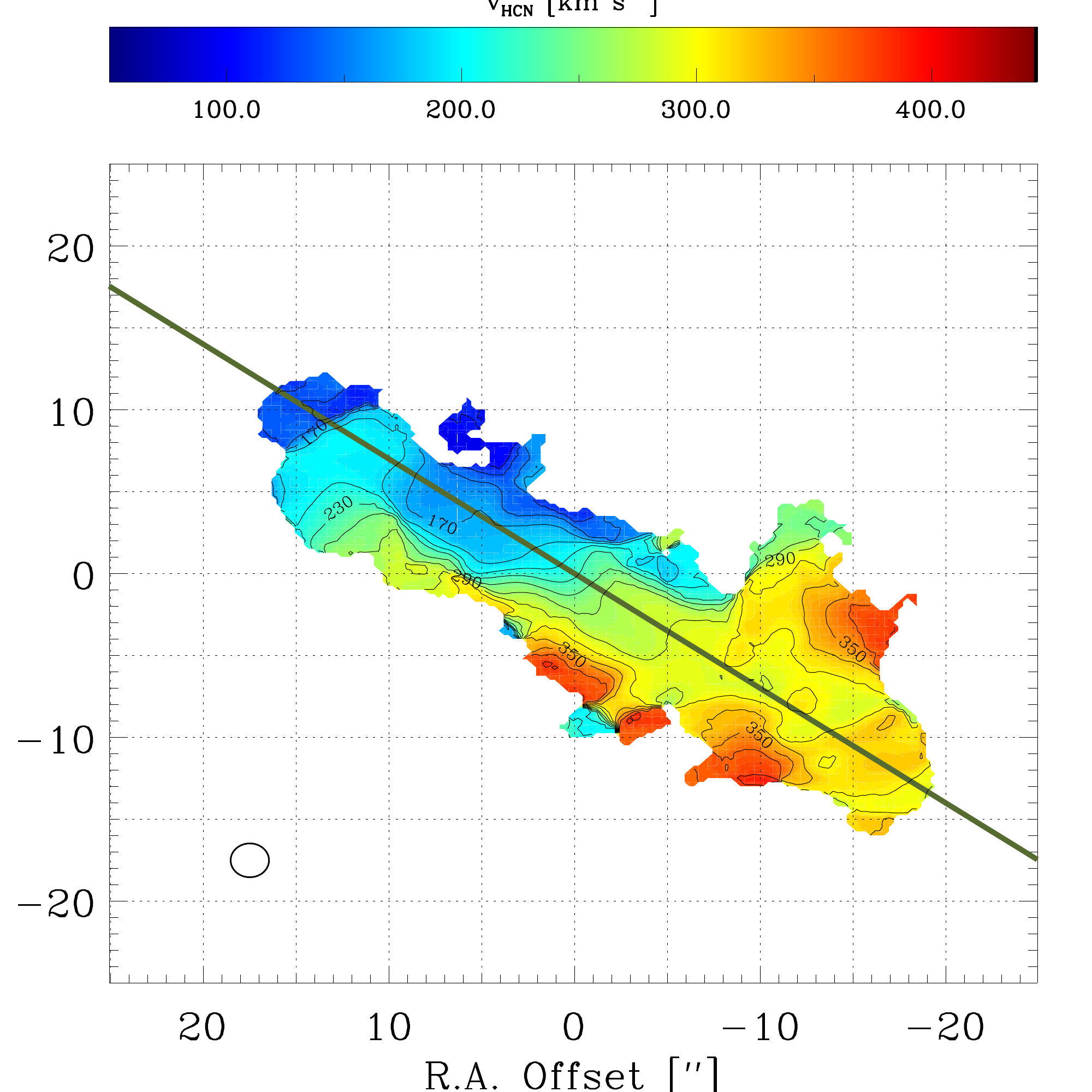}
\caption{ 
\label{fig:moment_maps}
Integrated intensity (``moment 0'', left) and intensity-weighted mean velocity (``moment 1'', right) maps for $^{12}$CO (1-0) (top), a bulk tracer of the molecular gas,
and HCN (1-0) (bottom), one of our three main tracers of the dense molecular gas. Stars label the location of individual molecular peaks. The integrated intensity 
maps show direct integrals of the data cube without masking, while the intensity-weighted moment maps are derived from masked cubes.
}
\end{figure*}

\begin{figure*}
\epsscale{0.8}
\plotone{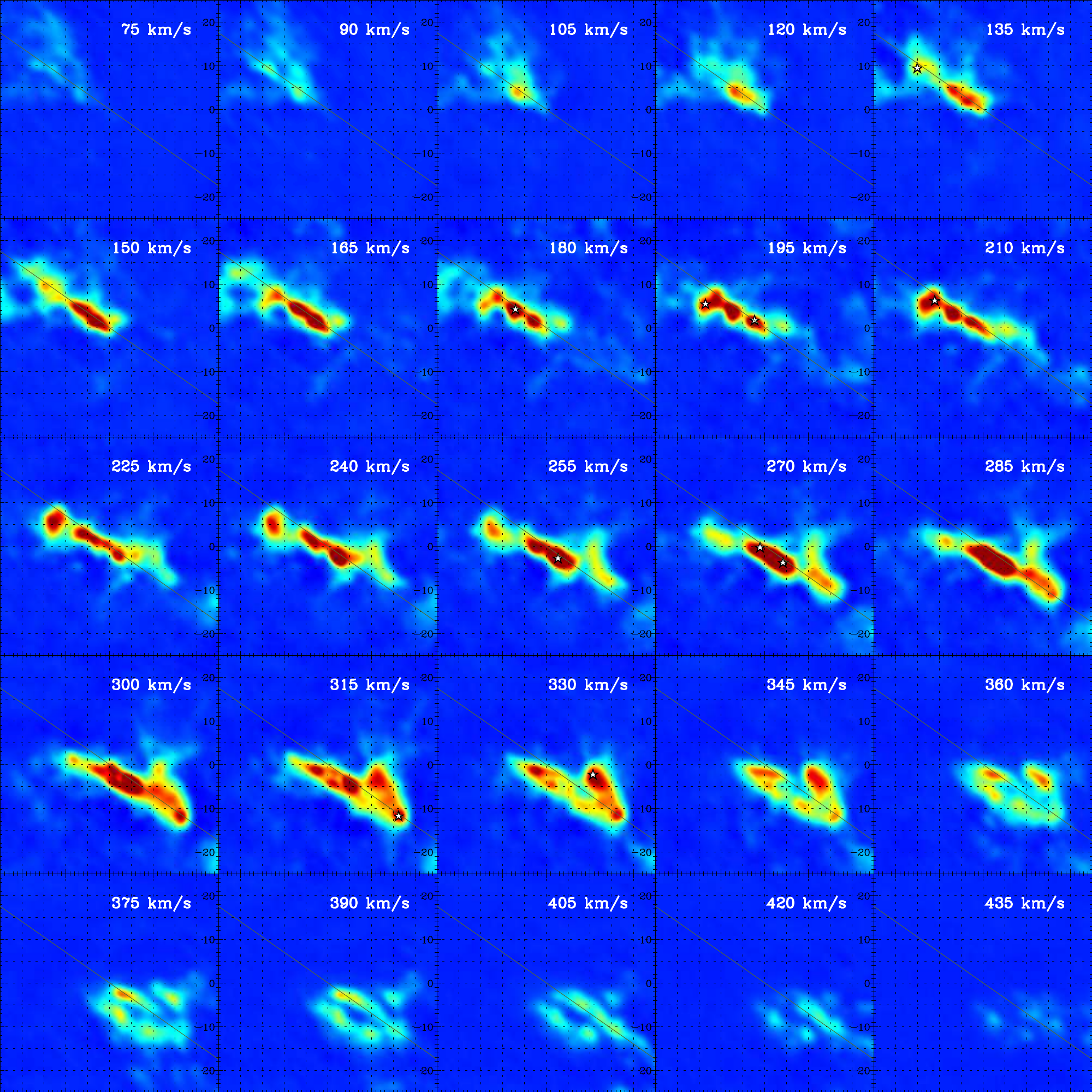}
\caption{ 
\label{fig:channel_co}
CO (1--0) emission in individual 5~km~s$^{-1}$ channels, stepping 15~km~s$^{-1}$ channels between each map. Stars indicate peaks of molecular emission identified
as part of our cloud analysis.
}
\end{figure*}

\begin{figure*}
\epsscale{0.8}
\plotone{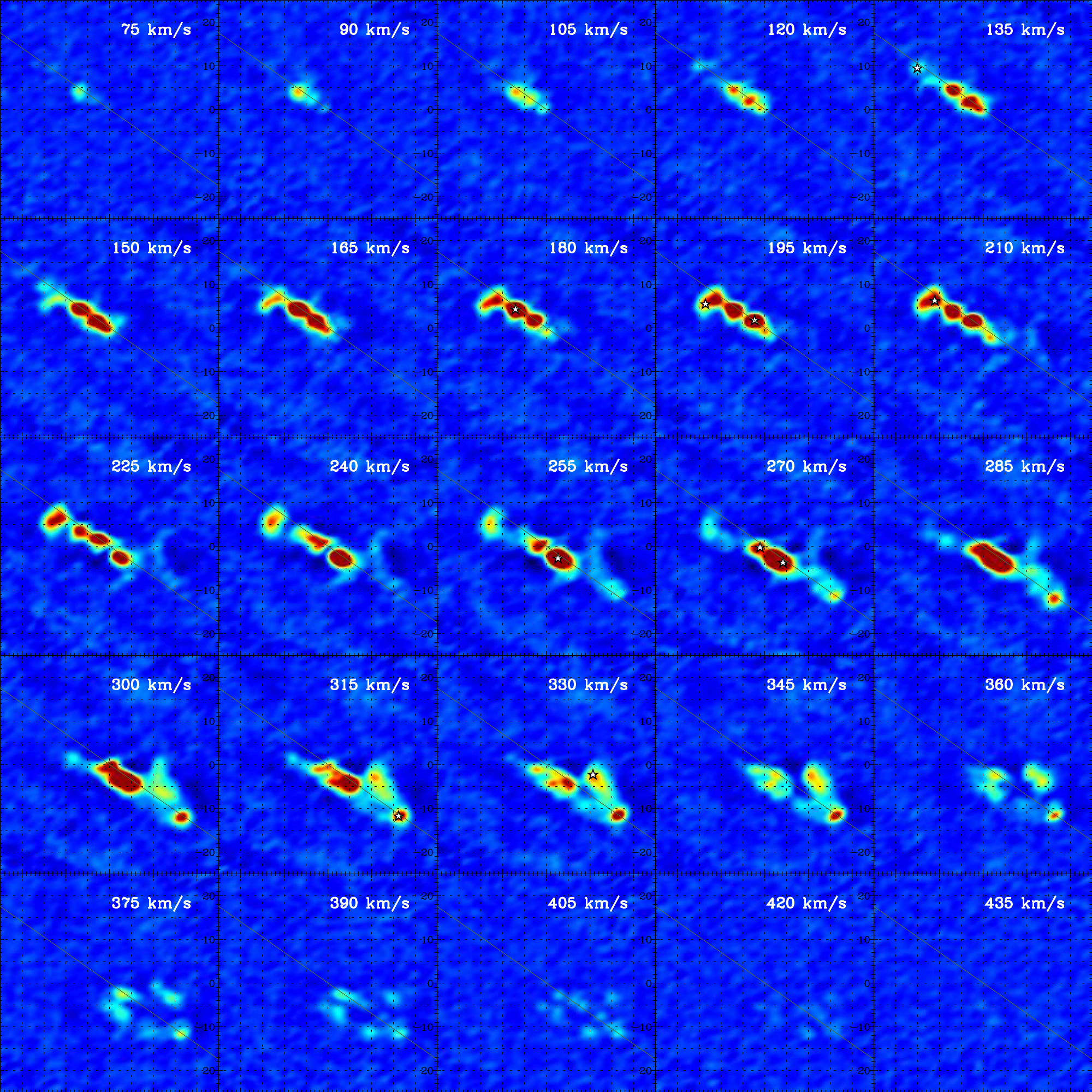}
\caption{ 
\label{fig:channel_hcn}
HCN (1--0) emission in individual 5~km~s$^{-1}$ channels, stepping 15~km~s$^{-1}$ channels between each map. Stars indicate peaks of molecular emission identified
as part of our cloud analysis.
}
\end{figure*}

Figures \ref{fig:moment_maps} -- \ref{fig:channel_hcn} show two of the resulting data cubes. In Figure \ref{fig:moment_maps}, we plot the integrated intensity and 
intensity-weighted velocity maps for $^{12}$CO, a bulk tracer of molecular gas, and HCN, a tracer of dense molecular gas. Figures \ref{fig:channel_co} and 
\ref{fig:channel_hcn} show channel maps for the same two transitions, plotting every third channel (i.e., stepping by $15$~km~s$^{-1}$). In the integrated intensity 
and channel maps, no masking has been applied in order to show the quality of the data as directly as possible. In the intensity-weighted velocity maps and 
most of the subsequent maps in this paper, we do apply masking designed to identify significant emission. 

\begin{figure*}
\plottwo{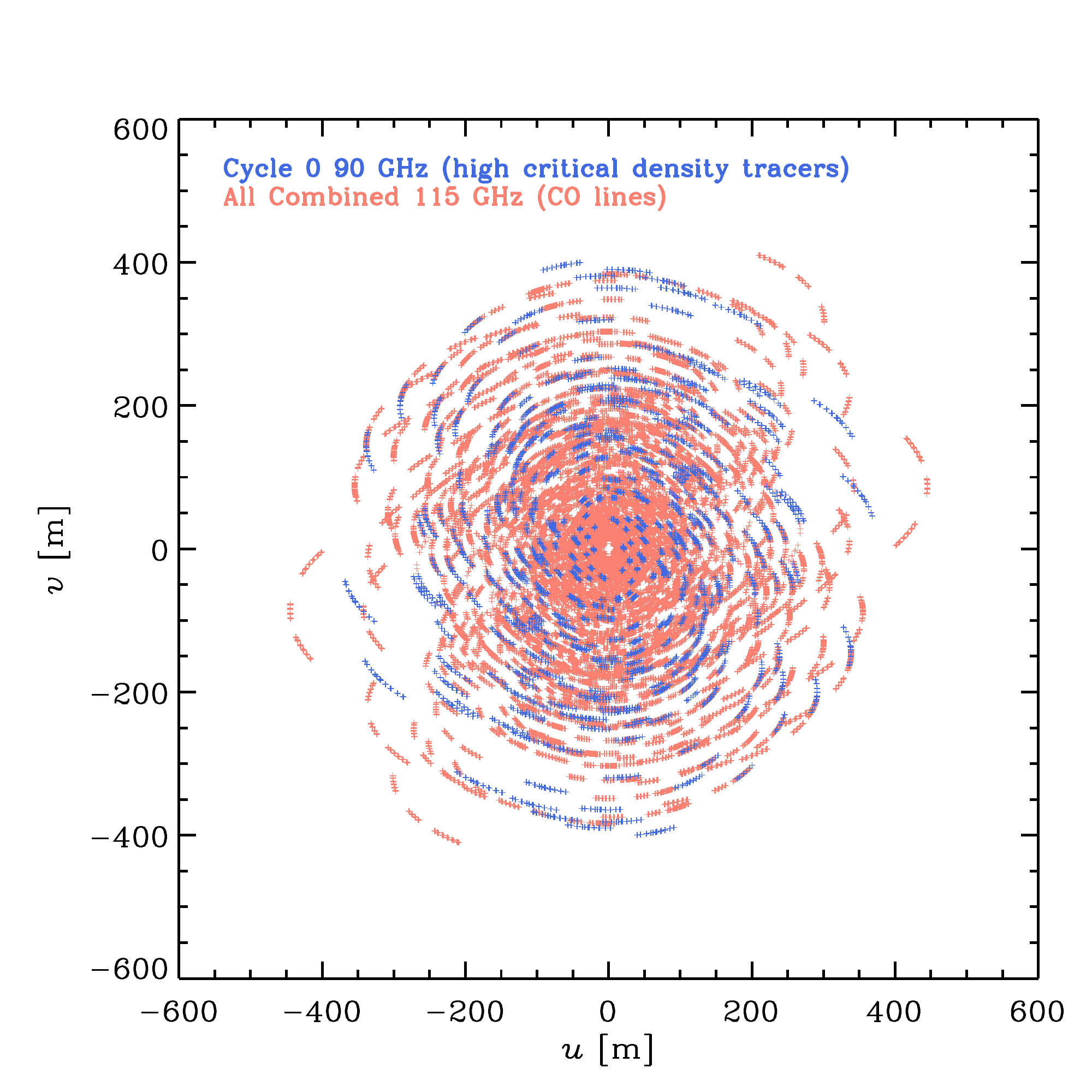}{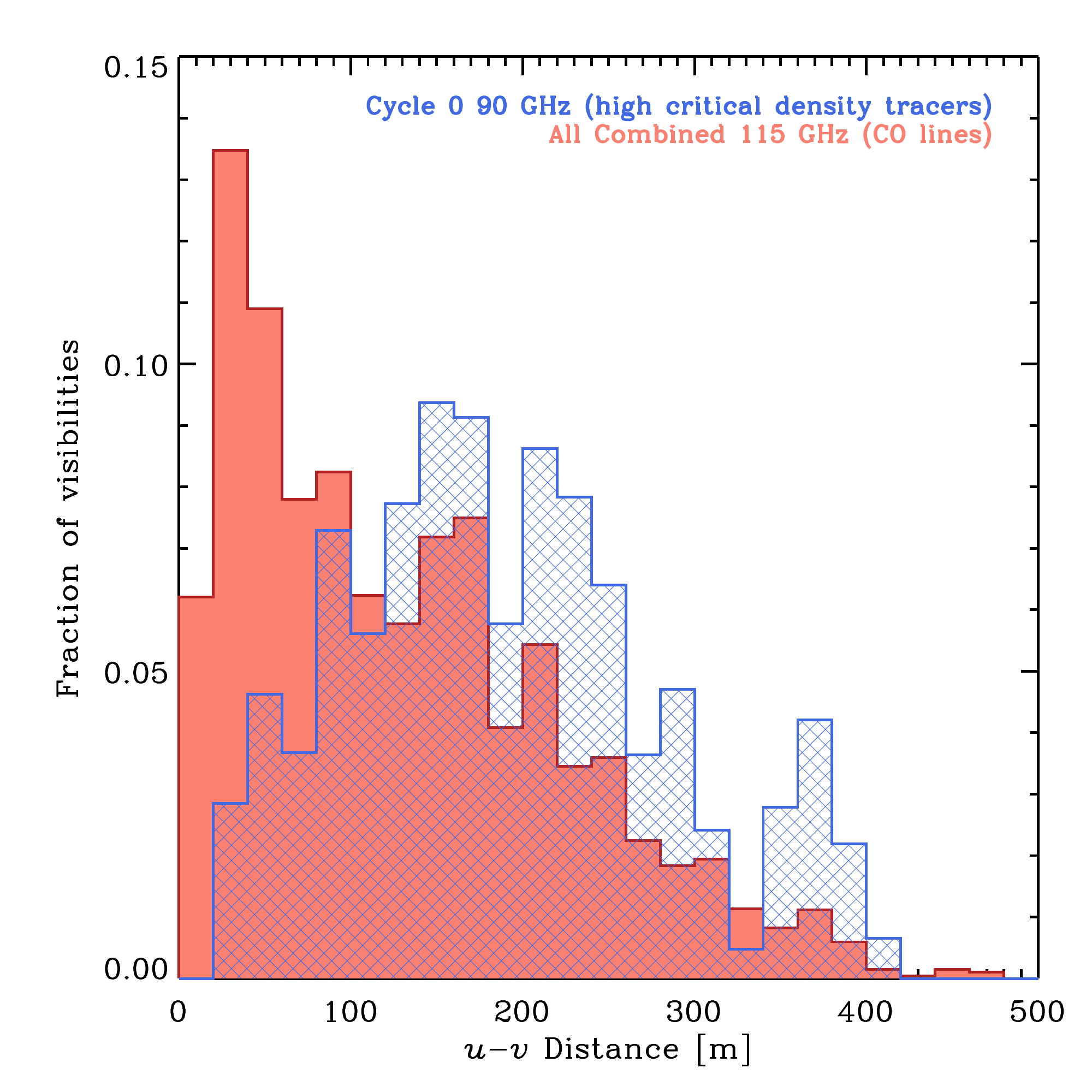}
\caption{Coverage of our observations ({\em left}) in the $u-v$ plane and ({\em right}) as a histogram of projected separations.
Blue points show the coverage for the high critical density tracers (HCN, HCO$^{+}$, etc.) observed in the ``extended'' configuration
in Cycle 0. Red points show the combined data for the CO lines ($^{12}$CO and C$^{17}$O), which combine Cycle 0, Cycle 1, and
Atacama Compact Array observations (as well as single dish data for the $^{12}$CO, not plotted). The synthesized beam is well-matched
across the two data sets (see Table \ref{tab:data}). The CO observations have better sensitivity to extended emission (and indeed CO is expected to be more extended),
but the flux recovered in our high density tracer cubes matches previous observations well. 
\label{fig:uv_coverage}}
\end{figure*}

{\em Flux Recovery and $u-v$ Coverage:} An important caveat affecting all of our data except the CO (1-0) is that they lack zero spacing information.
In the case of the high critical density observations, our coverage of short $u-v$ spacings is also limited. Figure \ref{fig:uv_coverage} summarizes the
$u-v$ coverage of our interferometric observations. The CO line observations have sensitivity to almost all spatial scales, while we expect that 
spatial structure with scale $\gtrsim 10.5\arcsec$ within an individual channel will be missed by our extended configuration observations. Fortunately, the
large velocity gradient in NGC 253 helps our recovery of the overall source structure and comparison to previous observations suggests that we recover
most of the flux. Based on OVRO observations, \citet{KNUDSEN07} quote HCN luminosities of $9.4 \times 10^6$~($8.9 \times 10^6$) and
$1.1 \times 10^7$~K~km~s$^{-1}$~pc$^{2}$ within a $20\arcsec$ and $27\arcsec$ aperture centered on the nucleus, which they claim
to match previous single dish observations \citep[e.g.,][]{NGUYEN89} well. In the same apertures, we recover $9.1 \times 10^6$ and 
$1.1 \times 10^7$~K~km~s$^{-1}$~pc$^{2}$, consistent with our fluxes matching the \citet{KNUDSEN07} values within
the calibration uncertainties; a similar case holds for HCO$^+$. Similarly, \citet{PAGLIONE95} quote an integrated HCN flux of 128~Jy~km~s$^{-1}$ within a 
20$\arcsec$ aperture, which they claim to be $\approx 70$\% of their measured single dish flux. For comparison, we convolve our HCN cube to 
$20\arcsec$ (FWHM) resolution and measure the flux from the spectrum at the center of the galaxy, which should match the calculation of \citet{PAGLIONE95}. 
We obtain $152$~Jy~km~s$^{-1}$, consistent with recovering
more of the flux than the \citet{PAGLIONE95} maps, though still perhaps $15\%$ lower than the single dish flux. For comparison to the numbers above, the HCN 
complexes in \citet{PAGLIONE95} include $\approx 0.6 \times 10^7$~K~km~s$^{-1}$~pc$^{2}$, all of which appears to be recovered in our map. Given a typical 
mm-wave flux calibration uncertainty of 15\%, our observations do not appear to be missing a large fraction of the HCN flux within the inner $\sim 30\arcsec$ of the 
map compared to previous observations. However, the rigorous portion of these comparison still only compares our data to previous interferometric observations and the lack 
of simple single dish maps of HCN and HCO$^+$ across the whole nucleus make this only an approximate check on the true flux recovery. We expect that our HCN and HCO$^+$ 
observations do miss some extended emission, which leads to visible bowling in the reconstructed images, but we do not expect the missing flux should not be a 
major issue for the individual clouds, which have typical sizes a few arcseconds. Meanwhile we focus our analysis of the starburst structure on 
CO or C$^{17}$O whenever possible.

{\em Submillimeter Array Data:} We supplement the ALMA data with maps of C$^{18}$O (2-1) and the 1mm and 850$\mu$m continuum from the 
Submillimeter Array (SMA) published by \citet{SAKAMOTO06,SAKAMOTO11}. These help constrain the mass of our clouds 
and the excitation temperature of high column density gas. The C$^{18}$O (2-1) data have velocity resolution of only $20$~km~s$^{-1}$. Given the broad 
lines that we observe, we do not expect that the lower velocity resolution substantially affects our results. Note that these data lack a short spacing correction, 
but several whole tracks are included so that the data do have appreciable rotation synthesis; see the original paper for details \citep{SAKAMOTO06,SAKAMOTO11}.

\begin{deluxetable}{lccc}
\tablecaption{Molecular Line Maps of the NGC 253 Starburst} 
\tablehead{ 
\colhead{Transition} & 
\colhead{Telescope} & 
\colhead{Resolution\tablenotemark{a}} & 
\colhead{Noise\tablenotemark{b}}
}
\startdata
CO (1-0) & ALMA & $2.0 \times 1.4\arcsec$ & 0.1~K\tablenotemark{c} \\
C$^{17}$O (1-0) & ALMA & $2.0 \times 1.5\arcsec$ & 0.05~K \\
C$^{18}$O (2-1) & SMA & $1.6 \times 1.5\arcsec$ & 0.5~K \\
\hline
HCN (1-0) & ALMA & $1.9 \times 1.3\arcsec$ & 0.2~K\tablenotemark{c} \\
H$^{13}$CN (1-0) & ALMA & $2.0 \times 1.4\arcsec$ & 0.2~K \\
HCO$^+$ (1-0) & ALMA & $1.9 \times 1.3\arcsec$ & 0.2~K \\
H$^{13}$CO$^+$ (1-0) & ALMA & $2.0 \times 1.4\arcsec$ & 0.2~K \\
CS (2-1) & ALMA & $1.7 \times 1.2\arcsec$ & 0.2~K \\
\hline
850$\mu$m\tablenotemark{d} & SMA & $1.8 \times 1.1\arcsec$ & 470 MJy/sr \\
1mm\tablenotemark{d} & SMA & $1.1 \times 1.1\arcsec$ & 100 MJy/sr \\
\enddata
\tablecomments{First version of CO data presented by \citet{BOLATTO13A}. SMA data from \citet{SAKAMOTO06,SAKAMOTO11}.}
\tablenotetext{a}{For $d=3.5$~Mpc $1\arcsec = 17$~pc.}
\tablenotetext{b}{In a 5 km~s$^{-1}$ wide channel (except C$^{18}$O, see text).}
\tablenotetext{c}{For CO at its native resolution, 1 Jy~beam$^{-1} = 32.7$~K; for HCN 1 Jy~beam$^{-1} = 61.3$~K.
At our common $2 \times 2\arcsec$ working resolution, for CO 1 Jy~beam$^{-1} = 21.6$~K and for HCN 1 Jy~beam$^{-1} = 36.5$~K.}
\tablenotetext{d}{Broadband continuum.}
\label{tab:data}
\end{deluxetable}

\begin{figure*}
\plotone{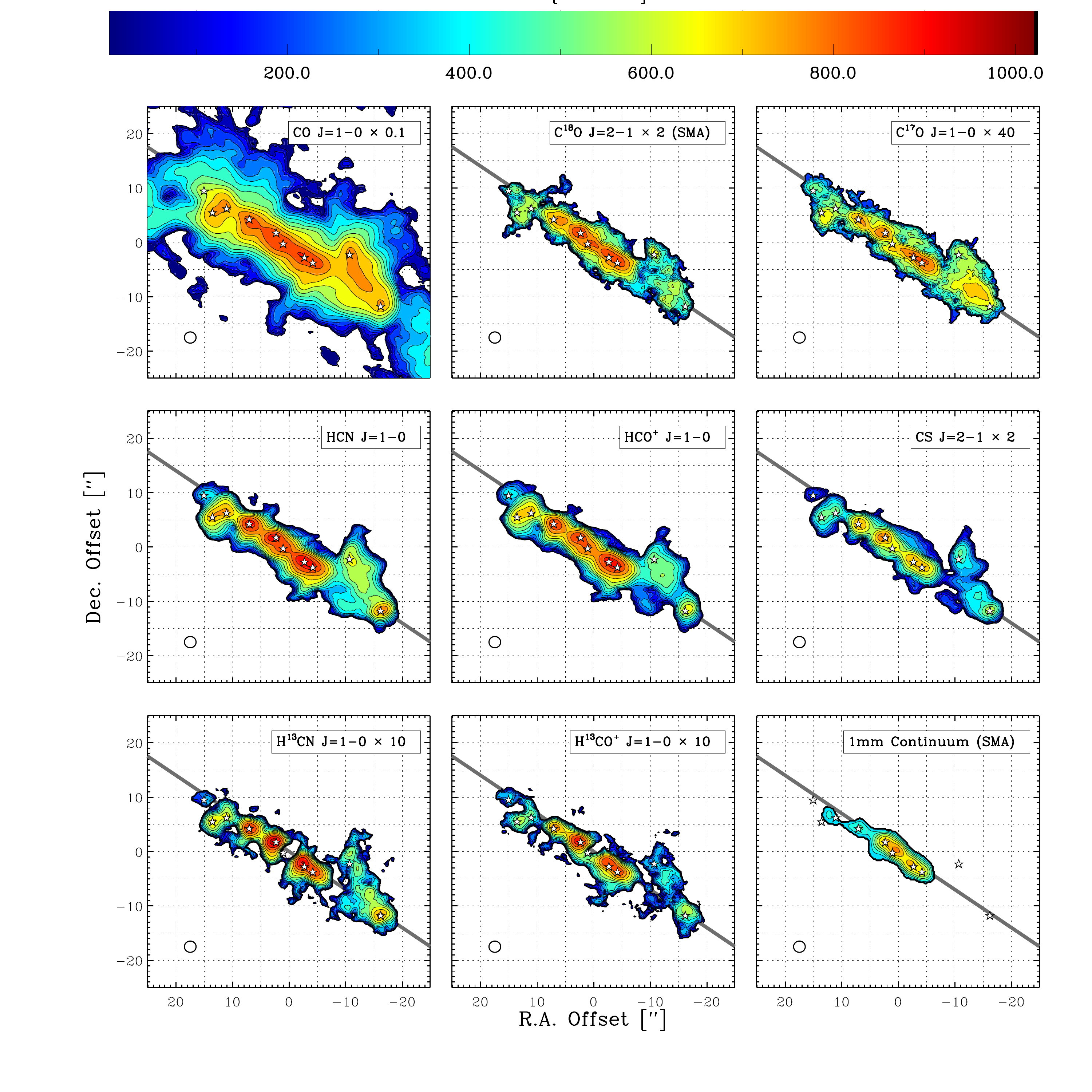}
\caption{Molecular line emission in the nuclear region of NGC 253.
Each map shows integrated emission from a different transition mapped by ALMA
or the SMA \citep[the C$^{18}$O (2-1) and 1mm continuum][]{SAKAMOTO11}.
Stars indicate peaks of molecular clouds (Table \ref{tab:cloudprops} and Figure \ref{fig:assign}). 
A circle in the bottom left indicates the resolution. Successive contours are spaced by factors of $\sqrt{2}$ and
run over the following ranges (in K km s$^{-1}$): $^{12}$CO 50$\rightarrow$3200;
C$^{18}$O: $1\rightarrow512$; C$^{17}$O: $1\rightarrow11$;
HCN, HCO$^{+}$, CS: 2$\rightarrow$1024; H$^{13}$CN, H$^{13}$CO$^{+}$: 3$\rightarrow$136; 
1mm: 24$\rightarrow$6400~MJyr~sr$^{-1}$. Offsets are measured from $\alpha = 0^{\rm h} 47^{\rm m} 33.14^{\rm s}$, $\delta = -25\arcdeg 17\arcmin 17.52\arcsec$
with the gray line indicating the adopted position angle of the starburst.
\label{fig:pp_maps}}
\end{figure*}

\begin{figure*}[ht]
\plotone{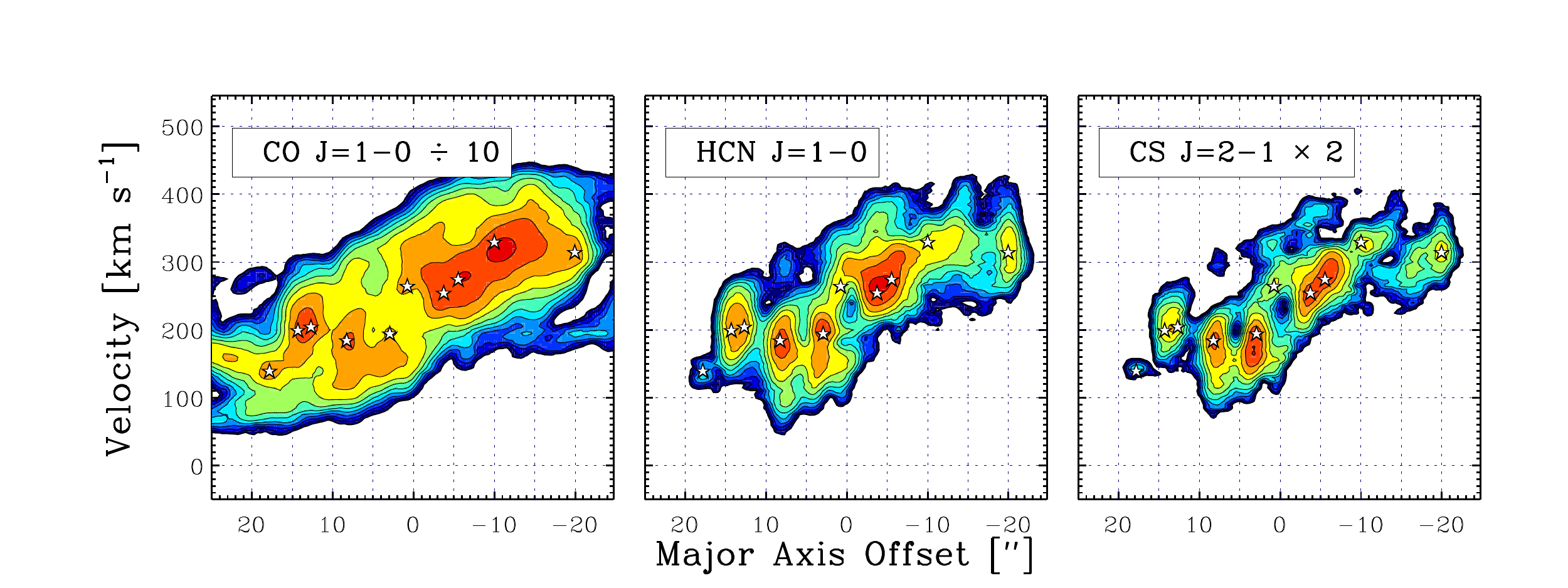}
\caption{Position-velocity diagrams for selected molecular lines adopting the position angle in Figure \ref{fig:pp_maps}. Stars indicate peaks of molecular
clouds (Table \ref{tab:cloudprops} and Figure \ref{fig:assign}).}
\label{fig:pv_maps}
\end{figure*}

\begin{figure}[ht]
\plotone{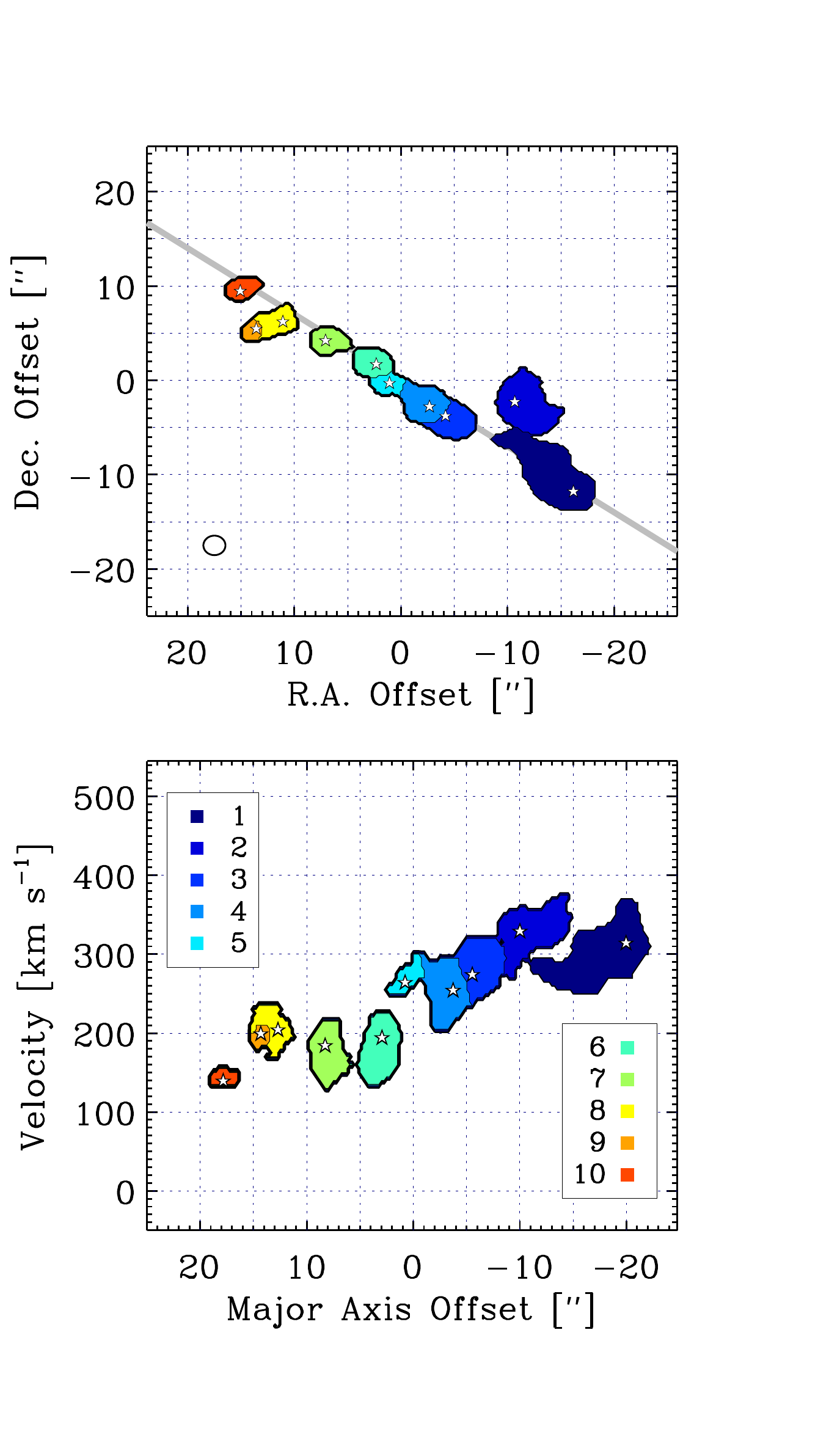}
\caption{Assignment of emission into candidate clouds in position-position ({\em top}) and position-velocity ({\em bottom}) space. Maps show the maximum
cloud assignment over the line of sight; comparing the upper and lower panels reveals some shadowing in velocity space.}
\label{fig:assign}
\end{figure}

\section{Giant Molecular Clouds in NGC 253}
\label{sec:clouds}

Figures \ref{fig:moment_maps} -- \ref{fig:pv_maps} provide the clearest view to date of the molecular medium fueling the starburst at the heart of
NGC 253. They show a clumpy, almost linear distribution of dense gas --- traced by HCN, HCO$^{+}$, and CS -- embedded in a more 
extended distribution of CO emission. The signatures of the shells and outflowing streamers studied by \citet{BOLATTO13A} are strikingly 
evident in the moment and channel maps. Our analysis focuses first on the properties of the compact structures seen in the dense gas tracers
(in this section) and then on the more general structure of the starburst (\S \ref{sec:starburst}).

Astronomers commonly label overdensities in the distribution of molecular line emission on $\sim 50$~pc scales 
as ``giant molecular clouds'' (GMCs). Assuming spherical symmetry, resolved observations of these structures yield luminosities, sizes, and line widths. These 
basic properties scale with one another in ways that reflect the turbulence, density, and mass-to-light ratio of the clouds 
\citep[][]{LARSON81,SOLOMON87,BOLATTO08,DONOVANMEYER13}. The high density tracer emission from NGC 253 exhibits  clear bright spots in 
the intensity cubes (Figure \ref{fig:channel_hcn}) with similar peaks appearing in the maps of all dense gas tracers (Figures \ref{fig:pp_maps} 
and \ref{fig:pv_maps}). Perhaps most strikingly, our peaks almost all correspond to clear local enhancements in the ratio of dense gas tracer
emission to CO (see the middle right panel in Figure \ref{fig:starburst}, which shows a map of the HCN-to-CO ratio, and details in \S \ref{sec:burstcalcs}), supporting the idea that these clouds 
indeed represent dense condensations of gas in a more extended molecular medium. We identify ten clouds, report their properties in Table \ref{tab:cloudprops}, contrast them with 
a large compilation of GMC properties from other systems, infer key time, mass, and length scales for our systems, and compare these to the 
star formation properties of the nuclear burst.

\subsection{Calculation of Cloud Properties}
\label{sec:cloudcalcs}

\begin{deluxetable*}{lccccccccc}
\tablecaption{Molecular Clouds in NGC 253} 
\tablehead{ 
\colhead{Cloud} & 
\colhead{$\Delta$R.A.\tablenotemark{a}} & 
\colhead{$\Delta$Dec.\tablenotemark{a}} & 
\colhead{$v_{\rm LSR}$} & 
\colhead{Radius} & 
\colhead{Line Width} &
\colhead{Mass} &
\colhead{$L_{\rm HCN}$} &
\colhead{$L_{\rm HCO+}$} &
\colhead{$L_{\rm CS}$} \\
\colhead{} & 
\colhead{[\arcsec]} & 
\colhead{[\arcsec]} & 
\colhead{[km s$^{-1}$]} & 
\colhead{[pc]} & 
\colhead{[km s$^{-1}$]} &
\colhead{[$10^7$ M$_\odot$]} &
\multispan{3}{[$10^6$~K~km~s$^{-1}$~pc$^2$]}
}
\startdata
1 & -15.3 & -10.3 &  309. &   43. $\pm$  0.23~dex &   25. $\pm$  0.04~dex &  2.51 $\pm$  0.46~dex &  0.65 &  0.29 &  0.37 \\ 
2 & -11.8 &  -2.2 &  331. &   52. $\pm$  0.13~dex &   17. $\pm$  0.05~dex &  1.33 $\pm$  0.41~dex &  0.64 &  0.47 &  0.31 \\ 
3$^\star$ &  -4.6 &  -3.5 &  285. &   34. $\pm$  0.06~dex &   24. $\pm$  0.14~dex &  3.73 $\pm$  0.47~dex &  1.15 &  1.16 &  0.56 \\ 
4$^\star$ &  -2.3 &  -2.0 &  259. &   31. $\pm$  0.16~dex &   20. $\pm$  0.05~dex &  5.58 $\pm$  0.67~dex &  1.38 &  1.25 &  0.55 \\ 
5 &   0.9 &   0.1 &  271. &   28. $\pm$  0.02~dex &   20. $\pm$  0.32~dex &  0.63 $\pm$  0.65~dex &  0.22 &  0.24 &  0.09 \\ 
6$^\star$ &   2.5 &   2.3 &  182. &   22. $\pm$  0.12~dex &   27. $\pm$  0.14~dex &  3.85 $\pm$  0.63~dex &  0.83 &  0.72 &  0.52 \\ 
7$^\star$ &   6.8 &   4.7 &  177. &   30. $\pm$  0.07~dex &   31. $\pm$  0.31~dex &  1.47 $\pm$  0.48~dex &  0.89 &  0.74 &  0.40 \\ 
8 &  11.9 &   6.6 &  207. &   44. $\pm$  0.06~dex &   24. $\pm$  0.34~dex &  1.08 $\pm$  0.50~dex &  0.50 &  0.43 &  0.18 \\ 
9 &  13.9 &   5.7 &  193. &   11. $\pm$  0.05~dex &   13. $\pm$  0.44~dex &  0.20 $\pm$  0.53~dex &  0.06 &  0.05 &  0.03 \\ 
10 &  14.8 &  10.1 &  143. &   23. $\pm$  0.07~dex &    9. $\pm$  0.30~dex &  0.35 $\pm$  0.43~dex &  0.03 &  0.05 &  0.02 \\ 
\hline
\enddata
\tablenotetext{a}{Relative to the center at $\alpha = 0^{\rm h} 47\arcmin 33.14\arcsec$, $\delta = -25\arcdeg 17\arcmin 17.52\arcsec$.}
\tablenotetext{b}{Particularly for mass, the uncertainty is better described logarithmically than linearly. For convenience, we note that
0.1~dex corresponds to a factor of 1.26; 0.2 dex to 1.58; 0.3 dex to 2.0; and 0.5 dex to 3.16.}
\tablecomments{$\star$ indicates massive dense gas concentrations (\citet{SAKAMOTO11} peaks S1, S2, S4, and S5). Following
convention, radius is defined as $R = 1.91 \sqrt{\sigma_x \sigma_y}$, with $\sigma_x$ the RMS size in the $x$ direction \citep{SOLOMON87}. In this convention
for a Gaussian cloud $R = 0.81$ times the FWHM size. The line width quote is the one dimensional RMS dispersion along the line of sight,
$\sigma_{v}$. See \citet{ROSOLOWSKY06} for details.}
\label{tab:cloudprops}
\end{deluxetable*}

Our high density tracer observations reveal several bright regions with sizes comparable to our resolution. In other galaxies
structures with similar size, velocity width, mass, and contrast with the background are often labeled GMCs. Although these are
usually traced using CO or its isotopologues, the pervasive bright, optically thick $^{12}$CO in the NGC 253 starburst renders it a poor tool to
study the bound star-forming structures \citep[see also][]{SAKAMOTO11,ROSOLOWSKY05}. We identify possible GMCs in NGC 253
from our high critical density tracers, derive position-position-velocity assignments that map the data cube to discrete structures, and then measure their properties.
We discuss the legitimacy of identifying these structures as GMCs in our analysis of the starburst structure (\S \ref{sec:starburst}).

To identify molecular complexes in NGC 253, we identify local maxima based on the CS cube, selecting peaks that 
contrast with the local background by $>2\sigma$ and have both a unique associated area greater than a beam and velocity width of at least three 
channels. We use these local maxima to segment the cubes into discrete structures in position-position-velocity (ppv) space, labeling 
the emission uniquely associated with each peak (i.e., in an unshared isointensity contour) as a discrete cloud \citep[this is the segmentation
approach of CPROPS][]{ROSOLOWSKY06}.

After this procedure, we still found two cases which appeared as blended local maxima in the CS maps. That is, we could distinguish two peaks 
but the shared contour between these peaks was high (forming a ``peanut'' in the channel maps). We took these cases and broke the
single CS peak into multiple peaks using the H$^{13}$CN cubes, which show a higher contrast (but also have lower S/N) than the CS data. For this 
blended peak case, we identify H$^{13}$CN peaks inside the (blended) CPROPS CS cloud and then segment the emission inside that cloud into
sub-clouds employing a variant of the CLUMPFIND algorithm \citep{WILLIAMS94} that uses their distance criteria to assign blended emission within
an isocontour based on distance to the peak.

The output of this exercise is a position-position-velocity (ppv) cube in which the pixel values correspond to the cloud with which the pixel is associated.
Figure \ref{fig:assign} shows projections of this ``assignment cube'' in position-position and position-velocity space. In the end, we identified $10$ clouds 
that closely match what one would see by eye as either peaks in the data cube or enhancements in the dense gas fraction. These peaks appear
in Figures \ref{fig:moment_maps}, \ref{fig:channel_co}, \ref{fig:channel_hcn}, \ref{fig:pp_maps}, and \ref{fig:pv_maps}, where it is clear
that the local maxima identified from CS and H$^{13}$CN also correspond to peaks in the other high density and 
optically thin tracers and represent a good match to what one would pick out by eye. We will also see below that the peaks correspond to enhancements 
in the dense gas fraction, e.g., traced by the ratio HCN/CO.

We apply this assignment cube to each spectral line data cube and then measure the properties of each cloud (ppv region) in each tracer. Following 
an expanded version of the methodology laid out by \citet[][``CPROPS'']{ROSOLOWSKY06}, we measure size, line width, luminosity, and position of each cloud 
based on intensity-weighted moments. We carry out this calculation for each cloud using HCN (1-0), HCO$^{+}$ (1-0), and CS (2-1) 
emission. We use a modified version of the original CPROPS code (Leroy, Rosolowsky et al., in prep.). Key differences 
from CPROPS are: (1) we deconvolve the beam in two dimensions; (2) we employ a wider suite of size and line width 
measures, including the area and an ellipse fit at half max in addition to the second moment; (3) we employ 
additional ``extrapolation'' (aperture correction) approaches, which essentially assume a Gaussian distribution to extrapolate 
the area and ellipse fits. The spread among the suite of lines and approaches used to measure and aperture-correct the cloud properties is significant.
Indeed, given the high signal-to-noise of the data, this systematic uncertainty appears to represent the dominant uncertainty in the measurement. 
We use the range of measured parameters across transition and method as our most realistic gauge of the uncertainty in each cloud property. 
All of the calculations use an updated version of the CPROPS code that is publicly available\footnote{\url{https://github.com/akleroy/cpropstoo}}.

\begin{figure*}
\plottwo{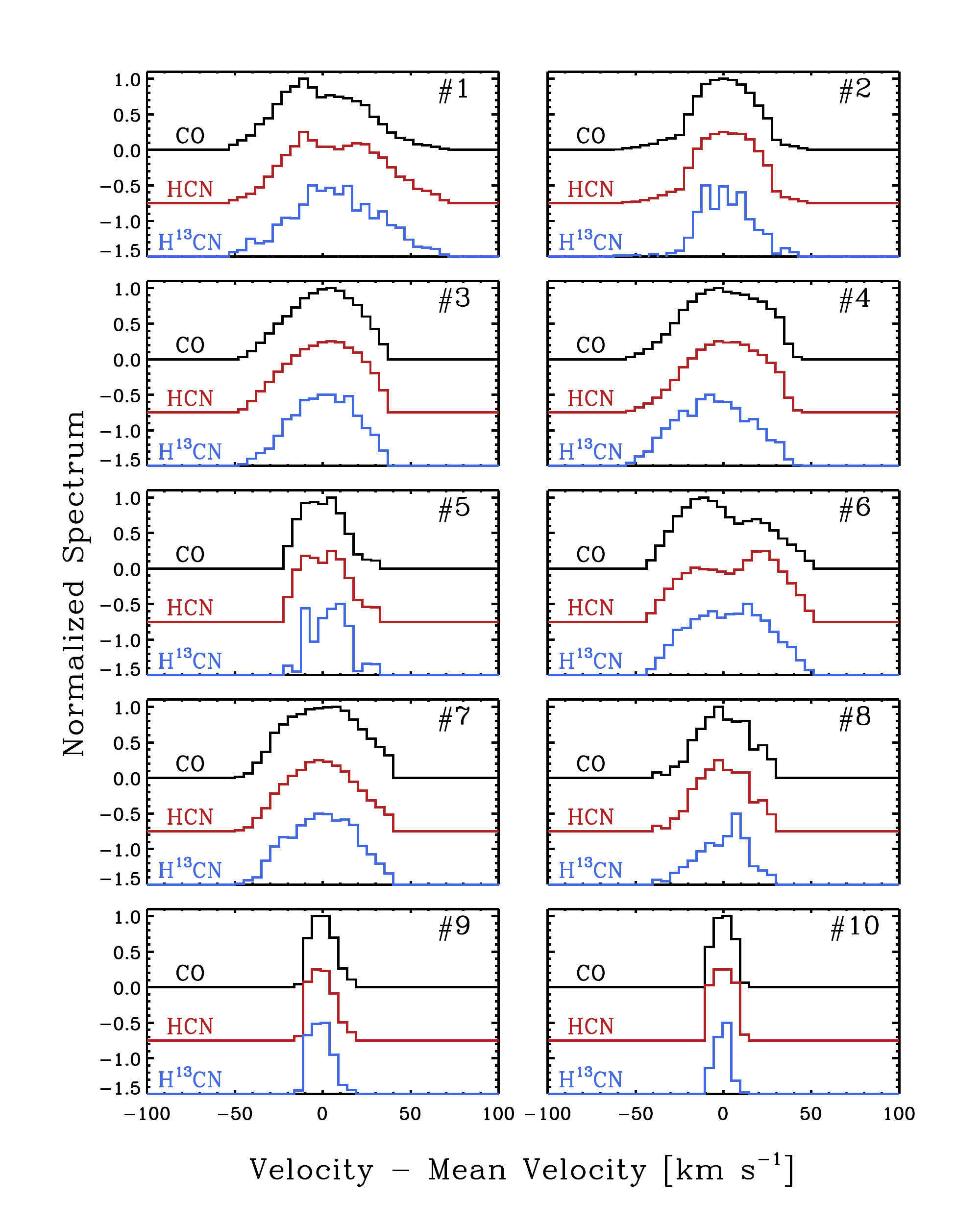}{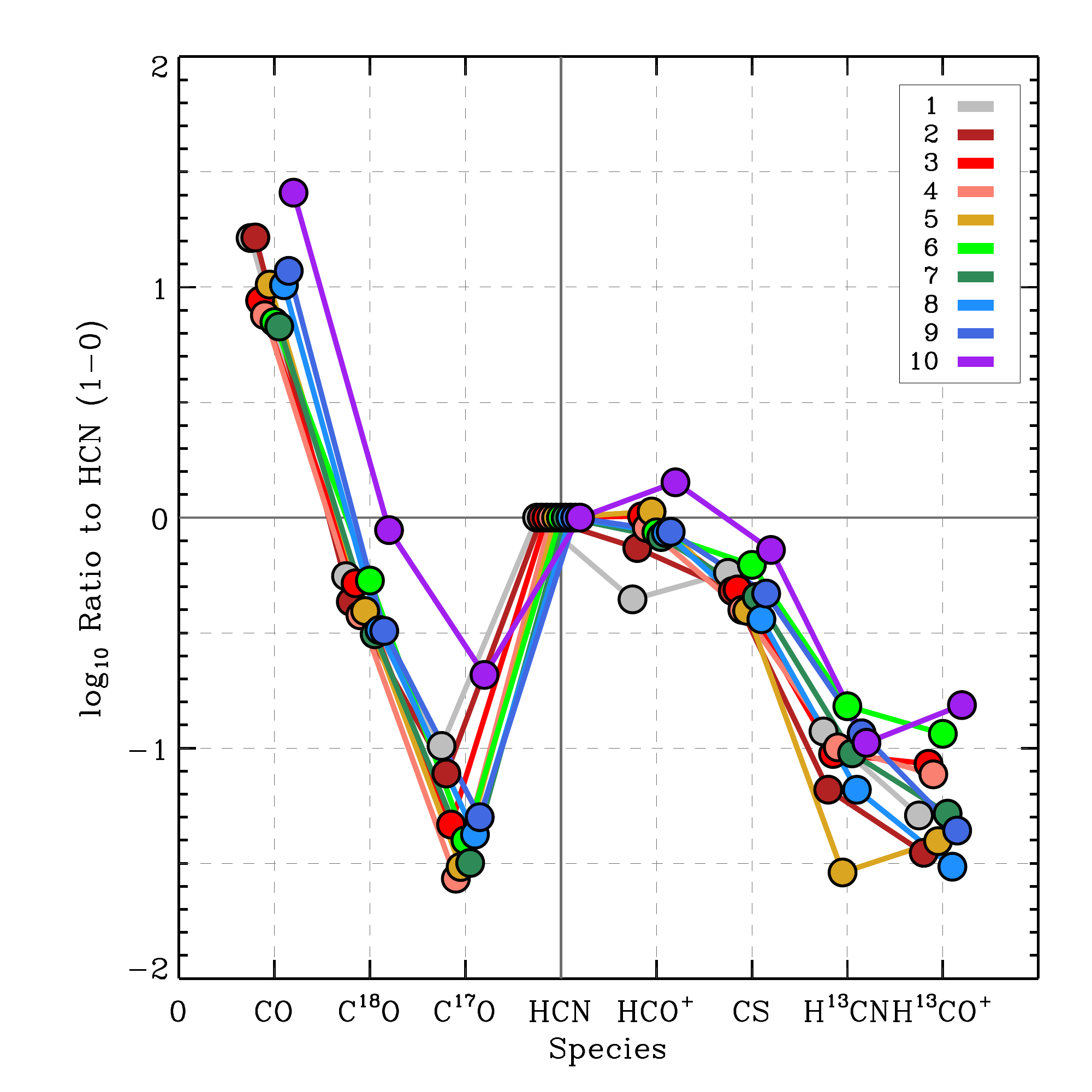}
\caption{({\em left}) Normalized spectra of each cloud in $^{12}$CO (1--0), HCN, and H$^{13}$CN; 
({\em right}) line ratios (taken in K) at the peak of each cloud after convolution to common resolution. Intensities are normalized to the HCN (1-0) intensity. 
The spectra and the line ratios demonstrate that our ``dense gas tracers'' HCN, HCO$^{+}$, and CS track one another well across the starburst
(except perhaps in the outermost clouds), meaning that chemical differences among clouds may not be a major concern. The spectra do show some 
signatures of multiple components and non-Gaussianity, suggesting that some of our sources may blend several more compact structures. The typical
H$^{13}$CN/HCN$\approx 10$, and H$^{13}$CO$^+$/HCO$^+ \approx 15$ contrast with the isotopic abundance of $\sim 40$ to demonstrate 
optical thickness even in the dense gas tracers HCN and HCO$^+$.
\label{fig:cloudspec}}
\end{figure*}

In the end, for each cloud we have measured sizes, line widths, positions, and luminosities in each of the transitions listed above. We report our
best estimate line width, size, and mass (see \S \ref{sec:masscalcs}) for each cloud, along with estimated uncertainties, in Table \ref{tab:cloudprops}. To assess the Gaussianity
of the resulting line profiles and help understand to what degree our clouds may blend several components, we plot the spectrum of each cloud 
in CO, HCN, and H$^{13}$CN in the left panel of Figure \ref{fig:cloudspec}. In order to understand the degree to which chemical or tracer variations 
affect our results, we also measure a set of line ratios measured at matched resolution at each local maximum, which appear in the right 
panel of Figure \ref{fig:cloudspec}; for much more detail on this topic and a more varied set of regions see Meier et al. (submitted).

\subsubsection{Mass Estimates}
\label{sec:masscalcs}

\begin{figure*}
\plottwo{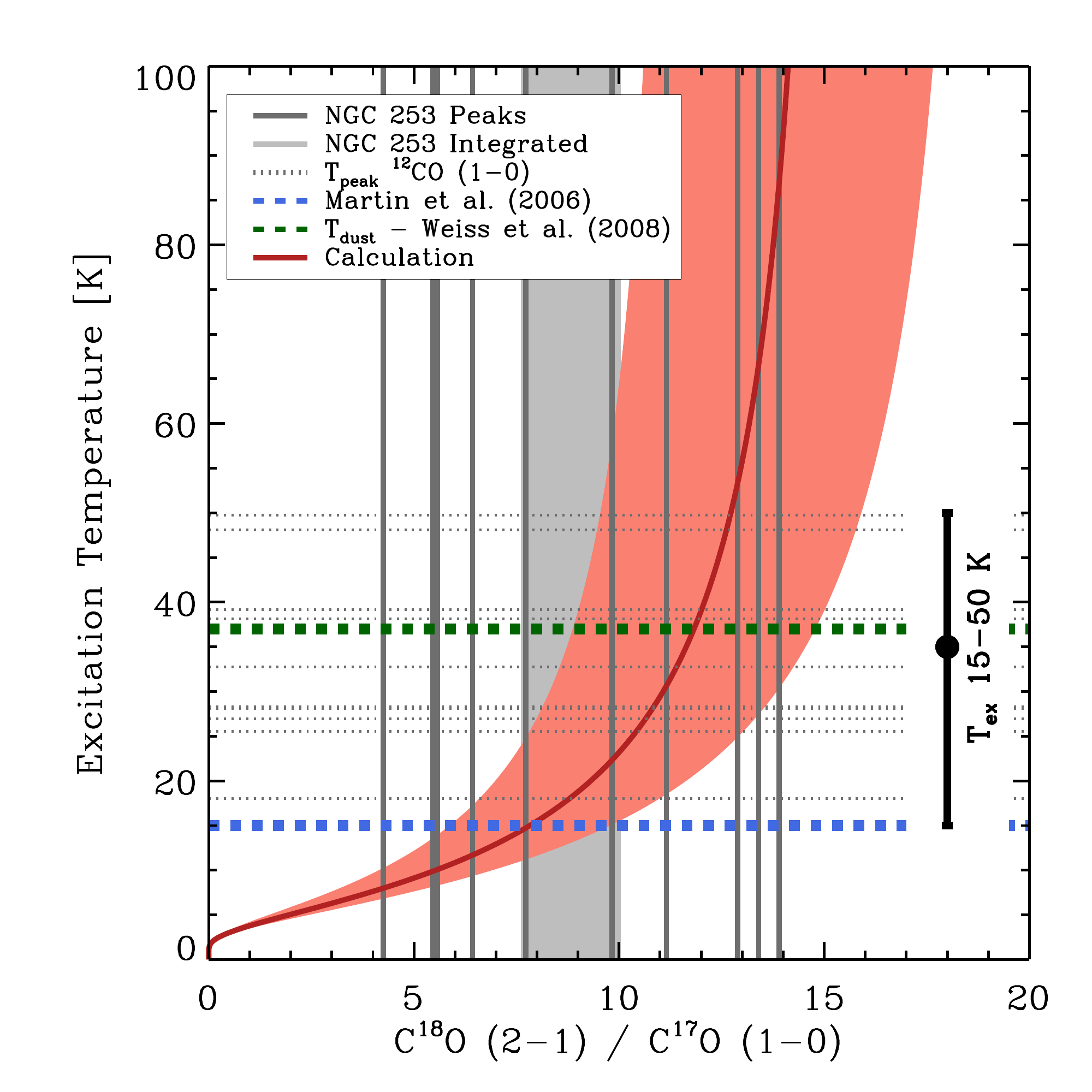}{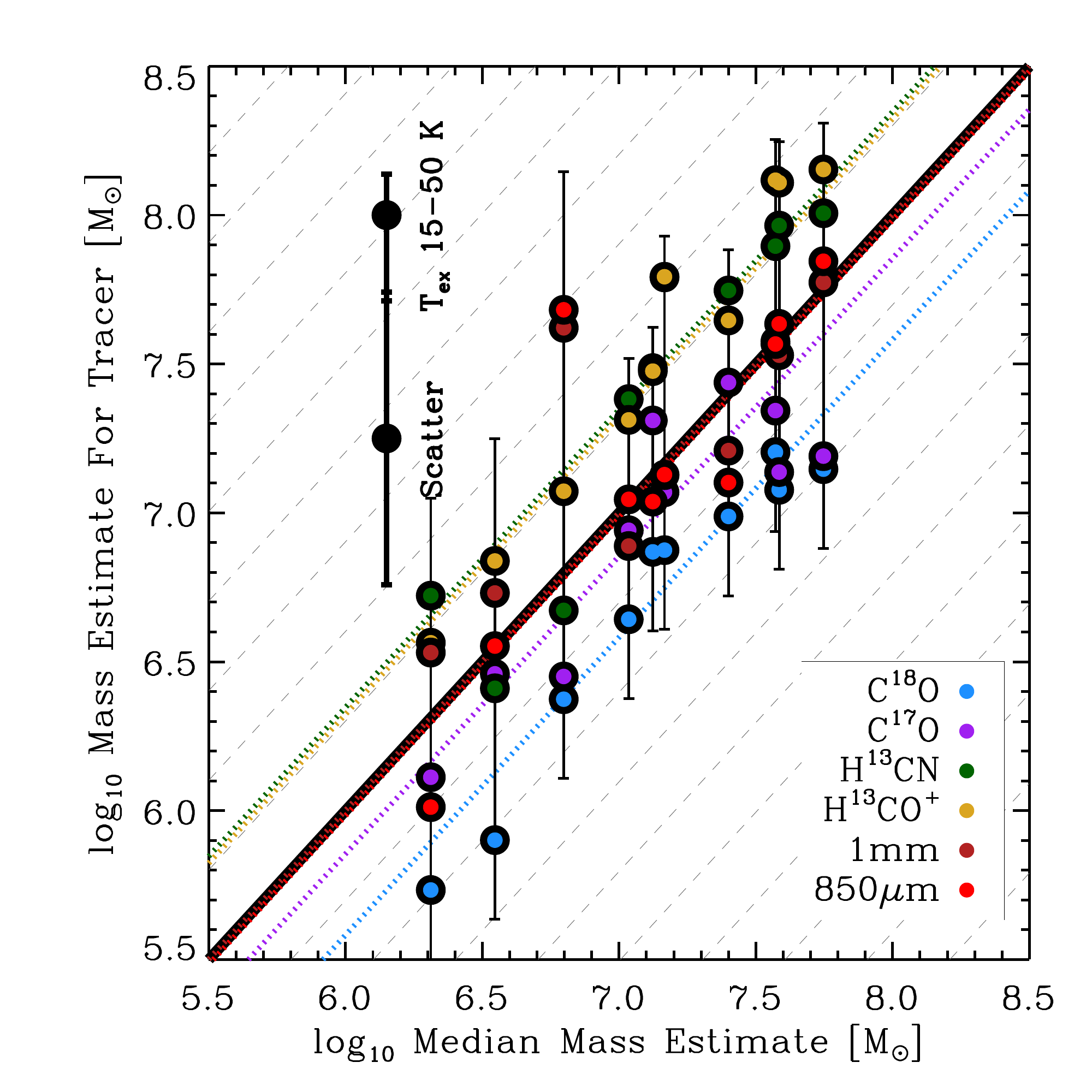}
\caption{Mass-related calculations. ({\em left}) Excitation temperature as a function of the C$^{18}$O/C$^{17}$O ratio for the assumptions
described in the text. The red curve shows the predicted relation; the width reflects the uncertain isotopic ratio. Gray vertical lines
show ratios measured for individual clouds and a light gray rectangle shows the range of ratios for the disk. Dashed horizontal gray lines
show peak CO brightness temperatures for each cloud. Green and blue lines show the temperatures found for the dust by \citet{WEISS08}
and adopted for the gas by \citet{MARTIN06}.  We adopt $T_{\rm ex} \sim 35$~K but repeat our calculations for $15$ and $50$~K to capture 
the uncertainty in the excitation temperature. ({\em right}) Mass estimates of our cloud complexes from optically thin emission --- C$^{18}$O, C$^{17}$O,
H$^{13}$CN, H$^{13}$CO$^{+}$ --- and dust continuum emission \citep{SAKAMOTO11}. See \S \ref{sec:masscalcs} for details 
of the calculations. The solid black line shows equality with dashed gray lines spaced by factors of 2. Vertical error bars indicate
the full spread of mass estimates when the calculations are carried across our temperature range. Colored dashed lines indicate the 
ratios of the median mass for that approach to the median mass for all approaches and thus show the systematic differences among the 
mass estimates. The order of the cloud masses remains mostly the same as our approach 
changes, but the range of plausible assumptions corresponds to a wide range of masses. We proceed using the median across all methods 
and infer a factor of three systematic uncertainty from the scatter among various estimates. \label{fig:massmass}}
\end{figure*}

We derive the mass of the structures that we observe in two ways: from optically thin tracers (H$^{13}$N, H$^{13}$CO$^{+}$,
C$^{18}$O, C$^{17}$O) and from the dust continuum mapped by the SMA. Note that we do not employ dynamical mass estimates
because we wish to estimate the dynamical state of the clouds, i.e., their degree of boundedness and virialization, rather than to assume
a dynamical state. 

Masses from the optically thin transitions rely on the abundances in Table \ref{tab:assumptions}, the assumption of local thermodynamic 
equilibrium (LTE), an estimate of the excitation temperature, and molecular information from the Leiden Atomic Molecular Database 
\citep[LAMDA,][]{SCHOIER05}. To derive these masses, we first calculate the column density in the upper state of the observed transition
by comparing the observed intensity to the probability of spontaneous emission, the Einstein A coefficient. We do so via

\begin{equation}
\label{eq:nupper}
N_{J} = \frac{8 \pi k_B \nu_{J\rightarrow J-1}^2}{h c^3 A_{J \rightarrow J-1}} I_{J \rightarrow J-1} \left[ {\rm K~cm~s}^{-1} \right]
\end{equation}
 
\noindent where $N_{J}$ is the column density of the upper state (e.g., $J=1$ for H$^{13}$CN 1-0), $I_{J \rightarrow J-1}$ is the observed 
intensity in K~cm~s$^{-1}$, $A_{J \rightarrow J-1}$ is the Einstein A coefficient, and $\nu_{J \rightarrow J-1}$ is the frequency. All constants 
are in CGS. The calculation simply combines the definition of the Einstein A coefficient, the Rayleigh Jeans law used to calculate the
brightness temperature, and a factor of $4\pi$ to integrate over all solid angle.

We relate the column density in the upper ($J$) state to the total column density by assuming local thermodynamic equilibrium (LTE).
We calculate the partition function, $Q_{rot}$ from an adopted excitation temperature, $T_{\rm ex}$, and information on the molecule 
energy levels from LAMDA via

\begin{equation}
\label{eq:partfn}
Q_{rot} = \sum_i \left(2 J_i + 1\right) \exp \left( \frac{-E_i}{k~T_{\rm ex}} \right)~.
\end{equation}

\noindent We use $Q_{rot}$ to translate $N_{J}$ to a total column density

\begin{equation}
\label{eq:ntot}
N_{tot} = \frac{N_{J}~Q_{rot}}{\left( 2 J + 1\right) \exp \frac{-E_{J}}{k_B~T_{\rm ex}}}
\end{equation}

\noindent where $E_{J}$ and $J$ again refer to the energy and rotational energy level for the upper state. Finally we apply the 
abundances in Table \ref{tab:assumptions} to derive a total H$_2$ mass.

$T_{\rm ex}$ plays a critical role in this calculation. Without a doubt, NGC 253 harbors gas at a 
wide range of temperatures and the observed emission results from a convolution of several populations
\citep[e.g., see studies of NGC 253 by][]{OTT05,MANGUM13}. The characteristic excitation temperature 
associated with emission from high critical density tracers in NGC 253 has been estimated at anywhere from 
$\sim 15$~K \citep{MARTIN06} based on multi-species, multi-transition modeling to 
$\sim 150$~K \citep[][]{KNUDSEN07} based on the HCN line SED. \citet{SAKAMOTO11} adopted $\sim 50$~K based 
on the peak brightness temperature of the apparently optically thick $^{12}$CO line.

With only low frequency data, we have a limited handle on $T_{\rm ex}$. We make two rough checks: first measuring
the peak CO (1--0) brightness temperature towards our peaks. CO emission appears to mildly optically thick (see Meier et al. submitted)
so that the brightness temperature should approach the temperature at least at the CO-emitting surface. Second, we compare
ALMA C$^{17}$O (1-0) emission to SMA C$^{18}$O (2-1) emission. Assuming both to be optically thin with the isotopic ratio 
in Table \ref{tab:assumptions}, the line ratio approximately relates to the excitation temperature via

\begin{equation}
\label{eq:trot}
\frac{I^{C18O} (2-1)}{I^{C17O} (1-0)} = R_{18/17}~\frac{A^{C18O}_{21}}{A^{C17O}_{10}}
\left(\frac{\nu^{C17O}_{10}}{\nu^{C18O}_{21}}\right)^2 \frac{5}{3} \exp \frac{- \Delta E}{k_B T_{\rm ex}}
\end{equation}

\noindent where $R_{18/17}$ is the C$^{18}$O/C$^{17}$O ratio, $\Delta E$ is the energy difference between C$^{18}$O (2-1)
and C$^{17}$O (1-0), $5/3$ reflects the degeneracies, and $\nu$ and $A$ refer to the frequencies and Einstein A coefficients 
for the relevant transition.

The left panel in Figure \ref{fig:massmass} shows the relationship between this line ratio 
and $T_{\rm ex}$ in red. The red range indicates the effect of a 25\% uncertainty in the isotopic ratio. Gray vertical lines show ratios for individual 
clouds. The large gray region shows the range for the disk. Overall, the calculation suggests a wide range of temperatures, with about
half the clouds showing high enough line ratios that we have a limited handle on the temperature. Based on 
this and on \citet{MARTIN06} and \citet{SAKAMOTO11} we adopt $T_{\rm ex} \sim 35$~K as our fiducial value and $15$--$50$~K as 
our plausible $T_{\rm ex}$ range to calculate uncertainties. This would correspond to the cooler component from the
unresolved study of NGC 253 by \citet{MANGUM13} and is roughly consistent with the excitation temperatures of the lower ammonia transitions 
in NGC 253 as seen by \citet{OTT05}, as well as with the ammonia formation temperature that they infer from the ortho to para ratio.
While hotter molecular gas is certainly present, we expect our low-$J$ lines to preferentially
sample a massive, cooler component rather than a low-mass hot component. This temperature also matches the characteristic 
$T_{\rm dust}$ \citep{WEISS08}; given the high densities that we find, a rough correspondence of $T_{\rm dust}$ and $T_{\rm gas}$ 
seems reasonable.

For each cloud, we calculate the mass for $T_{\rm ex} = 15$, $35$, and $50$~K. We derive the luminosity for the calculation using an aperture 
correction based on the peak-to-edge ratio within the cloud assignment and assuming 
the source and beam to be Gaussian. We carry out this calculation for C$^{18}$O, C$^{17}$O, H$^{13}$CN, and H$^{13}$CO$^+$. The 
range of temperatures and suite of molecules are used to calculate the uncertainty in the mass estimates.

We also use the submm dust continuum emission measured by the SMA to estimate the mass of our clouds. We associate 
all dust within the two-dimensional area of each cloud with the cloud and so simplify the calculation at the expense of including 
contamination from unassociated emission along the line of sight. We derive dust masses assuming a wavelength dependence 
of the opacity $\tau \propto \lambda^{-\beta}$ with $\beta = 1.5$ and mass absorption coefficient of dust at 250$\mu$m
$\kappa_{250} = 10$cm$^{2}$~g$^{-1}$. Using $\beta$, we calculate $\kappa$ at $\lambda=1$mm and $850\mu$m. We 
adopt $T_{\rm dust} = 35$~K and test a plausible range 15--50~K \citep{WEISS08}. This matches our adopted gas temperature 
but we note that the densities, while high, are not necessarily high enough that the temperatures must be coupled. From the intensity at 850$\mu$m and 1mm, 
we calculate the optical depth, $\tau$, and convert this to a mass surface density of dust using $\kappa$. We translate this to a surface 
density of gas assuming a dust-to-gas ratio 1-to-150.

Thus we have six mass estimates: four optically thin lines and two dust bands which we calculate for a representative range of temperatures. 
Figure \ref{fig:massmass} compares these estimates, which show a factor of $\sim 3$ scatter. This appears to 
represent a reasonable uncertainty, reflecting our underlying uncertainty in $T_{\rm ex}$, the molecular and dust abundances, and 
observational systematics (e.g., aperture corrections, resolution and projection effects, $(u, v)$ coverage). We take the scatter among mass 
estimates as our uncertainty and, lacking a strong reason to prefer one estimate, use the median among all estimates as our mass estimate.
Note that Figure \ref{fig:massmass} demonstrates that the uncertainty in the mass estimation is largely systematic; e.g., the optically
thin tracers H$^{13}$CO$^{+}$ and H$^{13}$CN return the highest masses for almost all clouds, so that adjusting their abundances
down by a factor of $\approx 2$ would bring most of our tracers into substantial agreement (leaving the SMA C$^{18}$O as the outlier). 
Future multi-transition studies will refine our knowledge of the abundances, excitation, deviations from LTE, and continuum SED and so improve these estimates.

\subsubsection{Notes on Length, Mass, and Time Scales}
\label{sec:length}

When comparing our cloud properties to literature measurements, we express our measurements in conventional terms for the GMC literature 
\citep[see Table \ref{tab:cloudprops} and][]{ROSOLOWSKY06}. We also find it useful to discuss cloud properties in terms of the full width at 
half maximum (FWHM) \footnote{For comparison, the conventional ``GMC'' definition of radius is $R = 0.81 FWHM$ \citep{SOLOMON87}.}
Doing so, we adopt the simplifying assumption of Gaussian clouds and also calculate surface density, volume density,
crossing times, and free fall times within the FWHM. For a three dimensional Gaussian, $\approx 1/3.4$ the mass is concentrated within the 
FWHM, and the average surface and volume densities within this FWHM radius are:

\begin{eqnarray}
\left< \Sigma \right>_{FWHM} &=& 0.77~\frac{M}{\pi~R^2} \\
\left< \rho \right>_{FWHM} &=& 1.26~\frac{M}{4/3~\pi~R^3}
\end{eqnarray}

\noindent with $M$ the mass of the cloud and $R$ the radius as conventionally defined for GMCs, $R = 1.91 \sqrt{\sigma_x \sigma_y}$ (Table \ref{tab:cloudprops}). Note that
the distinction between a two and three dimensional Gaussian matters for this correction (the enclosed mass is $1/2$ for a two dimensional Gaussian) and we employ
the three dimensional version.

To calculate the crossing time, we follow theoretical convention and use the three dimensional velocity dispersion, which we take to be $\sqrt{3}$ times the line of 
sight velocity dispersion. We quote crossing times relative to the FWHM of the structure in question, so that 

\begin{equation}
\tau_{\rm cross} = \frac{FWHM}{\sqrt{3} \sigma_{1D}} = \frac{1.23 R}{\sqrt{3} \sigma_{1D}}
\end{equation}

\noindent with $\sigma_{1D}$ the one dimensional velocity dispersion. Similarly, the free fall time is quoted relative to the density within the FWHM, so that 

\begin{equation}
\tau_{\rm ff} = \frac{\sqrt{3 \pi}}{4 \sqrt{2 G \rho_{FWHM}}}
\end{equation}

\noindent with $G$ the gravitational constant. Finally, we calculate the Mach number, $\mathcal{M} = \sqrt{3} \sigma_{1D} / c_s$, with $c_s = 0.45$~km~s$^{-1}$ calculated
assuming molecular hydrogen at $T=35$~K.

\begin{deluxetable}{lcc}
\tablecaption{Massive Star-Forming Complexes} 
\tablehead{ 
\colhead{Property} & 
\colhead{NGC 253} & 
\colhead{Disk GMC\tablenotemark{a}}
}
\startdata
$\left< R \right>$ & 30~pc & 19~pc \\
$\left< \sigma \right>\tablenotemark{b}$ & 22~km~s$^{-1}$ & 2~km~s$^{-1}$ \\
$\left< M \right>$ & $1.4~\times 10^7$~M$_\odot$ & $1.5 \times 10^5$~M$_\odot$\\
$\left< \Sigma \right>_{\rm FWHM}$ & 6,000~M$_\odot$~pc$^{-2}$ & 115~M$_\odot$~pc$^{-2}$ \\
$\left< n \right>_{\rm FWHM}$ & 2,000~cm$^{-3}$ & 75~cm$^{-3}$ \\
$\mathcal{M}$ & 85 ($T=35$~K) & 11 ($T=20$~K) \\
$\tau_{\rm ff}^{\rm FWHM}$ & 0.7~Myr & 2.5~Myr \\
$\tau_{\rm cross}^{\rm FWHM}$ & 1.0~Myr & 6.7~Myr \\ 
\enddata
\label{tab:sketchclouds}
\tablenotetext{a}{Here the ``Disk GMC'' reports the median properties of a cloud from the \citet{HEYER09} Milky Way sample taking $\alpha_{\rm CO} = 4.35$~\acounits .}
\tablenotetext{b}{Line of sight velocity dispersion.}
\tablecomments{Mass values systematically uncertain by $\sim 0.5$~dex. Considering only the four most massive complexes increases the typical NGC 253 cloud mass,
density, and surface density by a factor of $\approx 2$ (see the starred clouds in Table \ref{tab:cloudprops}).}
\end{deluxetable}

\subsection{Giant Molecular Clouds from Other Galaxies}
\label{sec:litgmc}

To understand the physics of the clouds in NGC 253, we find it useful to compare them to GMCs in more quiescent systems. To this end, 
we assemble a set of cloud property measurements from the recent literature. \citet{COLOMBO14} cataloged GMCs across the inner $9 \times 9$~kpc of M51. 
\citet{DONOVANMEYER13,DONOVANMEYER12} measured cloud properties in the inner regions of three nearby spirals. \citet{ROSOLOWSKY05} studied 
the clouds in the nucleus of NGC 4826. \citet{BOLATTO08} measured cloud properties from galaxies in the Local Group and nearby dwarf galaxies. 
\citet{WONG11} measured cloud properties from the regions of bright CO emission in the Large Magellanic Cloud. \citet{OKA01} observed cloud-like structures 
in the Galactic center, \citet{HEYER01} cataloged structures from the outer Milky Way, and \citet{HEYER09}
recalculated the properties of the \citet{SOLOMON87} large Milky Way clouds. Homogenizing the methodology and data from these diverse studies
represents a substantial research project on its own, beyond the scope of this paper. Here we focus on a broad brush comparison 
and take the literature data as presented and we will see that even this coarse comparison reveals an interesting contrast.
For those wishing to focus on the most rigorous comparison, the \citet{COLOMBO14}, \citet{BOLATTO08}, and \citet{WONG11}
clouds were all measured using a methodology that closely resembles our own. For the disk of the Milky Way and other galaxies except the LMC,
we apply a Milky Way CO-to-H$_2$ conversion factor, $\alpha_{\rm CO} = 4.35$~\acounits\ to estimate the mass. For the LMC we apply the 
recommendation of \citet{WONG11} and apply a conversion factor two times this value. For the Galactic center clouds of \citet{OKA01}, we
adopt $\alpha_{\rm CO} = 1$~\acounits ; see that paper and \citet{BOLATTO13B} for discussion.

\subsection{Results for the Cloud View}

We identify ten massive candidate GMCs in the nuclear region of NGC 253 from our high density tracer maps. The most massive 
four of these (\# 3, 4, 6, and 7) correspond well to four of the five  star-forming complexes identified by \citet{SAKAMOTO11}.
The \citet{SAKAMOTO11} cloud that we do not identify lies at the galaxy center,
where our analysis returns conflicting results. We do identify a cloud near the nucleus, \#5, but it corresponds to only a weak enhancement
in emission. The nucleus is a strong continuum source and shows wide CO lines but does not stand
out strongly in our dense gas tracers. This may reflect chemical or physical variations due to nuclear activity or it may reflect the difficulty
in disentangling broad lines from the strong continuum. In either case, the nucleus proper remains a subject for future investigation while
an AGN, if present is not expected to be energetically dominant or contribute much of the luminosity \citep[e.g., see][]{WEAVER02,BOLATTO13A}.

The other clouds lie at larger projected radii than the massive clouds, likely in regions of lower ambient density. They
may represent an intermediate population between the starburst clouds and the literature disk population. Note that one of our
clouds, \#2, lies at the base of the molecular outflow identified by \citet{BOLATTO13A}. It is displaced somewhat from the major axis and near one of the 
supershells identified by \citet{SAKAMOTO06}, and it may have been subjected to strong feedback. This entrainment of dense
gas in a starburst driven outflow has also been seen in M82 by \citet{KEPLEY14} and is discussed further by Walter et al. (in prep.).

{\em Substructure in the Line Profile:} Even with 35~pc resolution, our clouds  show evidence for substructure. The left 
panel of Figure \ref{fig:cloudspec} shows the average spectrum of each cloud $^{12}$CO, HCN, and H$^{13}$CN. Though vaguely Gaussian, almost all 
of the profiles show significant broadening or suggestions of multiple components. We have every expectation that higher resolution observations, especially 
of optically thin high critical density tracers, will resolve significant spatial substructure within the objects that we label clouds. However, we also show evidence below that the
structures that we have already identified are gravitationally bound at our $35$~pc resolution. Note that we do not see evidence for the self-absorption found by \citet{SAKAMOTO11}.

{\em Line Ratios:} The burst in NGC 253 exhibits clear chemical variations, which we explore in Meier et al. (submitted). However, the set of high density 
tracers studied in this paper do not appear to vary strongly among the clouds we study. Figure \ref{fig:pp_maps} shows this qualitatively;
the maps of different tracers resemble one another. The right panel of Figure \ref{fig:cloudspec} shows the line ratios 
(normalized to HCN) quantitatively. The clouds share a common pattern of line ratios. The two strongest outliers tend to be clouds \#1 and \#10, the
two clouds at largest radii (and so perhaps the two clouds in environments most distinct from the others). These appear to show somewhat 
high C$^{17}$O-to-HCN ratios, perhaps indicating lower volume densities than the other clouds, and deviate (in opposite directions)
from the other clouds in their average HCN-to-HCO$^{+}$ ratio.

Several of the line ratios plotted in Figure \ref{fig:cloudspec} are also of quantitative interest. We find CO-to-HCN ratio (in K) 
$\approx 10_{-3}^{+6}$, HCN-to-HCO$^{+} \approx 1.2_{-0.2}^{+0.2}$, HCN-to-CS$\approx 2.1_{-0.4}^{+0.4}$, and 
CO-to-C$^{17}$O$\approx 210_{-34}^{+65}$. This HCN-to-CO ratio resembles other starbursts \citep{GARCIABURILLO12,GAO04}. We find 
HCN/H$^{13}$CN$\sim 10^{+5}_{-2}$ and HCO$^{+}$/H$^{13}$CO$^{+}\sim 14^{+14}_{-5}$. For a typical isotopic ratio $^{12}$C/$^{13}$C 
of 40, these ratios indicate significant optical thickness for HCN and HCO$^{+}$ (see Meier et al. submitted). We only measure line ratios towards
the peaks; for a full discussion of the topic across the system see Meier et al. (submitted).

\begin{figure*}
\plotone{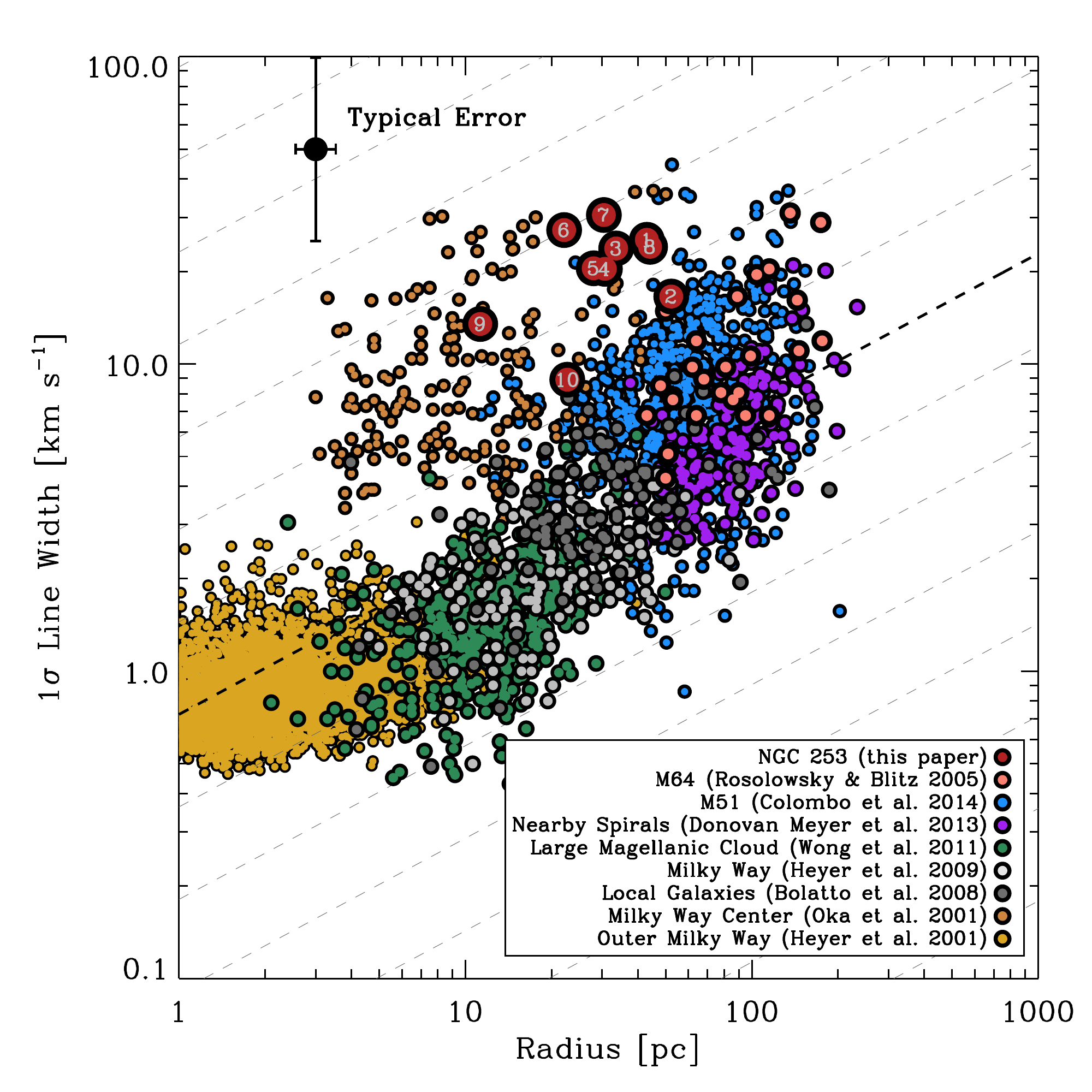}
\caption{The line width-size relationship for our NGC 253 complexes (red circles labeled with cloud number) and giant molecular clouds
from other nearby galaxies and the Milky Way. Dashed lines show $\sigma \propto R^{0.5}$, the relation expected for virialized clouds with fixed
surface density, spaced by a factor of two in $\sigma$ and a factor of four in $\Sigma$; the bold line corresponds to $\Sigma \approx 285$~M$_\odot$~pc$^{-2}$.
Clouds in NGC 253 show very high line widths relative to their size, similar to the enhancement in
line widths seen in clouds near the center of the Milky Way \citep{OKA01} and the largest clouds in the nucleus of M64 \citep{ROSOLOWSKY05}.
For $T = 35$~K, the median Mach number for the NGC 253 clouds is $\mathcal{M} \sim 85$ and median the turbulent crossing time (FWHM definition)
is $\sim 1$~Myr.
\label{fig:sizelw}
}
\end{figure*}

\begin{figure*}
\plotone{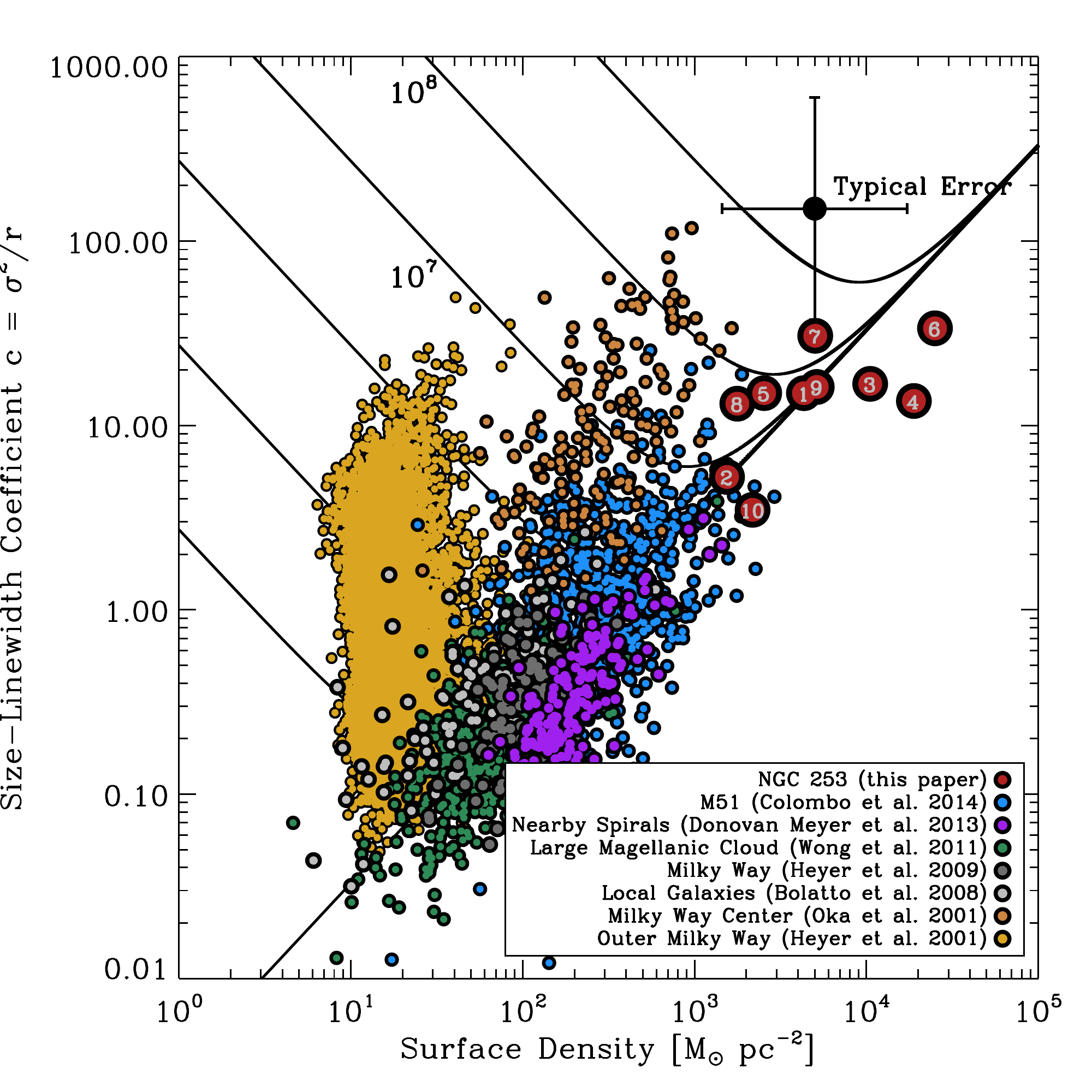}
\caption{Dynamical state of our NGC 253 complexes (red circles labeled with cloud number) and giant molecular clouds from other nearby
galaxies and the Milky Way. The $y$-axis shows $\sigma^2 / R$, which has a one-to-one relationship with surface density ($x$-axis) for
a virialized cloud with negligible surface pressure (heavy line). Confinement by external pressure allows a larger line width for a cloud. Based on simplifying 
assumptions about cloud structure, 
\citet{FIELD11} suggest that the plotted curves describe an equilibrium for external pressures ranging from $P / k_B = 10^3$~cm$^{-3}$~K 
to $10^{9}$~cm$^{-3}$~K; the curves are labeled with the corresponding external $P/k_B$. Most extragalactic clouds follow a relation consistent 
with subdominant external pressure and only surface density variations driving their line widths, but small clouds in the outer Milky Way
\citep{HEYER01} and clouds in the Milky Way center \citep{OKA01} exhibit wide line widths that require external pressure for confinement. The clouds that
we observe in NGC 253 appear consistent with self-gravitating confinement without the need to invoke external pressure. Their wide line widths simply reflect 
the high internal pressure required for equilibrium given their very high surface densities. The cloud entrained in the outflow (\#2) resembles
a normal disk galaxy GMC in this space.
\label{fig:csig}
}
\end{figure*}

\begin{figure*}
\plotone{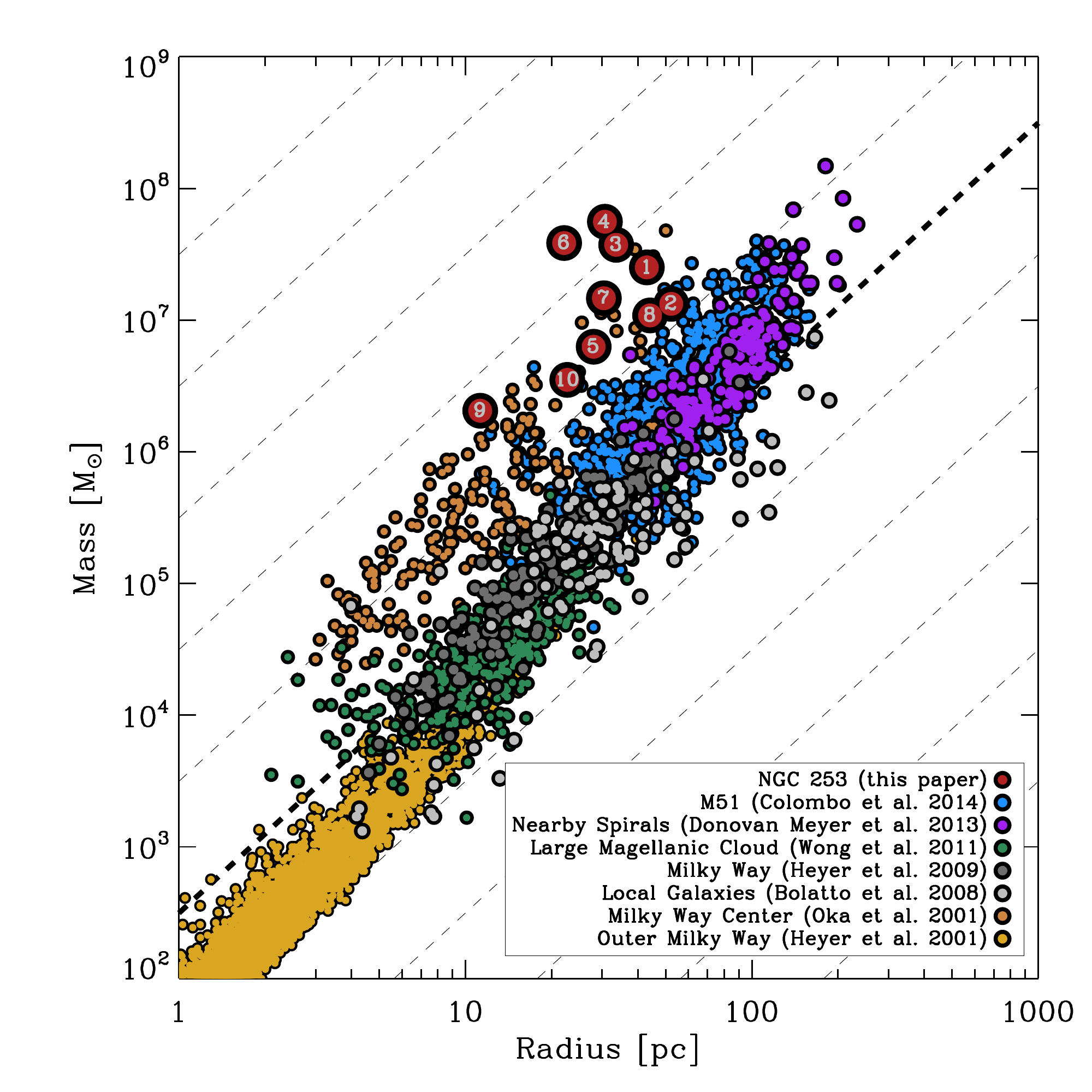}
\caption{Mass as a function of radius for our complexes in NGC 253 (red circles labeled with cloud number) and giant molecular clouds
from other nearby galaxies and the Milky Way. \S \ref{sec:masscalcs} describes our mass calculations; for the literature clouds we use a 
Milky Way conversion factor \citep[2.25 times this value for the LMC][]{HUGHES10}. Dashed lines show fixed surface density, with the 
bold line indicating $\Sigma = 100$~M$_\odot$~pc$^{-2}$. NGC 253 clouds, especially the four most clearly defined, massive complexes, show exceptionally high surface 
and volume densities; using the FWHM definition, for the four most massive clouds $\left< \Sigma \right>_{\rm FWHM} \approx 18,000$~M$_\odot$~pc$^{-2}$
and $\left< n \right>_{\rm FWHM} \approx 6,000$~cm$^{-3}$ ($\sim 500$~M$_\odot$~pc$^{-3}$).
In the Milky Way such conditions are associated with star-forming cores or protoclusters that typically make up a small part of a 
molecular cloud \citep[e.g.,][]{LADA12}. Even though these structures are massive and dense, note that young 
clusters still have radii $10\times$ smaller and densities $10\times$ higher than our complexes so that these do not represent true ``protoclusters.''
\label{fig:massrad}
}
\end{figure*}

{\em Cloud Scaling Relations:} Figures \ref{fig:sizelw} -- \ref{fig:massrad} compare the properties of the NGC 253 complexes to 
molecular structures measured in other environments. Collectively they demonstrate that the NGC 253 clouds are exceptionally 
massive with very large line widths although they are not unusually large in size. Table \ref{tab:sketchclouds} summarizes our
comparison of the NGC 253 clouds to a typical Milky Way GMC \citep{HEYER09}.

{\em Line Width-Size Relation:} Figure \ref{fig:sizelw} shows the line width of a cloud as a function of its size. This parameter space 
reflects the surface density of clouds in virial equilibrium, the character of turbulence when considered across a wide range of scales, and 
the confining pressure for clouds that are not gravitationally bound. All of these effects are evident in our literature compilation. Moving from low to high size, we see 
pressure-bound outer Milky Way clouds \citep{HEYER01} transition into the line width-size relation for dwarf and spiral galaxy GMCs 
\citep[][]{COLOMBO14,DONOVANMEYER13,WONG11,HEYER09, BOLATTO08}. Displaced to high line width for a given size, we see
the Milky Way center clumps observed by \citet{OKA01} \citep[see also][]{SHETTY12B}. These clumps show high line widths for a given size when compared to GMCs
in less extreme environments, an effect that has been attributed to the influence of external pressure \citep{OKA01,FIELD11} . Similar
high line widths at a fixed size, though of smaller magnitudes, appear in the central molecule-rich regions of NGC 6946 \citep[][purple]{DONOVANMEYER13} and 
NGC 4826 \citep[][light red]{ROSOLOWSKY05}. The NGC 253 clouds exhibit high line widths even though their sizes are not markedly different from a spiral galaxy GMC. 
Instead, our NGC 253 clouds appear to extend the Milky Way center linewidth-size population up to higher sizes and line widths or to present
more compact, higher line width versions of the extreme NGC 6946 and NGC 4826 clouds.

{\em Origin of the Line Widths:} Is self-gravity sufficient to confine clouds with these high line widths? The line width for a cloud in virial equilibrium with low 
external pressure can be calculated from its mass 
surface density and size. To examine the interplay of line width and surface density, Figure \ref{fig:csig} plots $\sigma^2 / r$ as a function of mass surface density, $\Sigma$
\citep[e.g.,][]{HEYER09}. The thick diagonal line shows the expectation for virialized, isolated clouds, $\Sigma \approx  330 \sigma^2 / R$
 \citep[though clouds in free fall look quite similar][]{BALLESTEROSPAREDES11}. Clouds clustered around this line, which includes much of the population 
of Milky Way and nearby galaxy GMCs, have line widths that can be explained by self gravity. Clouds above this line, including many of
the small outer Milky Way clouds and the Milky Way center clouds, have line widths that may reflect a high pressure, turbulent 
medium rather than gravitationally bound clouds; for clouds more than a factor of two above this line the kinetic energy of the cloud exceeds the gravitational
energy. \citet{FIELD11}, following \citet{KETO86}, considered clouds confined by a combination of  surface pressure and internal 
gravity and calculated the relationship between $\sigma^2/r$ and surface density for a range of external pressures. We plot their curves in Figure \ref{fig:csig}.

Figure \ref{fig:csig} shows that external pressure is not required to explain the very high line widths observed for NGC 253 clouds. Within the
uncertainties, the surface density of the clouds is high enough that the observed line widths are actually what one would expect from clouds 
with these masses and sizes in virial equilibrium. The plot also shows the ambient pressures that would be needed,
according to the \citet{FIELD11} calculations, in order for clouds at a given surface density and size to have a broader line width. For most of our 
clouds, the calculations suggest that a ISM pressure $P/k_B \gtrsim 10^8$~cm$^{-3}$~K. Though certainly far above the thermal pressure 
\citep[$P/k_B \sim 10^6$~K~cm$^{-3}$ according to ][]{SAKAMOTO11}, the turbulent pressure could approach this value in a region with 
$\log_{10} n$~[cm$^{-3}$] of $3$--$4$ and velocity dispersion $\sim 20$--$40$~km~s$^{-1}$, numbers 
that are plausible for NGC 253. We discuss an alternative origin for the line width --- line of sight blending of a rotating disk --- in the next section, but note that for our 
preferred ``bar'' geometry this will not dominate the observed line widths. The most straightforward explanation of Figure \ref{fig:csig} is that the peaks
that we identified indeed correspond to bound clouds with internal pressure placing them near the locus of virial equilibrium (or at least marginal boundedness).

{\em Density and Surface Density:} As one might infer from their ability to drive these high line widths or simply their appearance in an HCN and HCN-to-CO
map, the clouds in NGC 253 have high volume and surface densities. Figure \ref{fig:massrad} shows mass as a function of radius for NGC 253 clouds and 
literature data. At matched radius, the clouds in NGC 253 are more massive than similar size clouds in normal spirals 
by an order of magnitude. Even the lowest density NGC 253 cloud show surface densities comparable to Milky Way center clouds or the highest surface 
density clouds in M51, that is, at the extreme end of the population of GMCs known from more quiescent systems.

A corollary to this is that a typical surface density of clouds in NGC 253 within the FWHM is $\Sigma_{\rm FWHM} \sim 6,000$~M$_\odot$~pc$^{-2}$
and a typical volume density is $n_{\rm FWHM} \sim $2,000~cm$^{-3}$ or $\sim 200$~M$_\odot$~pc$^{-3}$ 
\citep[in good agreement with][]{PAGLIONE04,SAKAMOTO11}. In the  Milky Way, these high surface and volume densities would correspond 
only to the densest, immediately star-forming parts of a Galactic molecular cloud \citep[e.g.,][naively these clouds have $A_V \sim 400$~mag]{KAINULAINEN09}.

We lack an exhaustive knowledge of the cluster population in NGC 253, but observations do show a number of 
massive ($\gtrsim 5 \times 10^4$~M$_\odot$) young clusters and at least one super star cluster with $M_* \sim 10^6$~M$_\odot$ 
\citep{WATSON96,FERNANDEZ09,KORNEI09}. These clusters tend to have radii of a few pc so that both the size and the mass of
the clusters, even the super star clusters, are an order of magnitude smaller than the GMCs that we pick out. Clusters must form out
of even denser, smaller substructures within the candidate clouds that we observe. Given the substructure visible in the line profiles
of individual clouds, we expect that future ALMA observations will be able to reveal these individual regions.

The volume densities that we estimate represent averages over the sizable area of our $35$~pc beam. These will relate back to a more 
physical local volume density via some filling or clumping factor. We may estimate a lower limit to the local volume density by
comparing the observed brightness temperatures of CO and HCN. Line ratios with optically thin isotopologues argue that we may consider
both to be at least somewhat optically thick \citep[see above and Meier et al. submitted and][]{SAKAMOTO11}. Further assuming that the
CO more nearly fills the beam and that it captures the temperature of the HCN, the ratio $T_{\rm HCN} / T_{\rm CO}$ represents an approximate
two dimensional filling factor. For our clouds, the median peak HCN brightness temperature is $\approx 4$~K and the typical ratio of CO to
HCN peak brightness is $\approx 8$; simply assuming an intrinsic temperature of $35$~K we find a median ratio $\approx 9.5$. In either case,
the implied filling factor for the dense gas within our beam, assuming that it is optically thick with a brightness temperature similar to the CO, is
$\approx 9$. Accounting for this, the local volume density in our clouds would be $n \gtrsim 3 \times 10^4$~cm$^{-3}$ with the inequality reflecting
that this calculation treats only two dimensions. This value approaches the effective density for HCN to be bright, implying a degree of self constancy to
this overall picture.

{\em Masses and Jeans Mass:} Though we lack the sample size to form a meaningful mass spectrum, the characteristic mass of the objects that
we observe in NGC 253 is clearly much higher than what is found in spiral galaxies. Numerical simulations suggest a link between the two dimensional
Jeans mass in a disk and the characteristic mass of gaseous structures formed by large-scale gravitational instability \citep[e.g.,][though note that these simulations assumed an effective temperature rather
than explicitly modeling turbulence]{KIM02,KIM03,KIM06}. As a result, this quantity, $M_{\rm J, 2D} = \sigma^4 / (G^2 \Sigma)$, has been invoked as
a characteristic cloud mass \citep[e.g.,][]{KRUMHOLZ12B}. In \S \ref{sec:burstcalcs} we will estimate $\sigma \sim 30$~km~s$^{-1}$ and $\Sigma \sim 2,500$~M$_\odot$~pc$^{-2}$
for the burst region overall. For these numbers, the implied Jeans mass is $\sigma^4 / G^2 \Sigma \sim 2 \times 10^{7}$~M$_\odot$,
in good agreement with our cloud masses. In this case, the concentration of gas to the center of the galaxy and the high degree of turbulence naturally produce
the very massive clouds that we observe.

{\em Timescales, Mach Number, and Relation to Star Formation:} For $n_{\rm FWHM} = 2,000$~cm$^{-3}$, the implied free fall time
is $\tau_{\rm ff} \sim 0.7 \times 10^5$~yr, $\sim 4$ times shorter than is found for more normal GMCs (Table \ref{tab:sketchclouds}). Adopting $\tau_{\rm ff}$ 
as a characteristic timescale for star formation in clouds \citep[e.g.,][]{KRUMHOLZ12B,FEDERRATH12} and assuming an efficiency of $\sim 0.5\%$ per free fall 
time, the four massive clouds each form $\sim 0.15$~M$_\odot$~yr$^{-1}$ and the ensemble of clouds accounts for $\sim 1$--$2$~M$_\odot$~yr$^{-1}$, 
about the SFR of the nuclear burst (with very roughly $\sim 5 \times 10^4$~M$_\odot$, comparable to a massive cluster, formed per cloud per free fall time). 
\citet{SAKAMOTO11} have already shown the good correspondence between the molecular peaks 
and signatures of recent star formation in NGC 253; these clouds appear to be star formation engines each capable of accounting for a 
large part of the total star formation activity in the galaxy. On its own, the higher density of the star-forming structures can explain much of the enhanced
star formation efficiency in the NGC 253 starburst; the denser clouds simply collapse faster.  Given the measured velocity dispersions, 
the implied crossing time is $\tau_{\rm ff}^{\rm FWHM} \sim 1$~Myr; again $\sim 5$ times shorter than for normal "disk" GMCs.

Assuming a  thermal temperature of $T \sim 35$~K to compute the sound speed, the clouds have Mach number  $\mathcal{M} \sim 30$--$120$, 
with $\mathcal{M} \sim 85$ on average. The clouds are thus highly supersonic. High Mach numbers are expected to 
broaden the distribution of densities found within the cloud \citep{PADOAN02}. Relevant to star formation, the additional compression to high densities from such
broadening should increase the fraction of gas that reaches critical density necessary for the onset of star formation. That is, the broader 
density distribution will yield a larger set of high density regions that should collapse and form stars. For discussion of the analytic predictions
of the effect of Mach number on the critical density for star formation and the overall efficiency with which a cloud forms stars see \citet{FEDERRATH12}.
The line width of our clouds exceeds that of typical GMCs by a factor of $\sim 15$, though the gas in NGC 253 is also hotter than in a typical GMC. 
At face value, the Mach number in NGC 253 likely exceeds that in a typical GMC by a factor of $\sim 10$. This not only implies a greater efficiency of star formation, but
the higher degree of turbulence could plausibly lead to a flatter core mass function \citep{HOPKINS13}, which might in turn affect the
initial mass function of stars or clusters.

\subsection{Synthesis of the Cloud View} 

This cloud view of NGC 253 reveals massive star-forming structures with high density and high surface density. These
structures emerge as peaks in data cubes of dense gas tracers but also appear naturally in line ratio maps contrasting HCN and CO. That
is, they appear to be peaks in both the overall gas density and perhaps the dense gas fraction. Compared to a large
ensemble of literature cloud measurements based on CO observations, these clouds have large line widths, high masses, high surface densities, and high volume densities.
There is some tendency for the clouds at larger radii to have less extreme properties, so that they tend to overlap extreme clouds from the
Milky Way center and M51. The surface densities of the NGC 253 clouds are large enough that the line widths can be explained from self gravity alone,
leading us to conclude that we have indeed identified bound, star-forming structures. Assuming that our mass estimates, which are not dynamical, 
hold, then even if these clouds break apart into smaller structures in future observations they appear likely to remain bound.

The NGC 253 clouds are highly supersonic with free fall and crossing times $\sim 5$ times shorter than their more quiescent ``disk'' counterparts 
and Mach numbers $\sim 10$ times higher than in normal disks. These shorter timescales and enhanced turbulence likely combine to produce the 
more efficient star formation observed in the burst. Indeed, each of these clouds is capable of producing $\sim 0.15$~M$_{\odot}$~yr$^{-1}$ for plausible assumptions. 
The clouds far exceed individual clusters in scale and mass and still have lower densities than observed stellar clusters, 
so that these clouds still do not  represent true ``protoclusters,'' but a class of dense, massive GMC that is not seen in the  disks of normal galaxies,
including our own. Both these basic arguments, the already visible substructure in the line profiles, and filling factor arguments based on brightness temperatures 
lead us to expect that future, even higher resolution, studies with ALMA have the potential to reveal the immediate sites of star formation within the clouds.

\section{The ``Disk'' View of the Nuclear Starburst}
\label{sec:starburst}

One can also view the starburst in NGC 253 as a more continuous molecular structure, a mixture of star-forming clouds and 
diffuse, still mostly molecular, interstellar material, within which self-gravitating structures grow, form stars, and then
disperse \citep[e.g.,][]{OSTRIKER11,SHETTY12}. In this section we measure the properties of the nuclear gas distribution
making no distinction between bound and unbound gas. As in the cloud view, our focus is on  measuring the key length, mass,
and time scales in the burst and comparing these to simple expectations.

\subsection{Structure Calculations}
\label{sec:burstcalcs}

\begin{deluxetable}{lc}
\tablecaption{Structural Properties of the Nuclear Gas Disk} 
\tablehead{ 
\colhead{Property} & 
\colhead{Value} 
}
\startdata
Major Axis $r_{50}$ [$r_{90}$]\tablenotemark{a} --- CO & 150 [400]~pc \\
Major Axis $r_{50}$ [$r_{90}$] --- dense gas & 100 [300]~pc \\
Minor Axis FWHM\tablenotemark{b} --- CO & 100~pc \\
Minor Axis FWHM --- dense gas & 55~pc \\
Total Gas Mass & $\approx 3.5 \times 10^8$~M$_\odot$ \\
CO Luminosity & $\approx 3.3 \times 10^8$~K~km~s$^{2}$ \\
\enddata
\label{tab:structdisk}
\tablenotetext{a}{Radius encompassing 50 and 90\% of the flux along the major axis profile. CO averages }
\tablenotetext{b}{Minor axis FWHM measured from a Gaussian fit to the stacked profile with the beam size deconvolved.}
\tablenotetext{c}{Median from H$^{13}$CN, H$^{13}$CO$^{+}$, C$^{17}$O, C$^{18}$O, and dust emission \citep{WEISS08}. Scatter is $\approx 0.3$~dex.}
\tablecomments{``Dense gas'' measurements average HCN, HCO$^{+}$, and CS after normalizing the profiles. CO from averaging $^{12}$CO and C$^{17}$O.}
\end{deluxetable}

\begin{figure*}
\epsscale{0.8}
\plottwo{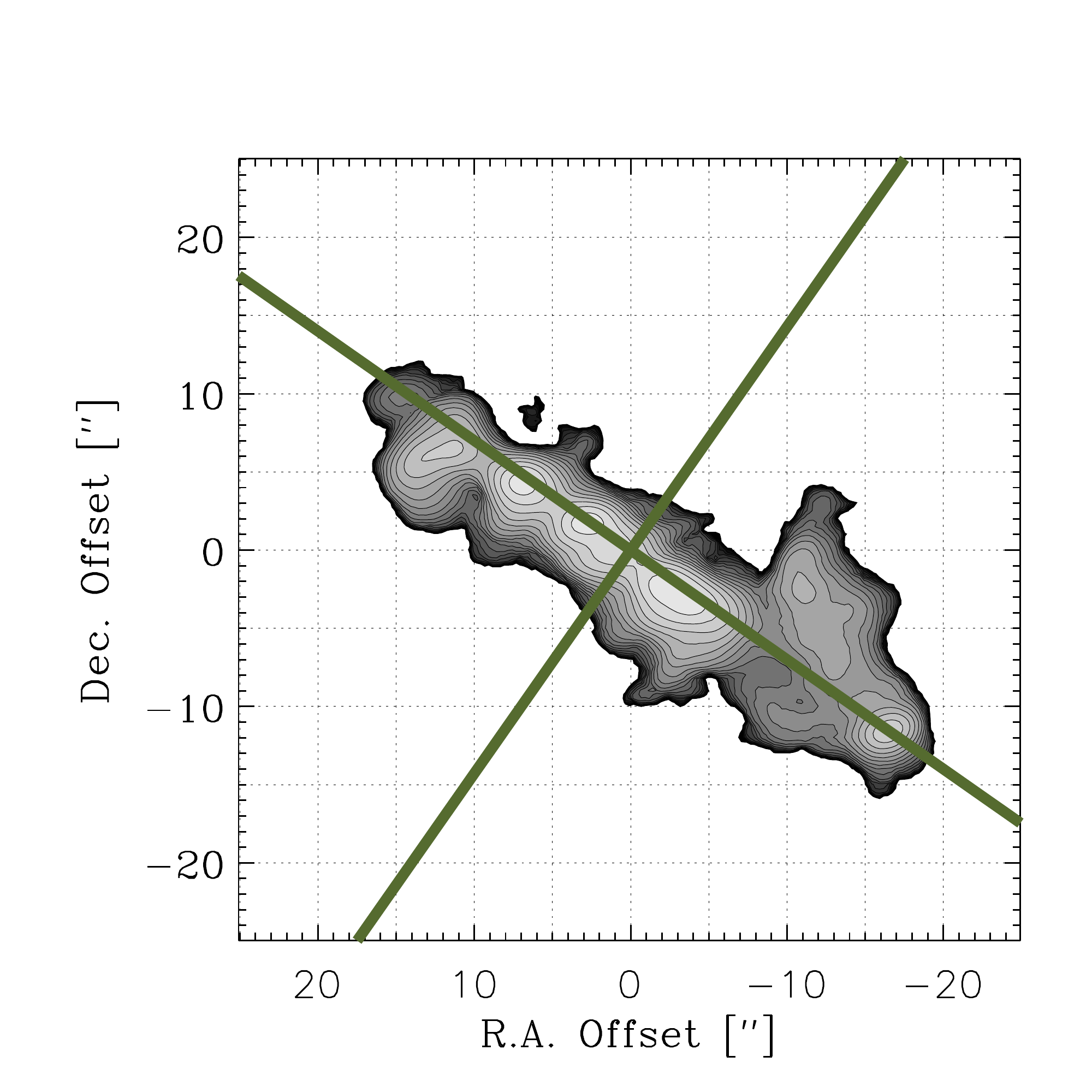}{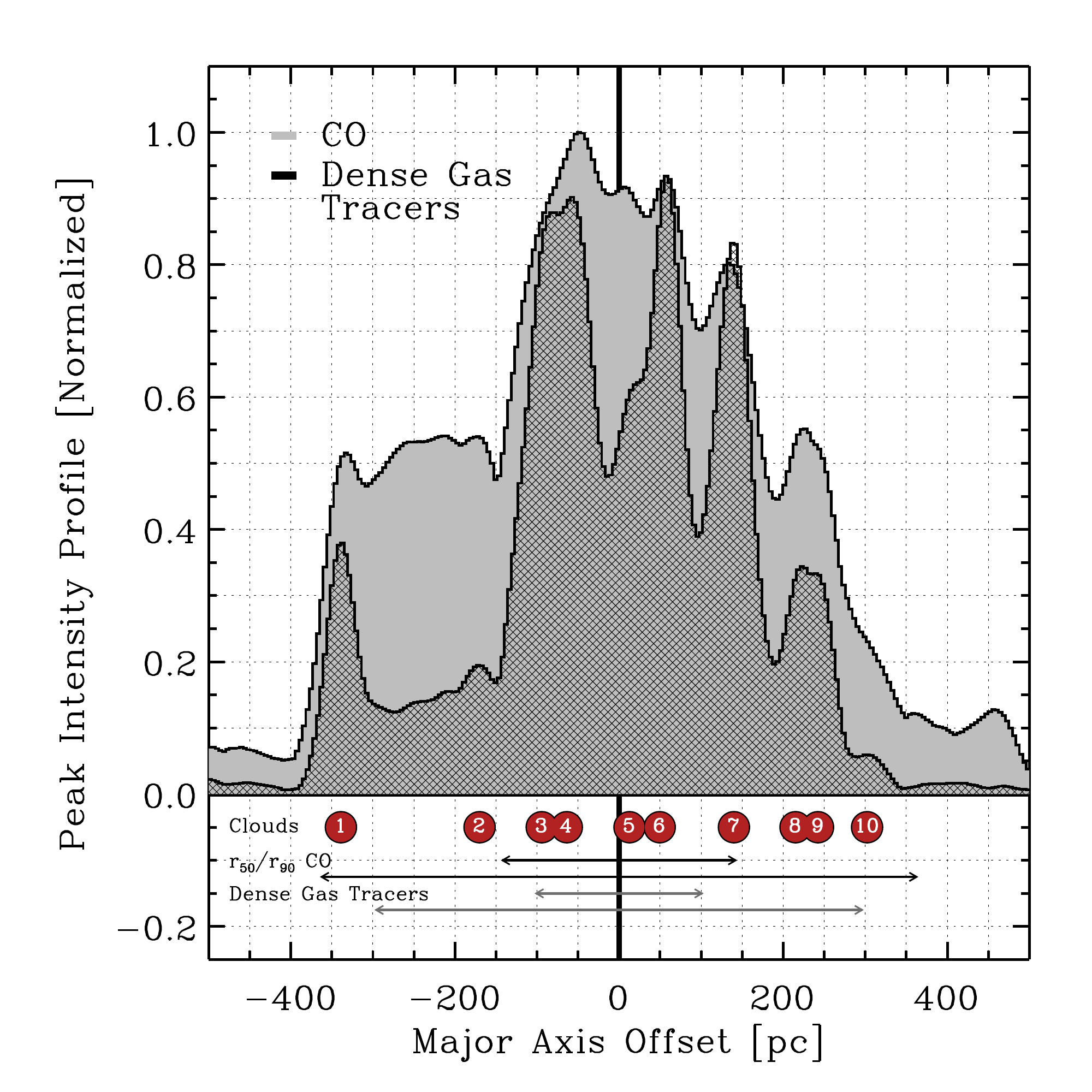}
\plottwo{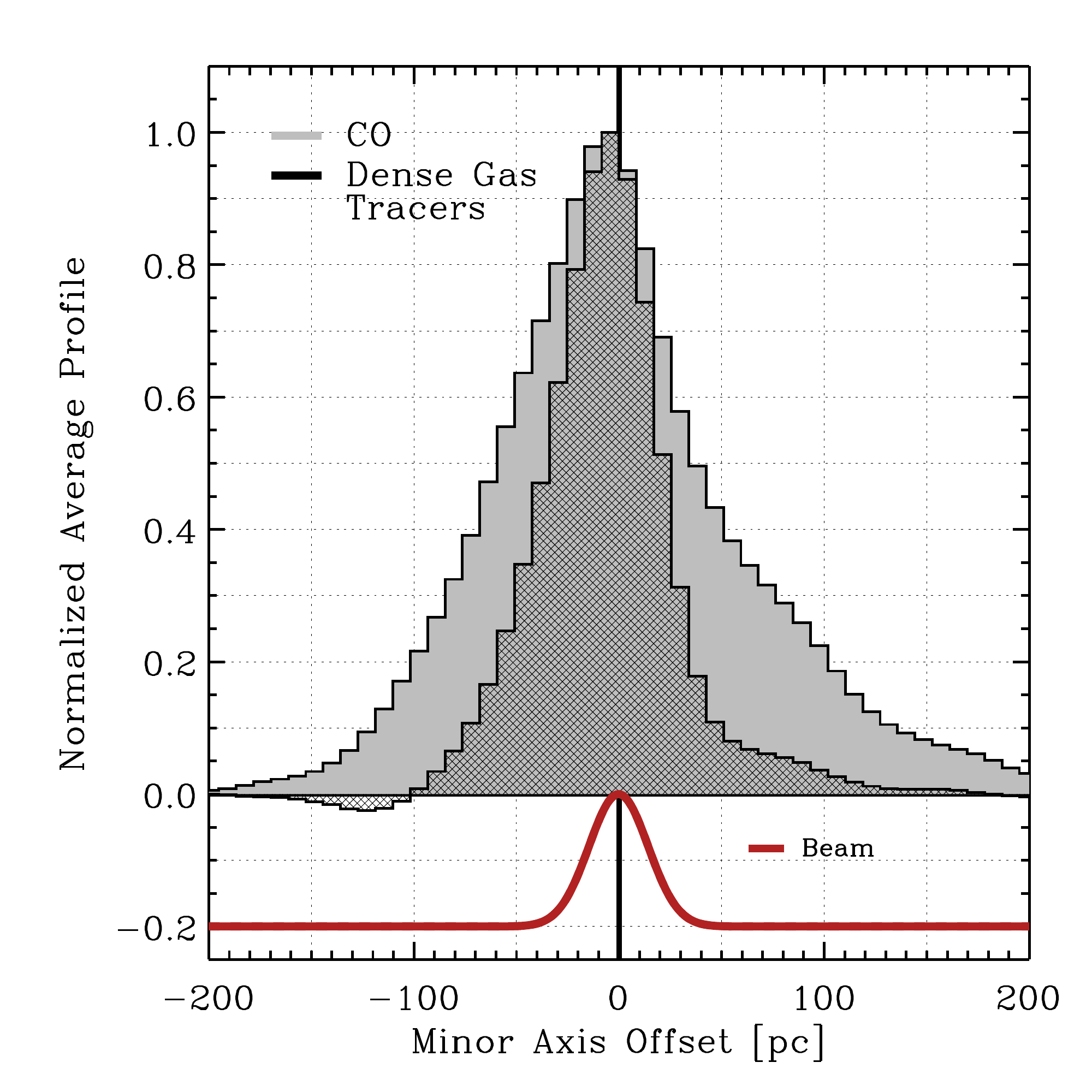}{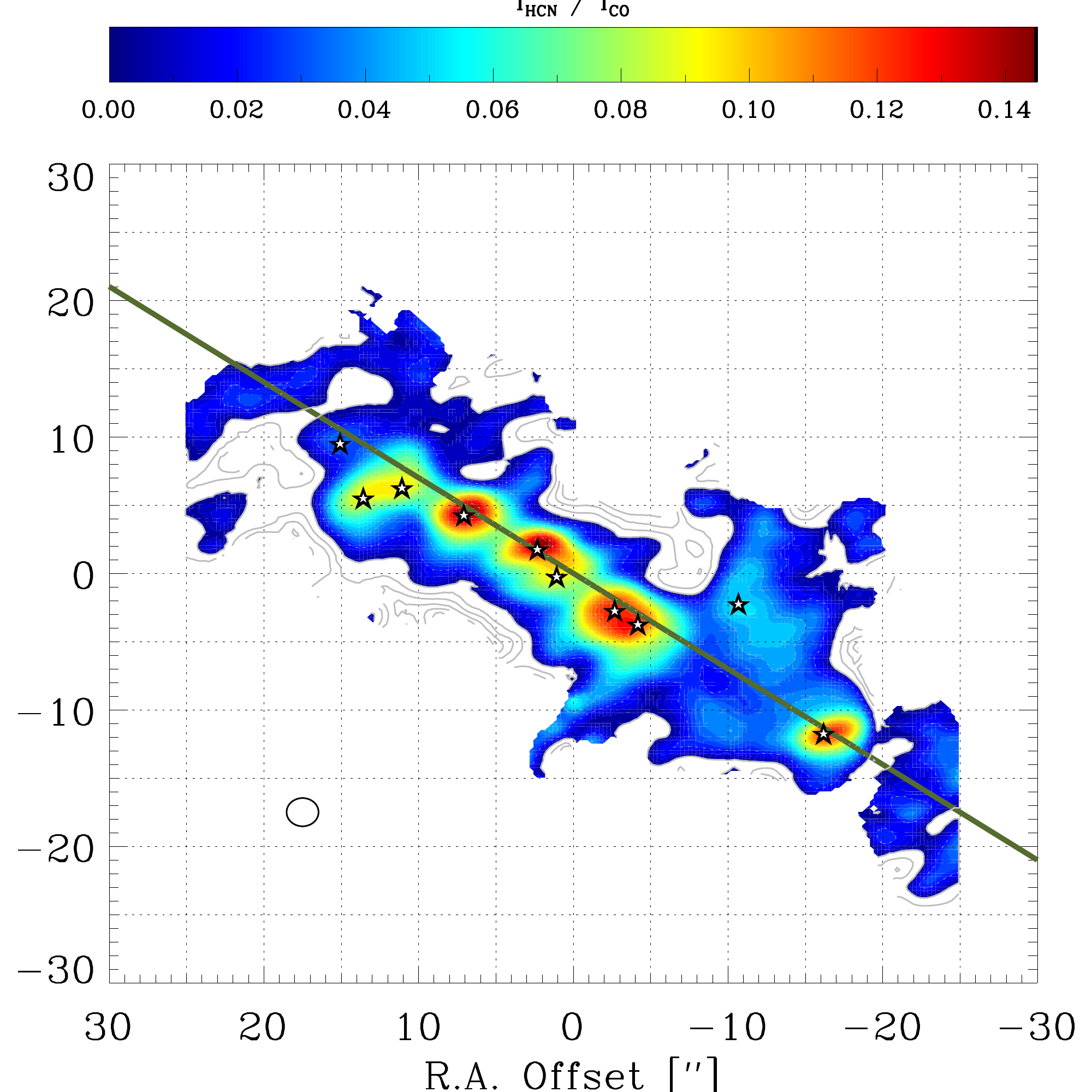}
\plottwo{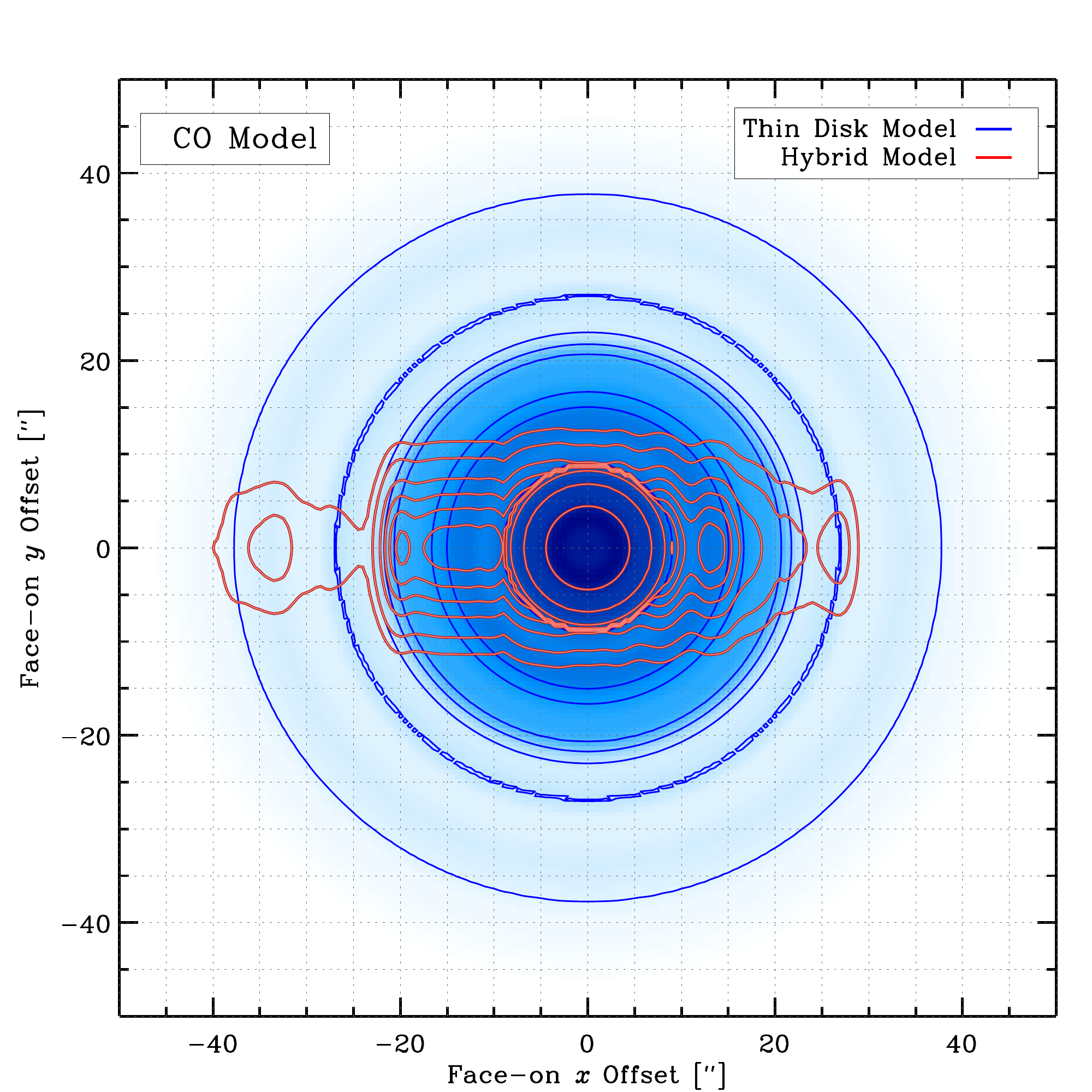}{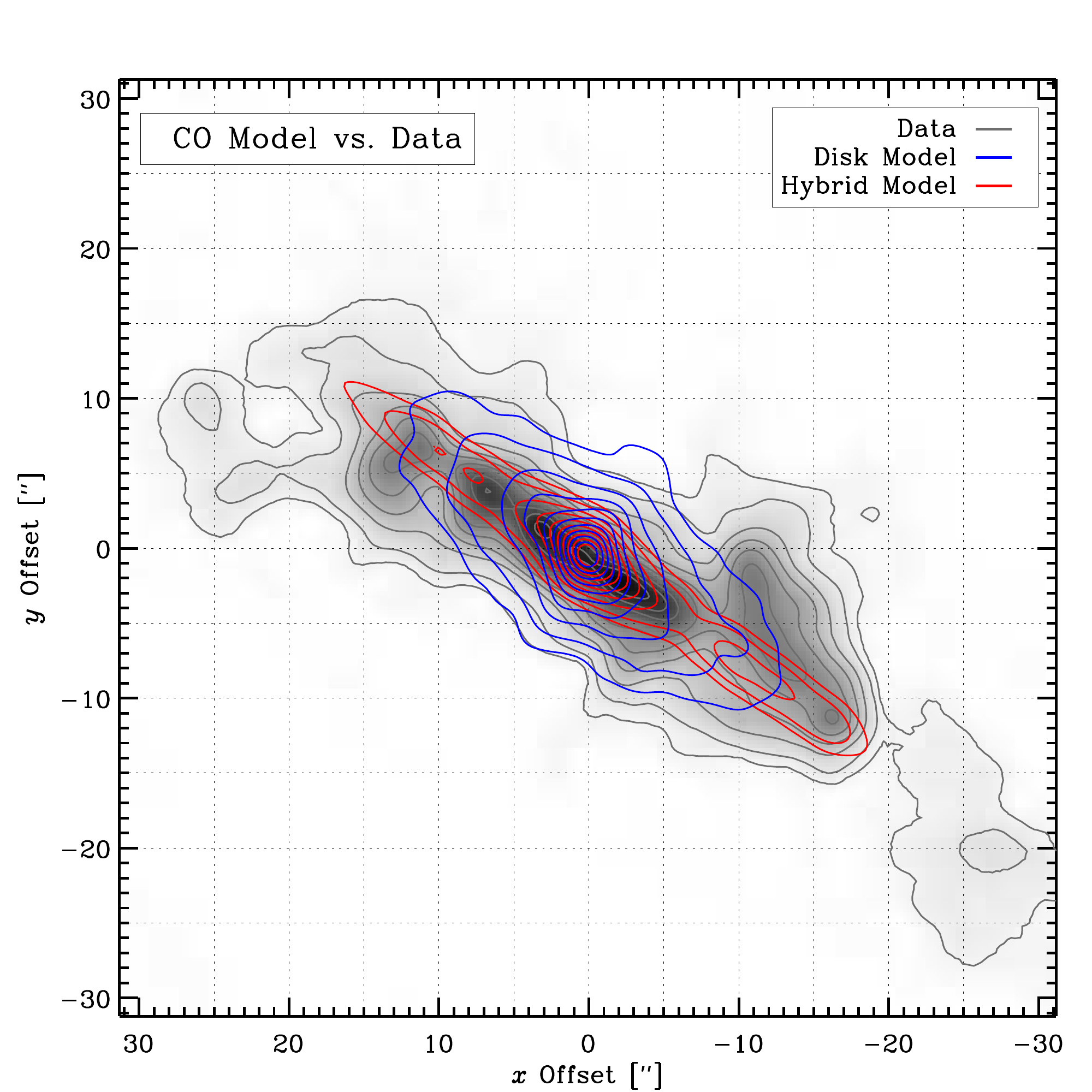}
\caption{Structure of the starburst considered as a whole. ({\em top left}) Illustration of orientation with major and minor axis noted. ({\em top right}) 
Normalized peak intensity profiles along the major axis $^{12}$CO (gray) and the average of our three dense gas tracers HCN, HCO$^{+}$, and CS (black).
We indicate peaks and the one dimensional extent that encompasses 50 and 90\% of the emission along the $x$-axis. ({\em middle left}) Normalized minor axis
profile for $^{12}$CO and dense gas tracers from summing emission along the minor axis (see text). The beam is indicated below the profile. The width of the profile in this figure
convolves true vertical extent with foreshortened minor axis extent. ({\em middle right}) Ratio of HCN to CO emission across the burst, with peaks (GMCs) indicated
via stars. This view shows a compact, clumpy, central distribution of dense gas embedded in a wider distribution of CO.
({\em bottom left}) Two models of the face on distribution of CO gas based on the major axis profile. Blue shows an azimuthally symmetric model over the whole range of
radii; light red shows a disk-like geometry out to $r_{50}$ but more linear structure outside this core ({\em bottom right}) The same two model distributions projected
back onto the sky and shown in contour over the actual CO emission. In all three cases, the contour lines reflect the same fraction of the peak, so that the contours should
match between data and model if the model was a good description of the data. Note how azimuthal symmetry predicts too much emission along the minor axis compared to
our observations.
\label{fig:starburst}}
\end{figure*}

\begin{figure*}
\epsscale{0.8}
\plottwo{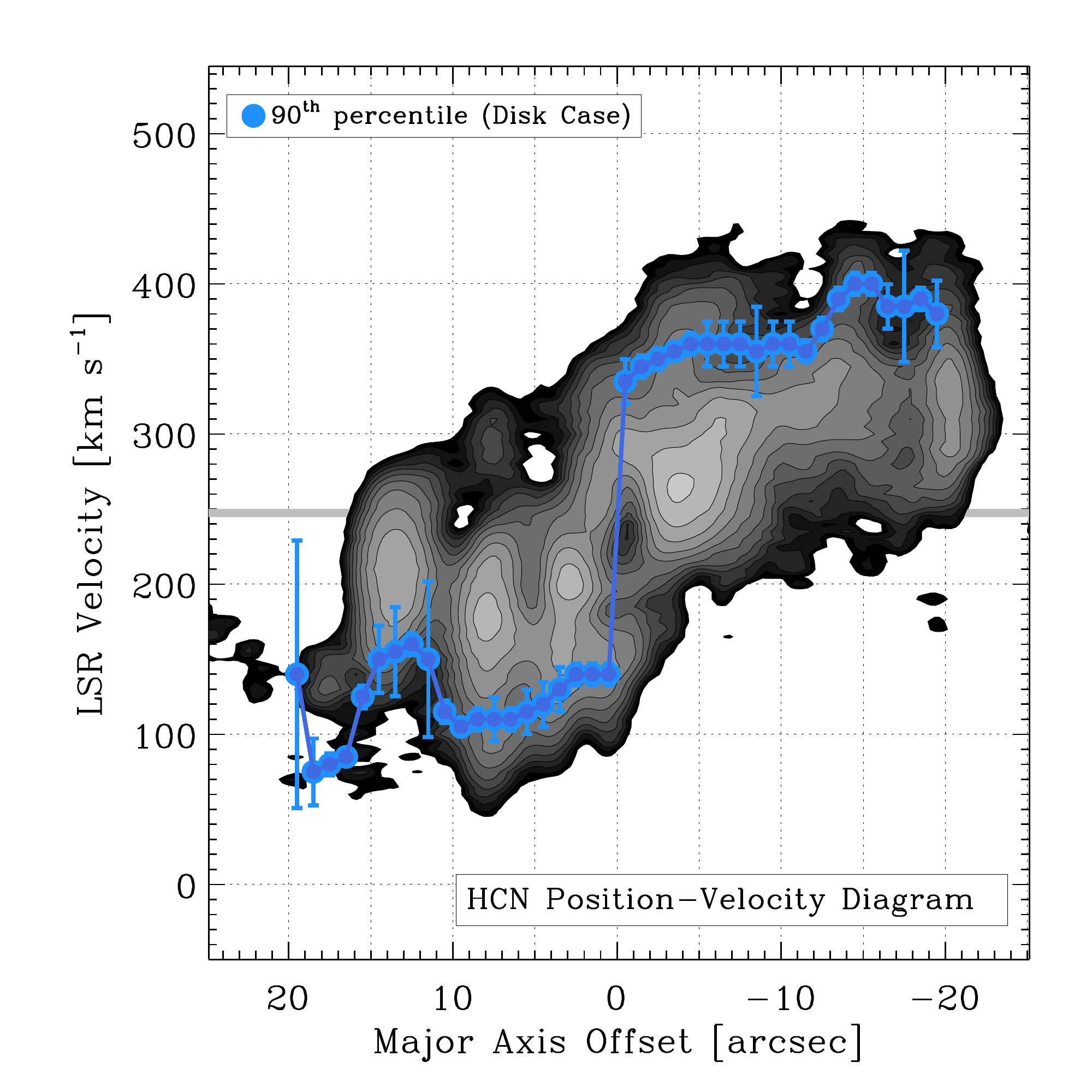}{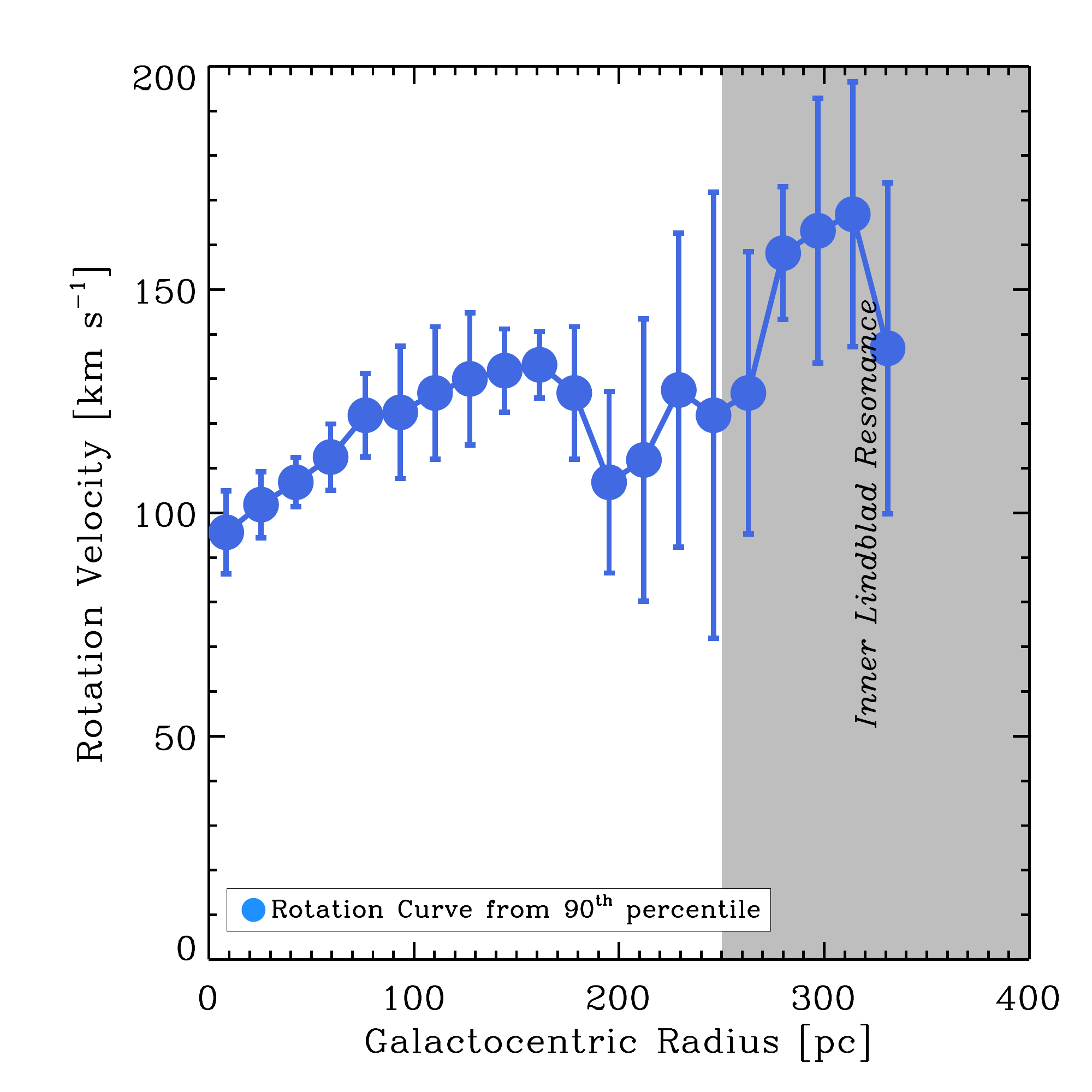}
\plottwo{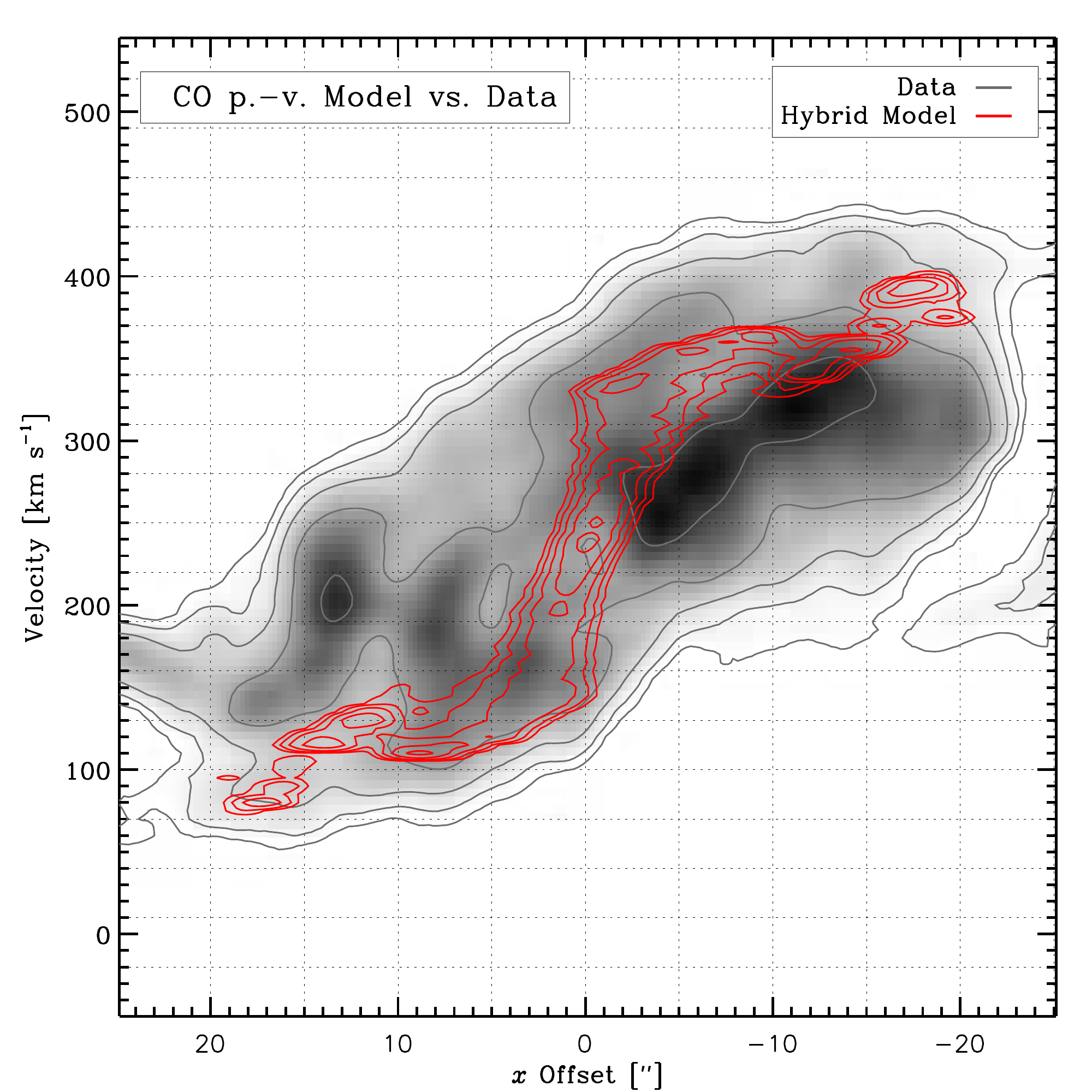}{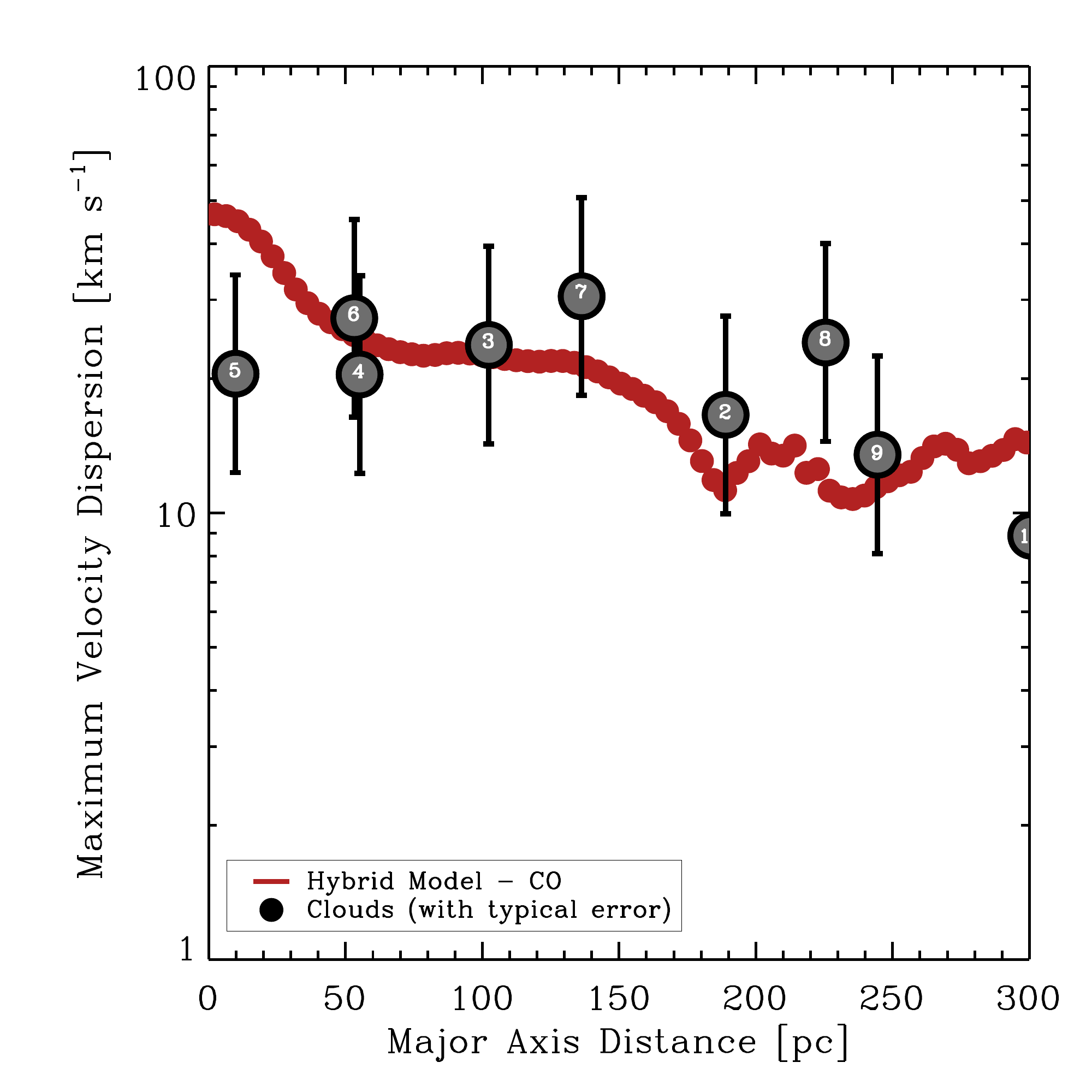}
\caption{Kinematics of the starburst. ({\em top left}) Position velocity diagram of HCN along the major axis showing the envelope of emission 
(blue, appropriate for a disk with emission at all position angles) averaged across all tracers. ({\em top right}) Implied rotation 
curve, along with literature estimates of the location of the Inner lindblad Resonance. ({\em bottom left}) The CO position velocity diagram
with our model position velocity diagram, obtained from combining the model intensity distribution and estimated rotation curve, overplotted. 
({\em bottom right}) The maximum velocity dispersion from the model at each radius (red) and the measured velocity dispersions for our
clouds plotted (gray points) with typical uncertainties. If the disk is a continuous structure then blending has the potential to contribute to 
our line widths, though the spectra of our clouds do not generally show clear evidence for multiple components (Figure \ref{fig:cloudspec}).
\label{fig:dynamics}}
\end{figure*}

{\em Major and Minor Axis Profiles:} As our most basic structural measurement, we derive major and minor axis profiles for $^{12}$CO (1--0), C$^{17}$O,
and the dense gas tracers HCN, HCO$^{+}$ and CS. We plot the results in the second and third panels of Figure \ref{fig:starburst}. For 
the major axis profile, we rotate the cube  by the position angle and then record the peak intensity at each position along the major axis.
After normalizing by the peak of the profile, we construct an average ``dense gas tracer'' profile, which we plot in black over the CO profile (gray).
The figure also indicates the position of the candidate GMCs and the one dimensional extents, $r_{50}$ and $r_{90}$, that contain 50 and 90\% of the integrated 
area under the profile (but note that the two dimensional $r_{50}$ and $r_{90}$ could differ from this depending on the geometry of the emission).

As a complement to the major axis profile, we create an average minor axis profile, which appears in the third panel of Figure \ref{fig:starburst}. 
We expect this profile to convolve the (heavily foreshortened; $1/\sin 14\arcdeg \approx 4.1$) in-plane structure of the galaxy with its vertical extent. To construct the minor axis profile, 
we step along the major axis and at each location (and for each tracer) we calculate the intensity-weighted mean position, which we take to reflect the midplane. 
We measure the vertical (minor axis) profile about this mid plane and keep a running sum as we step through strips from $-18\arcsec$ to $+18\arcsec$ 
($-300$~pc to $+300$~pc) in bins $1.5\arcsec$ wide along the minor axis. The resulting profile for CO and the average of the dense gas tracers appear 
in the Figure, though we measure the profiles for each tracer. Fitting Gaussians to this distribution, we take advantage of the high (but not edge-on) inclination 
of the galaxy to place upper limits on the vertical extent of the central burst.

Both the major and minor axis profiles show dense gas to be more concentrated and clumpier than CO emission and prompted by this difference,
we plot a map of the ratio of our brightest dense gas tracer, HCN, to CO emission in the fourth panel of Figure \ref{fig:starburst}. We have
already seen how this Figure highlights the locations of the peaks that we study in Section \ref{sec:clouds}.

{\em Geometry of the Burst:} The high inclination and small scales (compared to our beam) of the burst make it hard to infer the 
geometry of the system from simple inspection of the images. To explore this quantitatively, we build a series of three-dimensional models
which we then rotated to the plane of the sky. We do this following the approach by \citet{DAVIS13}, which we modified to consider
an arbitrary two dimensional distribution. We construct a face-on model of the galaxy and then use a Monte Carlo approach to draw
a sample of points in the face-on plane that reflect this distribution. The probability of drawing a point at a location is set by the intensity 
of that position in the face-on image  (potentially modified by a vertical distribution). These data are projected to the plane of the sky according 
to the inclination and position angle of the galaxy, gridded, and then convolved with the telescope beam. We also calculate the velocity of 
each particle and so create a model data cube.

We test models with a disk geometry and a more linear geometry, constructing images based on the major axis profile.  In the 
case of the disk geometry, we average the intensity on the two sides of the disk and impose azimuthal symmetry. In the case of the linear geometry, we 
simply impose the major axis profile as the model major axis distribution and then multiply it by a Gaussian with width $\sqrt{2}$ times the FWHM
of the minor axis profile corrected for the effects of inclination, thus assuming that half the minor axis extent comes from foreshortened emission 
in the plane of the disk. Comparing the results to the observed distribution show that azimuthal symmetry over the full radius of our image 
would predict a much larger azimuthal extent than we observe so that a disk-like structure out to many hundreds of parsecs
does not appear to match the data well. 

We also test a model with a strong linear dynamical feature (i.e., a bar). These may suppress collapse, however, and so be at odds with
the vigorous star formation observed in the NGC 253 nucleus. Our best guess is that in the inner regions, $\sim 100$--$150$~pc or $r_{50}$ 
for both the dense gas and CO, the molecular gas may have a more ring-like structure. Our observations do not strongly constrain the 
geometry this close to the center, but such structures are seen in simulations, e.g., see \citet{KIM12} and observations of the inner parts of 
galaxies \citep{MARTINI03,PEEPLES06}. In particular the simulations of \citet{KIM12} show fainter more linear features feeding in to such 
nuclear ring-like structures. Such a scenario loosely resembles the suggestion by \citet{PAGLIONE04} that NGC 253 host the kind of 
``twin peaks'' morphology observed in several barred galaxies by \citet{KENNEY92}.

In practice, this means that we will treat the emission inside $\sim r_{50}$ as azimuthally symmetric for purposes of coarse stability,
surface and volume density calculations. That is, we adopt the ``hybrid model'' illustrated in the bottom panels Figure \ref{fig:starburst}.
While certainly an oversimplification, treating the inner region as a disk allows ready comparison to analytic models 
\citep[e.g.,][]{OSTRIKER11,KRUMHOLZ13} that make similar simplifying assumptions and do not treat substructure or dynamical effects.
Calculating such average quantities we do not worry about a more extended azimuthally symmetric component confusing our observations 
because our modeling shows that such a component, if present, would disagree with the observed minor axis extent of both the CO and
high density emission. That is, the nuclear region must be highly asymmetric at radii $r \gtrsim 150$~pc, which also means that the mass is concentrated
roughly in line with what we see in the major axis profile. For the purposes of our calculations, we take $r_{50} \approx 150$~pc for the mass
in the starburst, motivated particularly by the extent of C$^{17}$O emission, which shows $r_{50} = 155$~pc along the major axis and should represent an 
optically thin tracer of the total mass distribution.

{\em Rotation Curve and Velocity Field:} The rotation curve allows us to estimate the orbital timescale, make basic stability arguments, and evaluate
the origin of the wide line widths measured for the clouds. We estimate the rotation curve from the combination of CO, C$^{17}$O, C$^{18}$O, HCN, and 
HCO$^{+}$ emission. Again, geometry complicates our measurement. The high inclination of the galaxy and significant non-circular motions 
\citep[e.g., see][and the velocity fields in Figure \ref{fig:moment_maps}]{SAKAMOTO06,BOLATTO13A} mean that a fit to the two dimensional velocity field is 
not straightforward. Instead, we constrain the rotation of the galaxy using the major axis position velocity diagram. Depending on whether the geometry is 
bar-like, disk-like, or a sparsely sampled set of clouds, the translation from position-velocity diagram into rotation curve can also vary. 
Following \citet{SOFUE99}, which remains to our knowledge the most complete study of the nuclear rotation curves of spiral galaxies using molecular gas, 
we assess rotation using the envelope of emission technique. We measure this as the velocity associated with the $90^{\rm th}$  percentile of emission sorted by 
velocity \citep[similar to][studying NGC 253]{SORAI00,PAGLIONE04}. This envelope will correspond to the tangent point velocity in a fully-sampled disk, but 
the intrinsic velocity dispersion in the gas, a low filling factor of molecular gas, and non-circular motions will all introduce some inaccuracy into the approach. 
Still, after trying several methods, we found that this approach yields a rotation curve that most closely resembles results for other massive spiral galaxies 
\citep[e.g.,][]{SOFUE99}. We plot the resulting envelope on top of the major axis position-velocity diagram for HCN in the bottom left panel of 
Figure \ref{fig:dynamics} and the implied rotation curve in the bottom right panel of the same figure. Note that the difference between 
$i=76\arcdeg$ and $i=90\arcdeg$ is negligible for this calculation.

For both the kinematics and the geometry of the burst, we emphasize that our goal is a basic structural characterization. The detailed kinematics
have been considered before \citep[e.g.,][in addition to the references already mentioned]{ANANTHARAMAIAH96,PENG96,DAS01}. Several of
these studies have hypothesized that the nuclear starburst may reflect with $x_2$ orbits associated with the bar, while others 
have hypothesized the existence of a second, orthogonal nuclear gas disk. Our data can be used to address these topics but this is outside 
the scope of this paper.

{\em Line of Sight Velocity Dispersion:} In a continuous disk observed at high inclination, the rotation of the gas convolved with the surface brightness distribution 
implies some line of sight velocity dispersion. The comparison of this disk velocity dispersion with the measured dispersion for clouds will prove 
helpful to interpret our results. Therefore we calculate the velocity dispersion implied by bulk motion alone for each model and plot this as a function
of radius, along with the positions and velocities of our measured clouds, in the bottom right panel of Figure \ref{fig:dynamics}. 

{\em Vertical Velocity Dispersion:} We require an estimate of the vertical velocity dispersion, $\sigma$, of the gas for several calculations. Lacking 
any semblance of face on observations, we assume a value. Observations of this quantity are scarce, but we expect $\sigma$ in NGC 253 to be higher 
than that observed in the disks of face on galaxies, $\sigma \sim 10$~km~s$^{-1}$ \citep[][]{LEROY08,TAMBURRO09,CALDUPRIMO13}. 
At coarser resolution, \citet{JOGEE05} measured $\sigma = 10$--$40$~km~s$^{-1}$ for a collection of nearby galaxy nuclei. The literature
thus suggests $\sigma \sim 20$~km~s$^{-1}$ with a  factor of $\sim 2$ uncertainty.

Alternatively, we can estimate the vertical velocity dispersion needed to support the disk. We carry out the calculation in the simplest way possible,
considering only gas in a turbulent disk confined vertically by its own gravity. Hydrostatic equilibrium yields

\begin{equation}
\sigma \sim \sqrt{\frac{\pi G \Sigma^2}{2 \rho_0}}~.
\end{equation}

\noindent Here $\Sigma$ is the mass surface density of the gas, $G$ the gravitational constant, and $\rho_0$ the mid plane density. We will
see below that when treated as an azimuthally symmetric disk, the inner part of NGC 253 ($r < 150$~pc) exhibits $\left< \Sigma \right> \sim 2,500$~M$_\odot$~pc$^{-2}$ and
$\left< n\right> \gtrsim 300$~cm$^{-3}$  ($\left< \rho \right> \sim 1 \times 10^{-21}$~g~cm$^{-3}$), which together imply $\sigma \lesssim 45$~km~s$^{-1}$. The inequality
results from our inability to cleanly assess the vertical distribution from the minor axis profile. We will adopt $\sigma \approx 30$~km~s$^{-1}$ for the inner disk when 
comparing to theoretical expectation but note the substantial uncertainty.

{\em Mass in the Starburst:} We follow Section \ref{sec:masscalcs} to calculate the integrated mass over the nuclear region from
our optically thin isotopologues. The SMA data have uncertain spatial filtering on larger scales, so we prefer the dust-based estimate
from \citet{WEISS08}, who observed NGC 253 using the LABOCA instrument on APEX. From the nuclear region, they infer $\approx 3 \times 10^8$~M$_\odot$
of gas associated with a ``cool'' ($T \approx 35$~K) component in the central region. This mostly arises from a $\approx 30\arcsec \times 16\arcsec$ Gaussian, 
roughly our nuclear CO disk. To derive this, they adopt a dust-to-gas ratio of $1$-to-$150$, the same that we adopt above. Though they find no evidence
requiring a cooler component, note that the spectral energy distribution certainly would allow a cooler component that could substantially change the mass.

The integrated disk mass from the optically thin isotopologues ranges from $1.4 \times 10^8$~M$_\odot$ for C$^{18}$O to $4.9 \times 10^8$~M$_\odot$ for
$H^{13}$CO$^+$. As above, H$^{13}$CN and H$^{13}$CO$^{+}$ yield the highest masses, suggesting a possible internal inconsistency in our adopted abundances
with those two species a factor of $\sim 2$ too high. C$^{17}$O, which should be an optically thin tracer of the bulk of the gas mass and has our best sensitivity to
spatial scales, yields a total mass of $3.5 \times 10^{8}$~M$_\odot$, in good agreement with the \citet{WEISS08} estimate. This is also the median mass estimated
from our various approaches and we adopt $3.5 \times 10^8$~M$_\odot$ as the mass of the disk. A crude check (taking $M \sim R V^2 / G$) against the rotation velocity 
shows that this mass along with up to about twice this mass in stars can be accommodated by our adopted rotation curve. As an additional check, our integrated 
nuclear $^{12}$CO luminosity, $L_{\rm CO} \approx 3.3. \times 10^8$~K~km~s$^{2}$, combined with a ``starburst'' conversion factor, also yields $\sim 3 \times 10^8$~M$_\odot$.

\subsection{Results for the Whole Starburst View}

\begin{deluxetable}{lcc}
\tablecaption{Physical Conditions in the Nuclear Gas Disk} 
\tablehead{ 
\colhead{Property} & 
\colhead{NGC 253} &
\colhead{Galaxy Disk\tablenotemark{a}}
}
\startdata
$\left< \Sigma_{\rm mol} \right>_{r50}$ & $\approx 2,500$~M$_\odot$~pc$^{-2}$ & 14~M$_\odot$~pc$^{-2}$ ({\sc Hi}+H$_2$)\\
$\left< n \right>_{r50}$ & $\approx 300$~cm$^{-3}$ & $\sim 1$~cm$^{-3}$ \\
SFR & $\sim 2$~M$_\odot$~yr$^{-1}$\tablenotemark{b} & \nodata \\
$\Sigma_{\rm SFR}$ & $14 $~M$_\odot$~yr$^{-1}$~kpc$^{-2}$ & $4 \times 10^{-3}$~M$_\odot$~yr$^{-1}$~kpc$^{-2}$ \\
$\tau_{\rm rot}^{r50}$ & $\approx 7$ Myr & $\sim 200$~Myr \\
$Q_{\rm gas}$ & $\sim 1$\tablenotemark{c} & $\sim 4$ \\
$\tau_{\rm cross}$ & 3~Myr & \nodata \\
$\tau_{\rm ff}$ & 2~Myr & $\sim 4 \times 10^{7}$~yr \\
\enddata
\label{tab:sketchdisk}
\tablenotetext{a}{Disk values varied dramatically; characteristic values drawn here from \citet{LEROY08} for the region where
$\Sigma_{\rm H2} \approx \Sigma_{\rm HI}$, i.e., the ISM is an even mixture of atomic and molecular gas. This is also roughly
$r_{50}$ for a typical molecular exponential disk \citep{SCHRUBA11}.}
\tablenotetext{b}{SFR for the whole nuclear burst. We take SFR$\sim 1$~M$_\odot$~yr$^{-1}$ within $r_{50}$.}
\tablenotetext{c}{Assuming a constant surface density and velocity dispersion $30$~km~s$^{-1}$.}
\tablecomments{Specific quantities calculated assuming azimuthal symmetry within $r_{50} \approx 150$~pc and so reflect
the strong central nuclear gas concentration.}
\end{deluxetable}

Figure \ref{fig:starburst} presents a quantitative view of the overall structure of the NGC 253 starburst region. Our 
most basic measurements are the major and minor axis extent of emission in CO and dense gas tracers (here meaning the average
of HCN, HCO$^+$ ,and CS). Half of the major axis CO emission comes from a region with linear extent $\sim 300$~pc (diameter), while 
the dense gas is somewhat more compact, $\sim 200$~pc. The difference is more extreme along the minor axis, where
CO emission shows FWHM 115~pc (C$^{17}$O is similarly extended) compared to $\sim 55$~pc for the dense gas tracers.

The observed distribution of gas is quite elongated in the plane of the sky, $\sim 1:3$--$1:5$ depending
on how one measures it. This might reflect either foreshortening in a tilted disk or a real elongated structure. Our simple modeling
compares these two hypotheses and suggests that a bar-like structure offers a better match to the data at large radii (outside $r_{50}$) because
azimuthal symmetry would predict more emission than we observe along the minor axis. A true bar structure extending to very
small radii appears unlikely, especially given the strong star formation present in the region, and we consider a hybrid view:
assuming azimuthal symmetry within $\sim 150$~pc but more linear structures outside this region.

A fundamental ambiguity exists between the extent of the bar along the minor axis (which is approximately perpendicular to the
bar, also) and the vertical extent of the emission. Similarly, in the case of a disk the minor axis profile will capture both the foreshortened disk and the
vertical extent. In any case, the width of the minor axis profile will capture an upper limit to the vertical thickness, which we can set to a limit on the 
half-thickness of $H \lesssim 100$~pc for the CO and $H \lesssim 55$~pc for the dense gas.

{\em Fraction of Mass and Emission in Clouds:} The total mass of gas in the clouds discussed in Section \ref{sec:clouds} is $2 \times 10^8$~M$_\odot$, 
or $\approx 60\%$ of what we derive for the gas mass of the disk as a whole under the same assumptions. This cloud mass calculation includes aperture corrections
to account for the use of clipping in defining the clouds. In terms of the data only, 66\% of the total HCN emission, 30\% of the C$^{17}$O emission, and
only $10\%$ of the $^{12}$CO emission originates from within the three dimensional cloud assignments (Figure \ref{fig:assign}). As we have emphasized
and Figure \ref{fig:starburst} shows, the clouds represent enhancements in dense gas emission embedded in a broader sea of CO emission.

{\em Surface, and Volume Density:}  For our best-estimate mass and $r_{50} \approx 150$~pc, we estimate an average 
$\left< \Sigma \right> \sim 2,500$~M$_\odot$~pc$^{-2}$ over the inner portion of the disk \citep[very close to the  column density found by][]{MARTIN06}.  
Folding in our limit on the vertical thickness, we estimate the volume  density  $\left< n \right> \gtrsim 300$~cm$^{-3}$ for the inner 150~pc, equivalent to 
$\gtrsim 19$~M$_\odot$~pc$^{-3}$, with the inequality reflecting ambiguity in the interpretation of the minor axis extent as a vertical width. Over this
inner region, the average volume density resembles that of a typical disk galaxy GMC while the surface density exceeds this value by an order of magnitude.
Note that this inner region contains the massive clouds starred in Table \ref{tab:cloudprops} which have average surface density $\sim 20,000$~M$_\odot$~pc$^{-2}$,
still contrasting with the average by a factor of $\sim 10$.

{\em Timescales in the Disk:} The orbital timescale, taken to be the relevant dynamical timescale for the disk, has received much
attention as a potential driver for differences in gas depletion times \citep[][among many others]{KENNICUTT98A,DADDI10,GENZEL10}.
For our estimated rotation curve, the orbital time at $r_{50} \approx 150$~pc is $\tau_{\rm orb} \approx 7$~Myr. Contrasting this with a typical
$\tau_{\rm orb} \sim 200$~Myr at large radii in the disk of a spiral galaxy yields demonstrates that NGC 253 obeys the same behavior as the
\citet{DADDI10,GENZEL10} samples --- the contrast in H$_2$ depletion time between NGC 253 a typical disk matches the contrast in
orbital times. 

Other key timescales in the disk show similar contrast. Given our estimated velocity dispersion of $\sim 30$~km~s$^{-1}$, the vertical crossing time is $\sim 3$~Myr, somewhat 
longer than for the individual clouds (here using the 1-d velocity dispersion and the CO vertical extent).  Given the estimated average volume density, which is $\sim 100$
times higher than in a normal disk, the free fall time inside $r_{50}$ is $\sim 2$~Myr. As with the orbital timescale, these contrast starkly with normal ``disk'' conditions.

{\em Stability and Toomre's $Q$:} The most common physical argument for the orbital time as a driver stems
from treating the system as a self-regulated, marginally stable disk \citep[e.g.,][]{SILK97}. Indeed, many analytic theories treat starburst 
nuclei, or indeed whole galaxies, as marginally stable gas disks, $Q \approx 1$ \citep[e.g.,][among many others]{SILK97,TAN00,THOMPSON05,KRUMHOLZ12}. 
From our studies of bound clouds and the geometry of the starburst, the issues with such an assumption should be immediately apparent:
much of the gas has already collapsed and the geometry likely deviates substantially from a thin disk at large radii. Nonetheless, the calculation
is so common that we consider the comparison useful.  For our adopted rotation curve (approximated by a polynomial to allow analytic derivatives),
we calculate calculate $Q = \sigma \kappa / \left( \pi G \Sigma \right)$ at $r_{50}$ (within which we consider a disk-y geometry somewhat reasonable) 
with $\sigma$ the estimated velocity dispersion (assuming that the radial and vertical dispersion are similar), $\Sigma$ the estimated gas mass surface 
density, and $\kappa$ the epicyclic frequency given by $\kappa = 1.41  \sqrt{1+\beta} v / r$ where $\beta = d \log v / d \log r$.  For our adopted 
$\sigma = 30$~km~s$^{-1}$, the result at $r_{50} \sim 150$~pc is $Q \approx 1$, placing the nuclear disk in the marginally stable regime even 
before one considers the potential additional impact of the stellar potential well (naively interpreting the 3.6$\mu$m emission, $\rho_\star \sim \rho_{\rm gas}$ in this region). 

{\em Dynamical Environment:} \citet{SORAI00}, \citet{PAGLIONE04}, and others have suggested that the Inner Lindblad Resonance (ILR) with 
NGC 253's strong bar (Figure \ref{fig:coverage}) may set the scale for the nuclear burst. We have indicated the approximate ILR estimated 
from literature studies \citep{SORAI00,PAGLIONE04,IODICE14} in Figure \ref{fig:dynamics} and indeed this corresponds with the extent of 
bright molecular emission in our maps. Our ability to independently constrain the existence or location of the ILR without careful dynamical modeling is limited.
Essentially, our treatment of the rotation curve (using an envelope method or using the mean velocity) determines whether an ILR emerges from
the data or not. Therefore we refer the reader to the thorough study of \citet{SORAI00} for a discussion of molecular gas kinematics in NGC 253. They
conclude that NGC 253 has an ILR at $r_{\rm gal} \approx 240$~pc, similar to $r_{90}$ that we find, though subject to the same uncertainty in the
treatment of inner galaxy emission.

{\em Origin of High Velocity Dispersions:} The rotation curve, combined with our model emission distribution, allows us to test the degree to which 
line of sight blending of rotating material could produce the high velocity dispersions that we observe for our clouds. The bottom right panel of 
Figure \ref{fig:dynamics} shows the velocity dispersion along the line of sight for our simple model with the dispersions of individual clouds
on the plot. The plot shows that blending in the inner part of the galaxy, especially in the disk-like part of our model within 150~pc, can be severe.
If all position angles are populated by emission and the rotation is as fast as implied by the envelope velocities, then we would expect a significant
spread in velocities along the line of sight. The figure represents an upper limit to the amount of blending, because in a cloudy, clumpy view the
clouds will sample only one part of the rotation curve with the width due to internal turbulent dispersion. Further, we note that the spectra presented
in Figure \ref{fig:cloudspec} do not show clear signs of multiple components, though neither are they perfectly Gaussian \citep[the exception is the double-peaked
cloud \#1, which lies near the base of an outflow and near a shell;][]{SAKAMOTO06,BOLATTO13A}. Nonetheless, the caveat that geometry may broaden 
our linewidths is important to bear in mind. If the clouds that we see are blends, then 
we expect the line widths and likely also the radii to decrease. Note that this is unlikely to change our conclusions regarding the boundedness of the
clouds, as removing the cloud-cloud dispersion from the accounting while keeping the total mass of clouds the same is likely to increase
the amount of potential relative to kinetic energy. It will be interesting to see what substructure is revealed by higher resolution observations of optically 
thin high critical density tracers.

{\em Resemblance to an Equilibrium Disk:} We can compare the measured starburst structure to the expectations for a molecular disk in vertical 
equilibrium supported by star formation feedback from \citet{OSTRIKER11}. \citet{OSTRIKER11} and \citet{SHETTY12} discuss, in the context of numerical 
modeling, how this equilibrium view may still describe a disk that is dynamically forming stars and hosting GMCs. They assume hydrostatic equilibrium in the 
vertical direction, similar to our discussion above, and predict the half thickness to be:

\begin{equation}
H = \frac{1}{1+\chi}\frac{v_z^2}{\pi G \Sigma} \approx 16~{\rm pc}~\left( \frac{v_z}{30~{\rm km~s}^{-1} } \right)^2
\end{equation}

\noindent with $\Sigma \approx 2,500$~M$_\odot$~pc$^{-2}$ the gas surface density averaged over the burst, $v_z$ the vertical velocity dispersion
(though note the mild circularity as we are forced adopted $v_z$), and $\chi$ a dimensionless parameter that captures the ratio of 
stellar and gas self-gravity\footnote{Based on the IRAC 3.6$\mu$m maps and assuming spherical symmetry, we estimate $\chi \approx 0.4 \rho_\star / \rho_{\rm gas} \sim 0.55$
for the bulge of NGC 253, with $\rho_{\star}$ slightly exceeding $\rho_{\rm gas}$.}. The midplane gas volume density is predicted to be:

\begin{equation}
\label{eq:diskff}
\rho_0 = \frac{(1 + \chi ) \pi}{2} \frac{G \Sigma^2}{v_z^2} \approx 80~{\rm M}_\odot~{\rm pc}^{-3}~\left( \frac{v_z}{30~{\rm km~s}^{-1}} \right)^{-2}
\end{equation}

\noindent where 80~M$_\odot$~pc$^{-3}$ is equivalent to $n_{\rm H2} \sim 1,200$~cm$^{-3}$.

These structural calculations show that the vertical extent and surface density that we measure can be reconciled with dynamical equilibrium for
reasonable assumptions about the vertical velocity dispersion. Recall that we do not know the vertical velocity dispersion due to the inclination of the
burst, nor can we precisely disentangle the relative contributions of vertical extent and foreshortened emission along the minor axis. In order to
agree with the predictions from \citet{OSTRIKER11}, the disk must have vertical velocity dispersion either $v_z \sim 50$~km~s$^{-1}$ or a significant part
of the minor axis extent that we observe must be due to foreshortening of emission in place. Either case is perfectly plausible.

For a feedback momentum from supernovae per unit mass of star formation $p_\star / m_\star = 3000$~km~s$^{-1}$, \citet{OSTRIKER11} predict the star formation rate
needed to sustain such an equilibrium disk

\begin{eqnarray}
\Sigma_{\rm SFR} &=& 0.092~{\rm M}_{\odot}~yr^{-1}~{\rm kpc}^{-2}~\frac{(1+\chi)}{f_p} \left( \frac{\Sigma}{100~{\rm M}_\odot~{\rm pc}^{-2}} \right)^2 \\
&\approx& 90~{\rm M}_{\odot}~{\rm kpc}^{-2}
\end{eqnarray}

\noindent where $f_p \approx 1$.  The predicted SFR within the FWHM of the nuclear part ($r_{50} = 150$~pc) of the burst is then
$\sim 6$~M$_\odot$~yr$^{-1}$. Our simple estimates based on the IR luminosity imply values place the SFR of the central burst closer to $2$~M$_\odot$~yr$^{-1}$,
and an additional factor accounting for the fraction of the SFR inside $r_{50}$ would increase the discrepancy. In order to explain the discrepancy while
maintaining a feedback driven equilibrium, one would either need to increase the amount of momentum yielded per unit star formation (e.g., radiation and cosmic rays 
could add to the supernova contribution) or make significant adjustments to the geometrical ($f_p$) or stellar ($\chi$) terms. More generally, a lower star formation rate 
than the self-regulation prediction would imply that dynamical processes other than feedback maintain the turbulent gas pressure; given the strong inflows produced by a 
bar, this is not implausible. Alternatively, the strong dependence on 
surface density means that one could explain the discrepancy if we have overestimated the mass of the disk by a factor of $\sim 2$. We note, however, that 
considering the whole burst region ($r_{90}$), the agreement using this prediction appears much better. In this case, 
$\left< \Sigma_{\rm SFR} \right> \approx 4$~M$_\odot$~yr$^{-1}$~kpc$^2$ and the model predicts
approximately this value from the average surface density over this larger scale $\approx 700$~M$_\odot$~pc$^{-2}$.

{\em Efficiency Per Free Fall Time:} In an attempt to explain star formation timescales across a wide range of local conditions, \citet{KRUMHOLZ12B} posited
a universal efficiency per free fall time, $\epsilon_{\rm ff} \sim 0.01$, that holds across scales \citep[see also][]{KRUMHOLZ07}. We check their predictions
by comparing the SFR estimated for the burst with the mass and free fall time that we estimate for the burst, calculating

\begin{equation}
\epsilon_{\rm ff} = \frac{SFR~\tau_{\rm ff} (\rho)}{M(\rho)}
\end{equation}

\noindent where we have written $M(\rho)$ and $\tau_{\rm ff} (\rho)$ to indicate that we compare the mass of material at a characteristic density to the free fall
time for that density. We carry out the calculation for three cases: (1) considering only the cloud populations so that $M$ is the mass in clouds and $\rho$ is the average 
cloud density, (2) considering the burst within $r_{50}$ and using the lower limit on the density implied by our minor axis profile, and (3) calculating a midplane
density from the average surface density within $r_{50}$ and our assumed velocity dispersion. In case (1) we take $SFR \sim 2$~M$_\odot$~yr$^{-1}$. In the other
two cases, we assume that half of the star formation occurs within $r_{50}$ and so take $SFR \sim 1$~M$_\odot$~yr$^{-1}$. For case (1) we find $\epsilon_{\rm ff} \sim 0.007$,
for case (2) $\epsilon_{\rm ff} \lesssim 0.01$ (the inequality reflects the inequality on $n$), and for case (3) $\epsilon_{\rm ff} \sim 0.005$ for $v_z = 30$~km~s$^{-1}$ (see Equation
\ref{eq:diskff}). We conclude that, similar to the case for the self-regulation prediction, the constant $\epsilon_{\rm ff} = 0.01$ prediction agrees with the observations within a factor of a few.

If $\epsilon_{\rm ff}$ varies only weakly across the universe then the ability to predict $\tau_{\rm ff}$, or equivalently the density, becomes critical. \citet{KRUMHOLZ12B} suggest
that that two potential timescales combine to predict $\tau_{\rm ff}$ over a wide range of environments: a free fall time within GMCs and a 
disk-averaged free fall time calculated assuming the gas to be in a marginally stable disk.  In \citet{KRUMHOLZ12B}, the former is computed by assuming that the 
typical GMC mass is equal to the 2D Jeans mass in the disk, and adopting a value for the typical GMC surface density. The latter, their ``Toomre free-fall time,'' arises from taking a marginally stable gas 
disk so that $\sigma \sim \pi G \Sigma/\kappa$ using the same formula as Equation \ref{eq:diskff} above (with $v_z \rightarrow \sigma$ and $1+ \chi \rightarrow \phi_P$).
In \S \ref{sec:clouds} we showed that the NGC 253 burst appears to host clouds in a semblance of virial equilibrium, which should place the system in the ``GMC regime.''
We saw in \S \ref{sec:clouds} that the masses of the NGC 253 clouds do agree with the two dimensional Jeans mass, but without a 
method to predict their surface density, the \citet{KRUMHOLZ12B} model lacks a way to theoretically arrive at $\tau_{\rm ff}$. Were we to adopt their formula for $\tau_{\rm ff}$ 
in the ``Toomre'' regime, we find $\tau_{\rm ff,T} \approx 0.6$~Myr at $r_{50}$; this is essentially the same as the result for $\tau_{\rm ff}$ derived from Equation \ref{eq:diskff}.  Although these "disk-averaged" free-fall timescales are similar to the values we compute within clouds, they are subject to the uncertain velocity dispersion.  

{\em Synthesis of the Disk View:} Considered as a whole, the nuclear starburst has volume density comparable to a typical GMC ($n_{\rm H2} \gtrsim 300$~cm$^{-3}$) but over 
much larger area. Even assuming azimuthal symmetry, the surface density of the whole inner ($< r_{50} \approx 150$~pc) region appears very high,
$\left< \Sigma \right> \approx 2,500$~M$_\odot$~pc$^{-2}$. Viewed coarsely, its dimensions and stability over the inner region appear consistent with a disk in some semblance 
of equilibrium, both radial and vertical, which might be expected given that the burst has likely persisted for several times the $\sim$ Myr collapse and 
crossing times. In detail, our observations suggest a more linear structure to the outer part of the burst (inferred from the lack of minor axis extent), though
we assume azimuthal symmetry within the inner part for comparison to simple models. Calculating key timescales,
we find an an orbital time $\sim 7$~Myr (which is a lower limit due to our use of the envelope method) and crossing and free fall times both $\sim 2$--$3$~Myr.
The contrast of these timescales, the free fall time, and the crossing time with the same quantities in the disk of a star-forming 
galaxy can all be invoked to explain the more vigorous normalized star formation rate in the burst and contrasted with the massive (starred) clouds in Table
\ref{tab:cloudprops} to obtain a combined clouds and disk view of the nuclear burst. Table \ref{tab:sketchdisk} summarizes some of the 
properties of the starburst disk and compares them to typical conditions found in a galaxy disk \citep[drawn from][]{LEROY08}.

\section{Molecular Mass-to-Light Ratios}
\label{sec:aco_results}

Our calculations return mass-to-light ratios for CO, HCN, HCO$^{+}$, and CS. Though not the main topic of this paper, 
these quantities have wide application. We therefore note that {\em in the clouds}, our median masses imply 
$\alpha_{\rm CO} \sim 2.8$~\acounits , though with a large ($\approx 0.5$~dex) uncertainty due to the mass determination. In the densest
four clouds this value is even higher, $\alpha_{\rm CO} \approx 4.5$~\acounits. For the dense gas
tracers $\alpha_{\rm HCN} \sim 24$~\acounits , $\alpha_{\rm HCO+} \sim 28$~\acounits , and $\alpha_{\rm CS} \sim 58$~\acounits , all similarly
uncertain. 

{\em In the starburst overall}, we estimate a lower $\alpha_{\rm CO} \sim 1$~\acounits\ from combining our disk mass ($\approx 3.5 \times 10^8$~M$_\odot$) 
with $L_{\rm CO} = 3.3 \times 10^8$~K~km~s$^{-1}$~pc$^2$. The starburst-wide $\alpha_{\rm CO}$ matches that found by 
\citet[][$\alpha_{\rm CO} \sim 1$~\acounits]{SORAI00,PAGLIONE01} and is lower than the $\alpha_{\rm CO} \sim 2$ adopted by 
\citet{SAKAMOTO11}, but consistent within the uncertainties.

The main new contribution of this study comes from the contrast between these two numbers: ALMA's resolution allows us to
distinguish the integrated ``starburst'' $\alpha_{\rm CO}$ from the value associated with the individual peaks. Several studies, including our own estimates
or comparison of our $L_{\rm CO}$ to the \citet{WEISS08} dust-based gas mass, argue that the nuclear region of NGC 253 exhibits a lower-than-Galactic conversion
factor, $\alpha_{\rm CO} \approx 1$~\acounits . Our measurements towards the individual clouds, which should be only minimally affected
by missing flux, indicate a more nearly Galactic, though admittedly uncertain, conversion factor.  The difference, a factor of $\approx 3$ could be just accounted
for within our uncertainties, but we point out a simple observation metric showing the same effect: while $\approx 30\%$ of the C$^{17}$O emission
comes from the cloud assignment regions, $\approx 10\%$ of the CO emission seems to do so. Note that in addition to the optically
thin tracers and dust, the virialized appearance of our clouds (Figure \ref{fig:csig}) means that a dynamical mass estimate would have returned the same value.

Thus our observations add support to the idea of a two-component model for $\alpha_{\rm CO}$ in starbursts: a high-$\alpha_{\rm CO}$ dense component 
and a low $\alpha_{\rm CO}$ diffuse component. This has been discussed in the context of spectral line energy distribution modeling by \citet{PAPADOPOULOS12},
though they argue that the high $\alpha_{\rm CO}$ component dominates the mass while we find the opposite: the net $\alpha_{\rm CO}$ is lower than Galactic.
Here we find support for a two-phase situation based on resolved observations that provide several independent estimates of $\alpha_{\rm CO}$ in the NGC 253
clouds. Such a model may also help explain the apparent contradiction between the low $\alpha_{\rm CO}$ found using dust to trace all gas in galaxy 
centers \citep{SANDSTROM13} and the more nearly Galactic  conversion factor needed to place clouds in virial equilibrium \citep{DONOVANMEYER13}.
Though very uncertain, for reference we note that in such a scenario the implied conversion factor for the emission not in clouds is $\alpha_{\rm CO} \approx 0.5$~\acounits .

We also have the opportunity to carry out several simple checks against expectations. Basic arguments \citep[e.g.,][]{SOLOMON87,MALONEY88} suggest that in
optically thick, virialized clouds, $\alpha_{\rm CO} \propto n^{0.5}~T^{-1}$. In our clouds we find an approximately Galactic $\alpha_{\rm CO}$. We estimate 
temperatures about three times those found in Milky Way clouds, and we find densities $\sim 40$ times higher than in Milky Way clouds. This would actually 
predict that in our clouds we might expect conversion factors $\sim 3$ or more times higher than in the Milky Way.

\citet{NARAYANAN12} have carried out an extensive suite of simulations, which they use to predict $\alpha_{\rm CO}$ based on the intensity-weighted 
intensity of the system, with $\alpha_{\rm CO} = 10.7~\left< W_{\rm CO} \right>^{-0.32}$. With the heavily resolved NGC 253 we are in the rare position 
where we can measure $\left< W_{\rm CO} \right>$ directly, though inclination adds a degree of ambiguity to the calculation. The 
intensity-weighted CO intensity ($\int I^2 / \int I$) in Figure \ref{fig:moment_maps} is $\approx 1,300$~K~km~s$^{-1}$, implying 
$\alpha_{\rm CO} \approx 1$~\acounits, but if we consider a face-on geometry the value would be somewhat higher. Note that their calculation applies to
the ensemble of emission and would not predict $\alpha_{\rm CO}$ for the individual clouds.

Finally, \citet{BOLATTO13B} suggest that in the high surface density regime, appropriate for NGC 253, $\alpha_{\rm CO} \approx 2.9~\Sigma_{100}^{-0.5}$ 
where the $\Sigma_{100}$ is the total surface density of stars and gas in units of 100~M$_\odot$~pc$^{-2}$. Taking 
$\Sigma_{100} \sim 10$ for the gas alone \citep[the same dependence as][]{NARAYANAN12}, this suggests $\alpha_{\rm CO} \sim 1$~\acounits ;
accounting for some contribution from the stars includes, the prediction is closer to $\alpha_{\rm CO} \sim 0.5$~\acounits . Given that
the \citet{BOLATTO13B} relation emerges from a fit that includes galaxy centers, the agreement may indicate that these centers and NGC 253 include 
approximately the same mix of phases.

Synthesizing, the NGC 253 results agree qualitatively with expectations --- $\alpha_{\rm CO}$ should be high given the combination of density, virialization, and
temperature that we find, while prescriptive recipes for whole systems predict the low $\alpha_{\rm CO}$ that we estimate for the whole system.
Clearly both the recipes and the measurements themselves require refinement but this basic result points to both ALMA's ability to observe
mixed phases of molecular gas in starbursts and the ability to access the cloud conversion factor via spatially resolved observations.

\section{Discussion and Conclusions}
\label{sec:discussion}

We use new ALMA observations of CO emission and high critical density tracers, combined with previously
published ALMA and SMA data \citep{BOLATTO13A,SAKAMOTO11}, to disentangle the structure of the nearest
nuclear starburst, NGC 253. From peaks in the dense gas tracer maps and local enhancements in the HCN-to-CO
ratio we identify ten overdensities that represent good candidates to be considered ``starburst giant molecular clouds.'' We
characterize these in the context of a large sample of GMCs in other galaxies and show them to be highly turbulent with high densities
and surface densities compared to GMCs in more quiescent systems.

We complement this study of the individual peaks with a ``zoomed out'' view of the starburst as a whole. We characterize
the major and minor axis exents and use simple modeling to attempt to constrain the three dimensional geometry. We 
find the dense gas tracers to be embedded in a smoother, more extended distribution of bright CO emission. Considering
the nucleus as a single structure, we find average surface density far in excess of a typical disk galaxy GMC and high volume
densities. If the material in the nucleus was arranged into a smooth rotating disk, it would be marginally stable against collapse
($Q \sim 1$) and our surface densities and minor axis profiles can plausibly arise from a system in a semblance of vertical hydrostatic 
equilibrium. 

Tables \ref{tab:sketchclouds}, \ref{tab:structdisk}, and  \ref{tab:sketchdisk} summarize the key length, mass, and time scales
for both the cloud and the disk view, contrasting them with typical disk galaxy values. We highlight the following specific conclusions
from the ``cloud'' view of NGC 253.

\begin{enumerate}

\item Ten overdensities of dense gas can be seen in the data cubes of high critical density tracers. The same peaks emerge
naturally from maps of HCN-to-CO, arguing for their identification as peaks of dense gas.  We measure the properties of
these peaks --- size, luminosity, and line width --- from bright tracers of dense gas and estimate their masses from optically thin
line and continuum tracers.

\item Comparing the properties  of these clouds to a large compilation of literature GMCs, we find the NGC 253 clouds
to have sizes comparable to giant molecular clouds seen in other systems, $r \sim 30$~pc, but very high line widths, 
$\sigma \sim 20$~km~s$^{-1}$ and masses $\sim 2 \times 10^7$~M$_\odot$, far in excess of any spiral galaxy GMC
though likely comparable to the Jeans mass in the nuclear disk.

\item The high line widths do not require external pressure; given the masses of the clouds, they are consistent with the 
expectations for virialized objects. The remarkable thing about the NGC 253 clouds, if real, is their exceptional surface
density, $\sim 6,000$~M$_\odot$~pc$^{-2}$ and volume density, $\sim 2,000$~cm$^{-3}$ (corresponding 
to a free fall time $\tau_{\rm ff} \sim 0.7$~Myr), both of which are found only in the densest, immediately star-forming 
parts of nearby clouds. These values are even higher if we consider only the four brightest, most massive clouds.

\item The low crossing and free-fall times of the clouds compared to those found in disk galaxies offers a simple explanation
of the higher normalized rate of star formation (shorter gas depletion time) in the NGC 253 nucleus. The high Mach numbers
may also have interesting implications for both the efficiency of star formation in dense gas and, very speculatively, the
stellar output of the starburst.

\item Though uncertain, the implied CO-to-H$_2$ conversion factor in the clouds, $\alpha_{\rm CO}$, appears nearly
Galactic. This contrasts with several lines of evidence that point to a lower, ``starburst'' conversion factor in the nuclear 
region overall. We interpret this as resolved evidence for multiple phases of molecular gas mixing to produce the 
observed emission from starburst galaxies.
\end{enumerate}

We also highlight the following conclusions from the nuclear starburst.

\begin{enumerate}

\item Dense gas traced by HCN, HCO$^{+}$, and CS is concentrated into a nuclear disk with dimensions on sky of
$\sim 200 \times 50$~pc. This disk of dense gas is embedded in a larger distribution of CO emission with size
$\sim 300 \times 100$~pc and showing a less clumpy distribution, presumably as a result of optical depth effects.
The geometry of this nuclear region is ambiguous due to the high inclination of the system, but simple
models indicate that the structure at large radii must be non-axisymmetric or we would observe more emission along
the minor axis.

\item The average surface density over the inner part of the starburst ($r_{50} \approx 150$~cp) is 
$\sim 2,500$~M$_\odot$~pc$^{-2}$. Over this area, the average number density is $n_{\rm H2} \gtrsim 300$~cm$^{-3}$. 
As a result of these high values, the timescales in this nuclear disk are short: the vertical crossing 
time and free fall time are $\sim 3$~Myr. The orbital time, estimated from the rotation curve is $\sim 7$~Myr at $r_{50}$. 
Each of these timescales is $\sim 10\times$ shorter than the corresponding value in the disk of a spiral galaxy. 
 
 \item The geometry of the system complicates a comparison to analytic models, which often assume a disk-like shape. Still,
 if all the material within $r_{50}$ is arranged into a disk, that disk would be marginally stable, $Q \sim 1$, and its 
 structure would conform approximately to expectations from simple models. Modulo geometric uncertainties, the star 
 formation rate is comparable to the self-regulation prediction, and the star formation efficiency per free-fall time is similar 
 to that in a wide range of other systems.
 \end{enumerate}
 
The ALMA view of a nuclear starburst thus gives us a concrete picture of the mass, length, and time scales involved in
the nuclear starburst. It highlights the degree to which starbursts of the sort found in NGC 253 represent a physically distinct regime from the 
more quiescent star formation found in the Solar Neighborhood. It also shows the power of ALMA to resolve the
complex environments at the centers of galaxies into multiple phases and discrete structures.\\

\acknowledgments We thank the anonymous referee for a constructive report that led to substantial improvements
to the paper. We are very grateful to Kazushi Sakamoto for sharing the SMA data and providing
helpful feedback on a draft of the paper. AKL thanks the NRAO/University of Virginia star formation group (including
Crystal Brogan, Remy Indebetouw, Kelsey Johnson, Amanda Kepley, and Scott Schnee), Jeff Mangum, John Hibbard,
Kartik Sheth, Alison Peck, and Mark Heyer for useful discussions. The National Radio Astronomy Observatory is a facility of the National 
Science Foundation operated under cooperative agreement by Associated Universities, Inc. 
ADB acknowledges support by the National Science Foundation through a CAREER grant AST-0955836, as 
well as a Cottrell Scholar award from the Research Corporation for Science Advancement.
The work of ECO is supported by grant AST-1312006 from the National Science Foundation. 
EWR is supported by a Discovery Grant from the Natural Sciences and Engineering Research Council Canada.
Support for this work was provided by the National Science Foundation through grant AST-1009583 to SV. 
This paper makes use of the following ALMA data: ADS/JAO.ALMA \#2011.0.00172.S, \#2012.1.00108.S. ALMA is a 
partnership of ESO (representing its member states), NSF (USA) and NINS (Japan), together with 
NRC (Canada) and NSC and ASIAA (Taiwan), in cooperation with the Republic of Chile. The Joint 
ALMA Observatory is operated by ESO, AUI/NRAO and NAOJ.


\end{document}